\documentclass[a4paper,10pt]{article}

\usepackage[utf8]{inputenc}
\usepackage[T1]{fontenc}
\usepackage{xcolor,graphicx}
\usepackage[top=1in,bottom=1in,right=1in,left=1in]{geometry}

\usepackage{natbib}
\usepackage[english]{babel}
\usepackage{hyperref}
\usepackage[normalem]{ulem}
\newcommand\aap{A\&A}                % Astronomy and Astrophysics
\newcommand\aaps{A\&AS}              % Astronomy and Astrophysics Supplement Series
\newcommand\apj{ApJ}                 % Astrophysical Journal
\newcommand\apjl{ApJ}                % Astrophysical Journal, Letters
\newcommand\apjs{ApJS}               % Astrophysical Journal, Supplement
\newcommand\mnras{MNRAS}             % Monthly Notices of the Royal Astronomical Society
\newcommand\pasa{Publ. Astron. Soc. Australia}  % Publications of the Astronomical Society of Australia
\usepackage{authblk}
\usepackage{multicol}
\usepackage{hyperref}
\usepackage{graphicx}
\usepackage{amsmath}
\definecolor{blue-violet}{rgb}{0.54,0.17,0.89}
\hypersetup{
    colorlinks=true,
    linkcolor=blue,
    filecolor=magenta,      
    urlcolor=blue,
    pdftitle={Overleaf Example},
    pdfpagemode=FullScreen,
    citecolor=blue
    }
\begin{document}
\begin{titlepage}
% \pagecolor{blue!10}
\begin{center}
	\begin{minipage}{15cm}
	\begin{center}
		\includegraphics[width=2cm,height=2.2cm]{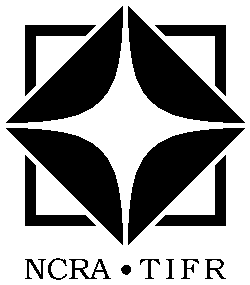}
	\end{center}
\end{minipage}\hfill\hfill\vspace{1cm}
%\begin{minipage}[c]{15cm}
	\begin{center}
	\textbf{\Large \color{blue-violet} National Centre for Radio Astrophysics}\\[0.1cm]
	\textbf{\large \color{blue-violet}  Tata Institute of Fundamental Research}
	\end{center}
%\end{minipage}\hfill
\vspace{0.6cm}
%\begin{minipage}[c]{10cm}
	%\begin{center}
\small Internal Technical Report: xxxx
	%\end{center}
%\end{minipage}

\vspace{5cm}
%\includegraphics[width=0.6\textwidth]{logo-isae-supaero}\\[1cm]

% Title

{ \Large \textbf{A detailed study of  the polarisation convention of the Giant Metrewave Radio Telescope (GMRT) \\[0.4cm]} }

{\large Poonam Chandra, S. Suresh Kumar,   Sanjay Kudale,   Devojyoti Kansabanik, Barnali Das, Preeti Kharb,   Silpa Sasikumar \& Biny Sebastian}\\[0.5cm]
\textit{Corresponding Author's email id:
 \textsf{\color{blue}pchandra@nrao.edu}
%\textsf{\color{blue}pchandra@nrao.edu}
and \textsf{\color{blue}poonam.chandra@gmail.com} }\\[1cm]
%Dec 5, 2022\\
2023\\

\vspace{5cm}
\begin{minipage}{20cm}
{\large \textbf{Objective: To check the polarisation convention of the GMRT and understand\\
 its causes and implications on the scientific outcome}} \\[0.5cm]
\end{minipage}{}

\vfill

\end{center}
\end{titlepage}
\normalsize

\section{Defining the problem}\label{sec:problem}
This work aims to investigate the polarisation convention of the u/GMRT and understand whether these follow the standard IAU/IEEE convention. These tests have been done on several strong and highly polarised pulsars with known polarisation properties at GMRT wavelengths. Before going into the tests and experiments, will discuss some basic ideas related to polarisation, polarisation convention, and polarisation calibration in Section \ref{sec:intro}.

\section{Introduction to Polarimetry}\label{sec:intro}
\subsection{Polarisation of electromagnetic radiation}

Electromagnetic (EM) waves from astronomical objects have preferred orientation or rotation of their electric field vectors; known as polarisation. The phenomenon of polarisation arises as a consequence of the fact that EM wave behaves as a two-dimensional transverse wave. Polarisation is a fundamental property of EM radiation and provides a rich source of information on the physical properties, such as magnetic fields, asymmetry in the astronomical phenomenon, etc. In radio bands, many processes can produce polarisation, which can be either circular, linear  or elliptical in nature. Generally, radio sources have low fractional polarisation, but some coherent emission mechanisms may produce up to 100\% polarisation.

The polarisation of an EM wave is defined by the motion of its electric field vector as a function of time within a plane perpendicular to the direction of propagation. This plane is known as the plane of polarisation. If the EM radiation propagates in the $z$-direction, the plane of polarisation will be the $xy$ plane, and  the electric field $\boldsymbol{E}$ in trigonometric notation can be written as
\begin{equation}
    \boldsymbol E(t,z)=\boldsymbol{E}(0,0) \cos (\omega t -k z - \phi)
    \label{eq:1}
\end{equation}
where $t$ is the time, $\omega$ is the angular frequency, $k = \omega/c$  is the wave vector, and $\phi$ is an arbitrary phase. The  $ \boldsymbol E(t,z)$ can be decomposed in two orthogonal components ($x$ and $y$).%{\sout{components}}.
\begin{equation}
\begin{split}
    E_x(t)&=E_x(0) \cos (\omega t - \phi_1)\\
    E_y(t)&=E_y(0) \cos (\omega t - \phi_2) 
\end{split}
\label{eq:2}
\end{equation}

The above notations are trigonometric notations of the electric field. They can also be written in complex notations as,
\begin{equation}
    \begin{split}
    E_x(t)&=E_x(0) e^{i(\omega t - \phi_1)}\\
    E_y(t)&=E_y(0) e^{i(\omega t - \phi_2)}
    \end{split}
    \label{eq:3}
\end{equation}

The angle between the plane of oscillation of the electric vector and the plane of polarisation is called the polarisation angle (PA). For elliptical polarisation, PA is the position angle of the major axis of the polarisation ellipse (see Fig. \ref{fig:PA1}).

\subsection{Linear and circular polarisation}
Polarisation of a EM wave can be defined by $E_x(0)$, $E_y(0)$, $\phi_1$ and $\phi_2$. The tip of the electric field vector generally follows an elliptical trajectory in the $xy$ plane; accordingly, the wave is denoted as elliptically polarised. Hence elliptical polarisation is the most general state of polarisation and linear and circular polarisation are special cases.

Linear polarisation occurs when the polarisation ellipse degenerates into a line, that is the phase difference $\delta =\phi_2-\phi_1=0$. For $\phi_1=\phi_2=0$, linear polarisation (without any loss of generality) is 
\begin{equation}
    \begin{split}
        E_x(t)&=E_x(0) \cos (\omega t )\\
        E_y(t)&=E_y(0) \cos (\omega t  )
    \end{split}\label{eq:4}
\end{equation}

The orientation of the ellipse in the plane of polarisation $xy$ plane depends only on ratio of $E_x(0) $ and $E_y(0)$, which are constant in time. Thus in other words the polarisation angle is constant with time for linear polarisation.

In circular polarisation, the EM field vector has a constant magnitude and is rotating at a constant rate in a plane perpendicular to the direction of the wave. Thus circular polarisation occurs when the polarisation ellipse degenerates into a circle, that is  $\phi_2=\phi_1\pm \pi/2$. For $\phi_1=0$, circularly polarised light is,
\begin{equation}
    \begin{split}
    E_x(t)&=E_x(0) \cos (\omega t )\\
    E_y(t)&=\pm E_y(0) \sin (\omega t  )
    \end{split}
    \label{eq:5}
\end{equation}

The tip of the electric field vector moves circularly in the $xy$ plane with a frequency, $\omega$, and the sign of $E_y(t)$ determines the sense of polarisation. A circularly polarised wave can rotate in one of two possible senses: clockwise or counterclockwise.

\begin{figure}
\centering
\includegraphics[width=0.45\textwidth]{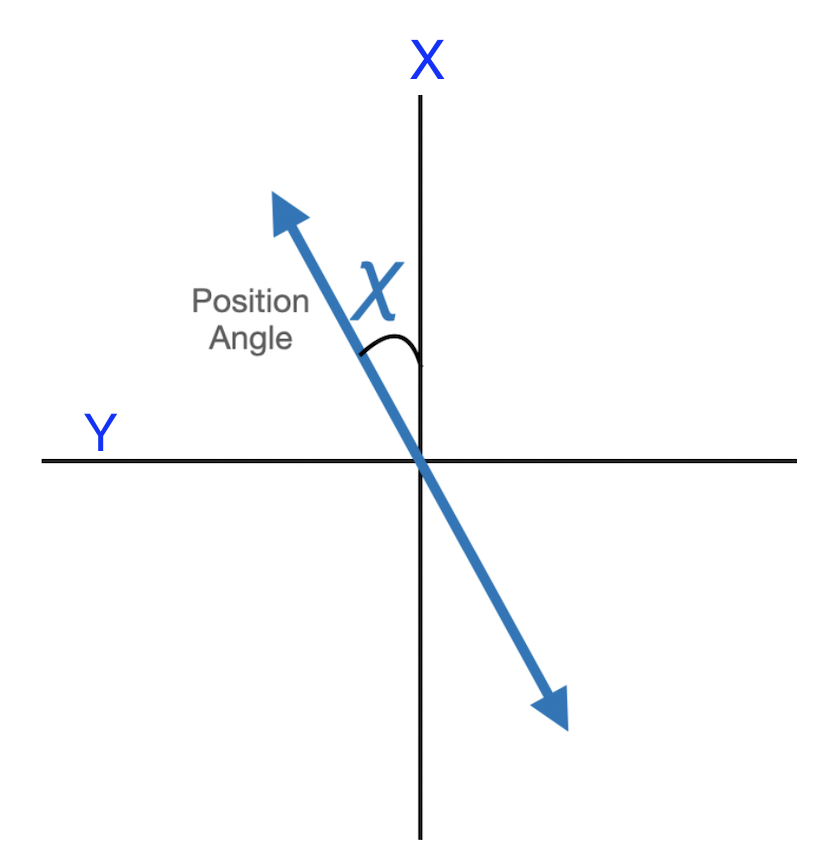}
\caption{Position angle of polarised light. X represents towards North and Y represent towards East.}
\label{fig:PA1}
\end{figure}

Polarisation angle of the plane of linear polarisation on the sky is measured from North towards East (Fig  \ref{fig:PA1}). In this figure, the major axis of the polarisation ellipse is oriented at an angle $\chi$  with respect to the  $X$ axis ($X$ axis represents towards North). The polarisation angle $\chi$ is zero at north; and $\chi$  positive while measured from North towards East. Thus represented on an image of the sky, a line segment representing polarisation rotates counterclockwise as $\chi$ increases, and $\chi=0^o$ corresponds to a vertical orientation \footnote{Researchers studying the polarisation of the Cosmic Microwave Background (CMB) have been defining polarisation angle to increase clockwise on the sky. This effectively swaps the sign of Stokes U and confuses astronomers studying Galactic polarisation using CMB satellite data.}.

\subsection{Stokes parameters}
The polarisation state of an EM wave is {\sout{be}} characterised by the amplitudes $E_x (0)$ and $E_y (0)$ and the phase difference $\delta = \phi_2 - \phi_1$. In general, astronomical observations deal with light intensities rather than with field amplitudes, thus it is convenient to quantify polarisation via characteristic intensities, i.e. the  Stokes parameters  $I$, $Q$, $U$, and $V$. Here $I$ is the total intensity, $Q$ and $U$ represent linear polarisation, and $V$ quantifies the amount of circular polarisation.

Stokes parameters are related as,
\begin{equation}
    I^2=Q^2+U^2+V^2
    \label{eq:6}
\end{equation}

The linearly polarised intensity $L$ can be defined as 
\begin{equation}
L=\sqrt{(Q^2+L^2)}
\label{eq:7}
\end{equation}

The polarisation angle $\chi$, is defined in terms of Stokes Q and U as,
\begin{equation}
\chi=\frac{1}{2} \tan^{-1} \left(\frac{U}{Q}\right); \,\, 0^o \le \chi \le 180^o
\label{eq:8}
\end{equation}

The Stokes parameters are defined in terms of auto and cross-correlations of the $x$ and $y$ components of the EM field vectors, i.e. $E_x(t)$, $E_y(t)$ in a Cartesian frame whose $z$ axis is along the direction of propagation. In trigonometric co-ordinate system  Stokes parameters are defined as,
\begin{align}
    I&=\langle|E_x|^2+|E_y|^2\rangle \label{eq:I1}\\
    Q&=\langle|E_x|^2 - |E_y|^2\rangle \label{eq:Q1}\\
    U&=2\langle|E_x||E_y| \cos \delta \rangle \label{eq:U1}\\
    V&=2\langle|E_x||E_y| \sin \delta \rangle \label{eq:V1}
\end{align}

In complex exponential notation defined in Equations \ref{eq:2} and \ref{eq:3}, the Stokes parameters  are defined as,
\begin{align}
    I&=\langle E_x E_x^*+ E_y E_y^*\rangle \label{eq:I2}\\
    Q&=\langle E_x E_x^* -  E_y E_y^*\rangle \label{eq:Q2}\\
        U&=\langle E_x E_y^* + E_y E_x^*\rangle \label{eq:U2}\\
    V&= -i[\langle E_x E_y^* -  E_y E_x^*\rangle ] \label{eq:V2}
\end{align}

\subsection{Jones and Muller matrices}
Polarised light can be represented by a Jones vector, $\mathbf{e}$, and linear optical elements can be represented by Jones matrices, $J$. More precisely, the Jones vector represents the amplitude and phase of the electric field in the $x$ and $y$ directions.
One can then define the Jones vector as,
\begin{equation}
\mathbf {e}\equiv \begin{bmatrix}
E_x(0) e^{i \phi_1}\\
E_y(0) e^{i \phi_2}
\end{bmatrix}
\label{eq:9}
\end{equation}

For example, the Jones vector for linear polarised light in the $x$ direction is $\begin{pmatrix} 1\\ 0 \end{pmatrix}$, Jones vector for linearly polarised at 45° from the $x $-axis is $\frac{1}{\sqrt{2}}\begin{pmatrix} 1\\ 1 \end{pmatrix}$, and Jones vector for right-hand circular polarised light is $\frac{1}{\sqrt{2}}\begin{pmatrix} 1\\ -i \end{pmatrix}$.

The Jones matrices are $2\times2$ linear matrix operators that represent the different instrumental and atmospheric effects. The measured Jones vectors then can be written as $\mathbf{e_{out}J\mathbf{e}}$.
 
Jones vectors and Jones matrices can be written in a four-vector format. Every Jones vector has its  four-vector counterpart in terms of Stokes parameters, ${\mathbf S}=\begin{pmatrix} I \\ Q\\U\\V\end{pmatrix}$. The instrumental response then represented in terms of a $4\times4$ matrix, known as Muller matrix ($M$). The Mueller matrix describes the modification of the Stokes vector, $\mathbf{S}$. The Mueller matrix relates the output Stokes vector, $\mathbf{S_\mathrm{out}}$ to the input Stokes vector, $\mathbf{S}$, as ${\mathbf S_{\rm out}}={\mathbf  M}. {\mathbf S_{\rm in}}$.
  
\subsection{Parallactic angle}
\begin{figure*}
\centering
\includegraphics[width=0.45\textwidth]{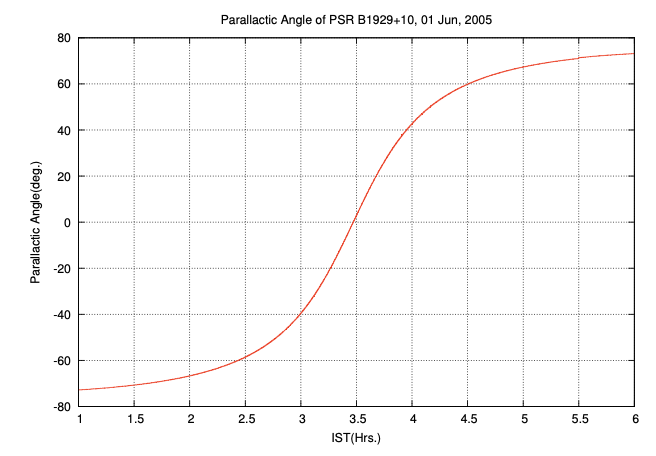}
\caption{Variation of parallactic angle with time for the source B1929+10 from rising to set. (Figure taken from MSc thesis of Sanjay Kudale).}
\label{fig:parallactic}
\end{figure*}

For the case of an alt-azimuth mounted telescope, the feed dipoles rotate in the plane of the sky, and hence with respect to the plane of polarisation of the incoming radiation. This rotation is characterized by an angle called the parallactic angle, $\psi_p$ given as, 
\begin{equation}
    \tan \psi_p = \frac{\cos \mathcal{L} \sin \mathcal{H}}{\sin \mathcal{L} \cos \delta - \cos \mathcal{L} \sin \delta \sin \mathcal{H}}
    \label{eq:parallactic_angle}
\end{equation}
where, $\mathcal{L}$, $\mathcal{H}$  are latitude and hour-angle of the telescope, respectively and $\delta$  is the declination of the source. As the telescope tracks a source across the sky the parallactic angle changes (Figure \ref{fig:parallactic}).

\subsection{IAU/IEEE Convention of polarisation}
Unfortunately, there are multiple conventions of polarisation based on coordinates systems and Stokes parameters. To alleviate this ambiguity, the Institute of Electrical \& Electronics Engineers (IEEE) defined a universal standard and published it in 1969. In 1974, the International Astronomical Union (IAU) endorsed the IEEE standard and supplemental with the definitions of Cartesian coordinates and the sign of Stokes parameter V.% {\color{red}(ref)}.

\begin{figure}
\centering
\includegraphics[width=0.44\textwidth]{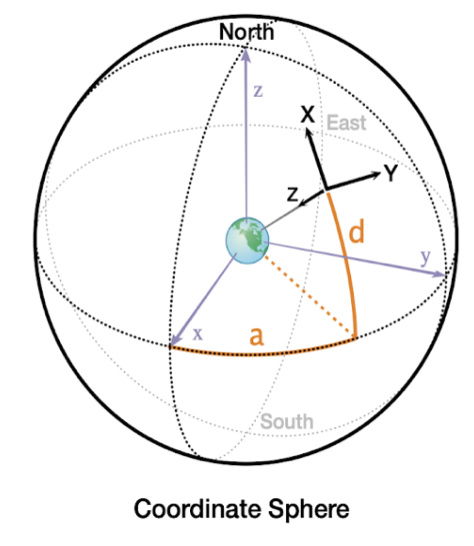}
\includegraphics[width=0.44\textwidth]{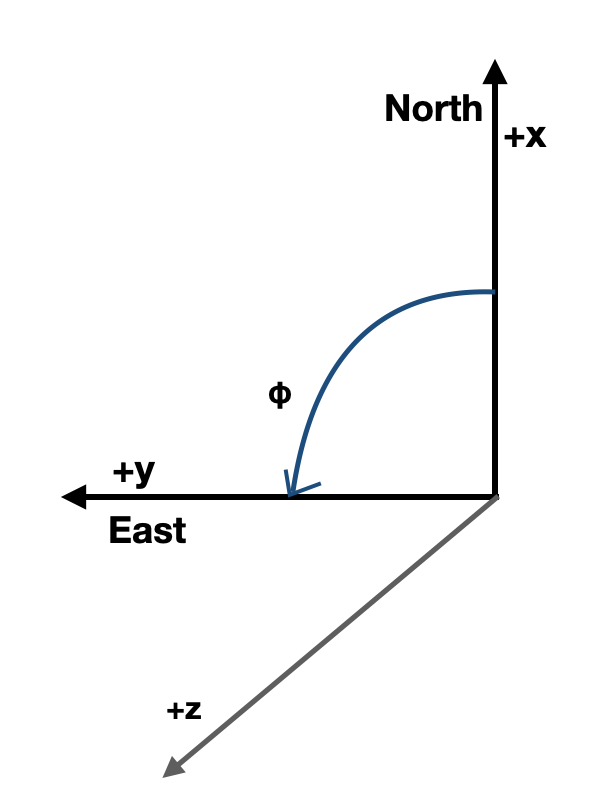}
\caption{Left figure shows the coordinate sphere and $XYZ$ right-handed coordinate system defined by the IAU. The perspectives can be different based on whether seen from the position of the source or the position of the observer. The right figure shows the right-handed coordinate system seen from the position of the observer. Here the radiation from the source propagates in the positive $Z$ direction.}
\label{fig:iau2}
\end{figure}

In the polarisation convention defined by the IAU, the local $X$-axis points towards North, the local $Y$-axis points towards East, and the local $Z$-axis points inwards to the observer for a right-handed system (Left panel of Figure \ref{fig:iau2}).

The position angle is measured starting from the North direction and going towards the East, i.e. increasing counter-clockwise, when looking at the source (Figure \ref{fig:PA1}). The orientation of the ellipse in the plane of polarisation $XY$ plane is constant in time; the polarisation angle $\chi$ corresponds to the angle between the positive $X$ axis and the semi-major axis of the ellipse (counted in the counterclockwise direction)
\footnote{The convention astronomers follow for the PA (Polarization Angle) goes back to the 19th century and it has been in use for observations going from radio to gamma rays: the PA increases counter-clockwise when looking at the source. This convention is consistent with the one used for the Position Angle and it has been enforced by the IAU with a Resolution by Commissions 25 and 40 at the IAU XVth General Assembly in Sydney in 1973 (see Transactions of the \href{https://www.iau.org/static/resolutions/IAU1973_French.pdf}{IAU, Vol. XVB, pg. 166}).}

\begin{figure}[h]
\centering
\includegraphics[width=0.55\textwidth]{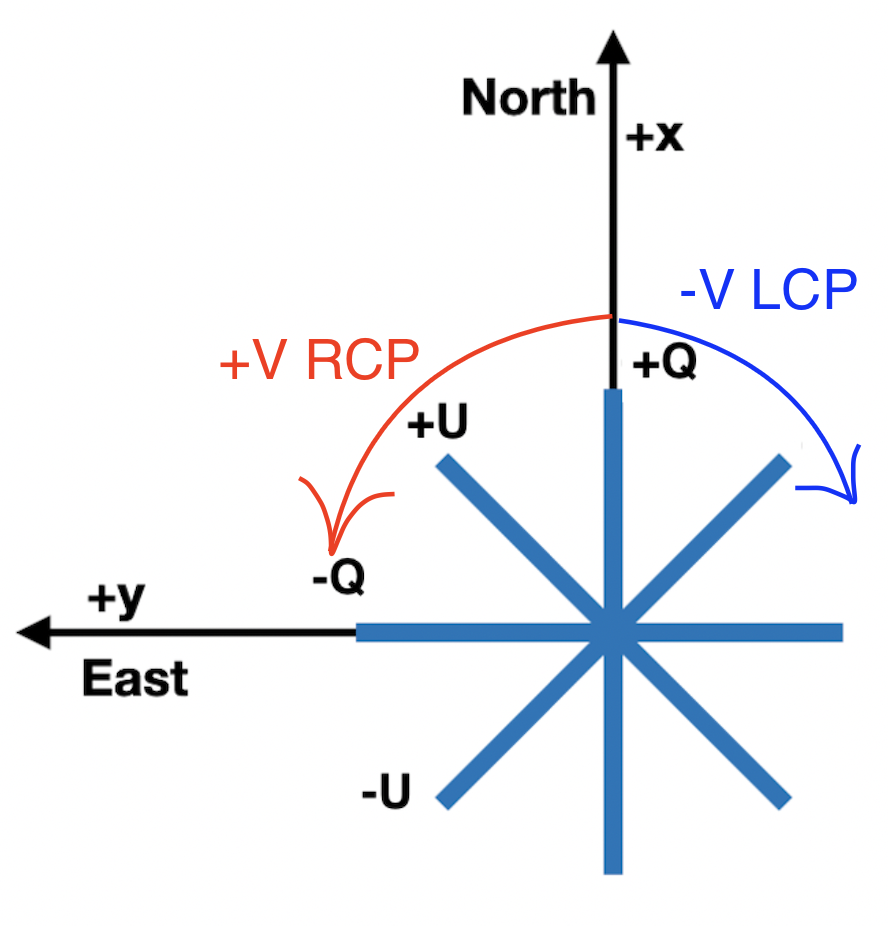}
\caption{The orientation of the electric vectors for Stokes $Q$ and $U$ on the $xy$ plane of the sky. The figure also shows the IAU/IEEE conventions for RCP and LCP and Stokes V.}
\label{fig:iau}
\end{figure}

IAU defines a right-handed coordinate system as shown in Figure \ref{fig:iau}. Once this convention is defined and $XY$ coordinate system is chosen, the Stokes parameters can be  defined unambiguously. In such a case from the Equations \ref{eq:I1} to \ref{eq:V2}, $+Q$ is pointing towards $X$ (E-W) and $-Q$ towards $Y$ (N-S). The $+U$ is 45 degrees from $X$ towards $Y$ (i.e. NE to SW line) and $-U$ towards 45 degrees from $X$ towards $-Y$ (i.e.
NW to SE line) (Figure \ref{fig:iau}).

The handedness of the elliptically polarised light describes the direction of rotation of the electric field vector as seen by this observer. It is more confusing to define Stokes $V$. As per IAU definition, for the {\bf right-handed circular polarisation}, the position angle of the electric vector at any point {\bf increases} with time. This implies that the $Y$ component of the field will lag behind the $X$ component, i.e. $E_y=\mp iE_x$.  In such a case Stokes $V$ is defined as positive for right-handed circular polarisation (IAU/IEEE), 
\begin{equation}
    {\rm Stokes}\,\, V=RR^*-LL^*
    \label{eq:19}
\end{equation}

Another way of putting this is that a right-hand circularly polarised radiation is the one where for the incoming radiation, the electric field vector rotates counter-clockwise. For left-hand circularly polarised radiation, the electric field vector rotates clockwise for the incoming radiation. Here one notes that the electric field vectors along the line of sight, at any instant of time, form a left-handed screw.

However the alternate way of defining Stokes V, i.e. $V=LL^*-RR^*$ is also prevalent, especially among the pulsar community. This convention, where Stokes $V$ is positive for left-hand circularly polarised radiation, is known as the {\bf ``PSR/IEEE''} convention. Both PSR/IEEE and IAU/IEEE conventions require that $RR^*$ and $LL^*$ are defined according to the IEEE convention, it is the definition of Stokes V that distinguishes these two conventions. This needs to be kept in mind for the discussion in the rest of the article.

\subsection{Polarisation for interferometers}
The polarisation ellipse can be decomposed in any orthonormal basis in the plane of polarisation. Generally in radio astronomy, the standard Cartesian linear basis and a basis of circularly
rotating unit vectors of opposite handedness are used.

Radio receivers in radio telescopes are polarimeters by construction. The receiver of a radio telescope is composed of two orthogonal dipoles. Each of these two dipoles receives the corresponding polarisation component and converts it into an electric voltage that can be recorded and processed electronically. The signals received by the dipoles can be auto-correlated as well as cross-correlated, and Stokes parameters can be measured.

The Stokes parameters $I$, $Q$, $U$, and $V$ are derived from the measured voltages (coming from the celestial object) recorded by two receivers in the telescope corresponding to two orthogonal polarisations: $X$ and $Y$ (representing $E_x$ and $E_y$, respectively) for telescopes with dipolar feeds; or $R$  and $L$ (representing $E_R$ and $E_L$, respectively). For IAU/IEE convention, the $R$ and $L$ are related to $X$ and $Y$ as,
\begin{equation}
    \begin{split}
        R=\frac{(X+iY)}{\sqrt{2}}\\
        L=\frac{(Y-iX)}{\sqrt{2}}
    \end{split}
    \label{eq:20}
\end{equation}

This implies: 
  \begin{align}
    I&=XX^*+YY^*\equiv RR^*+LL^*\label{eq:I}\\
    Q&=XX^*-YY^*\equiv RL^*+R^*L\label{eq:Q}\\
    U&=XY^*+X^*Y\equiv -i(RL^*-R^*L)\label{eq:U}\\
    V&=-i(XY^*-X^*Y)\equiv RR^*-LL^*\label{eq:V}
\end{align}

Here $X$ and $Y$ correspond to linear dipoles and $R$ and $L$ correspond to circular dipoles.

\section{The u/GMRT}\label{sec:uGMRT}
The GMRT antennas are 45m alt-azimuth mounted parabolic prime-focus dishes, with an elevation limit set to 17.5$^o$ (though the dishes can go down to 16$^o$). The different feeds for different observing bands are mounted on a rotating turret placed at the primary focus. The low noise receiving system of GMRT receives dual polarisations. For GMRT all feeds are linearly polarised. They are then converted to circular feeds for bands 2, 3 and 4. The L-band feed is kept as linearly polarised. The GMRT antennas have RF swap switch at the end of common box, which when set in swap mode, interchange the signal path with opposite polarisation electronics. Thus the polarisation channels can be swapped whenever required.

\begin{figure*}
\centering
\includegraphics[width=0.48\textwidth]{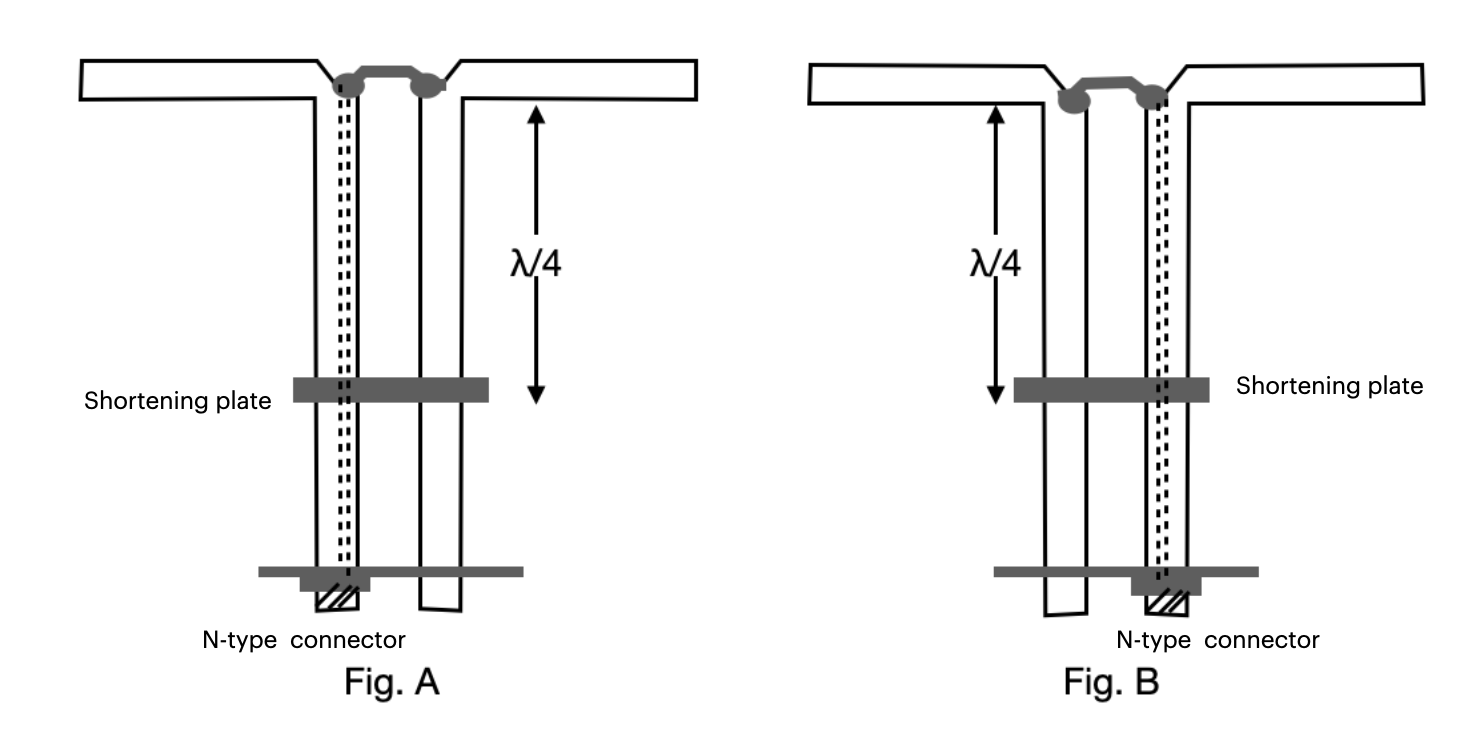}
\includegraphics[width=0.36\textwidth]{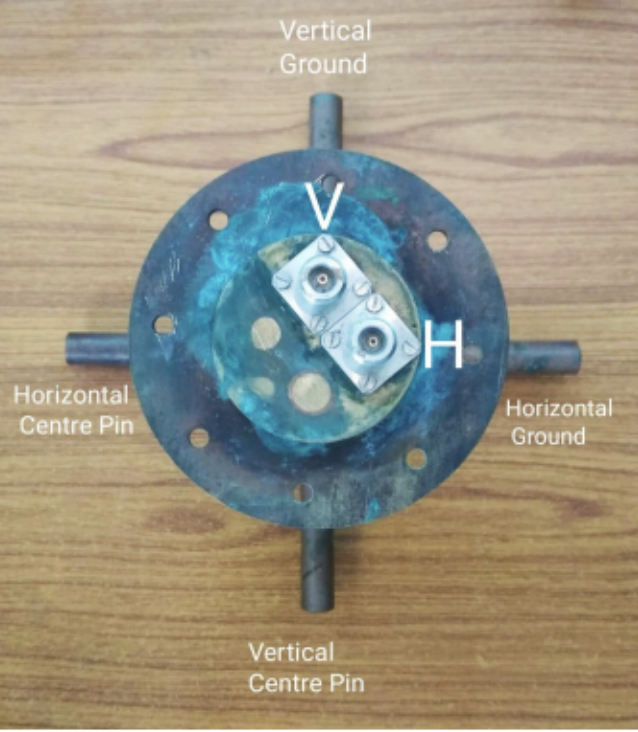}
\includegraphics[width=0.76\textwidth]{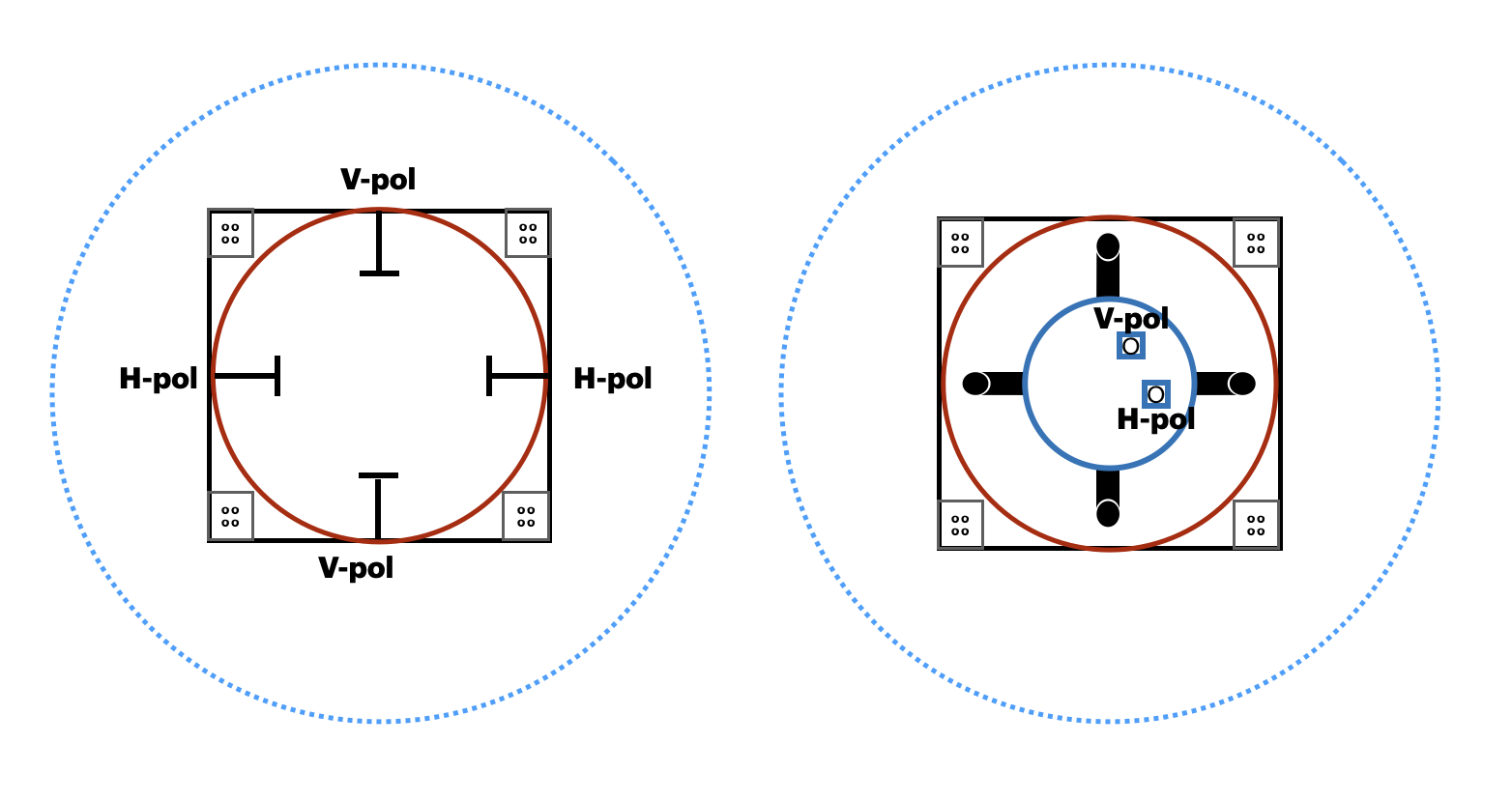}
\caption{{\textit Top Left:} The feed dipoles. One dipole element is connected to the center connector of the RF connector through a rod and the other is connected to the
ground of the RF connector. The Balun built with the feed is has a $\lambda/4$ short. {\textit Top Right:}  Back view of the cross-dipole.  {\textit Bottom:} View of the co-axial feed and the cross dipole feed focusing towards the dish (dashed blue circle). This is same for bands 2, 3 and 4.}
\label{fig:balun}
\end{figure*}

Each antenna has  two polarisers  named as V and H, with a 90 degree phase difference. For bands 2, 3 and 4, one dipole element is connected to the center of the RF connector through a rod and the other element is connected to the
ground of the RF connector. The Balun has a shortening plate at $\lambda/4$ to create 90 degrees phase difference. These are shown in Figure \ref{fig:balun}.

\begin{figure*}
\centering
\includegraphics[width=0.44\textwidth]{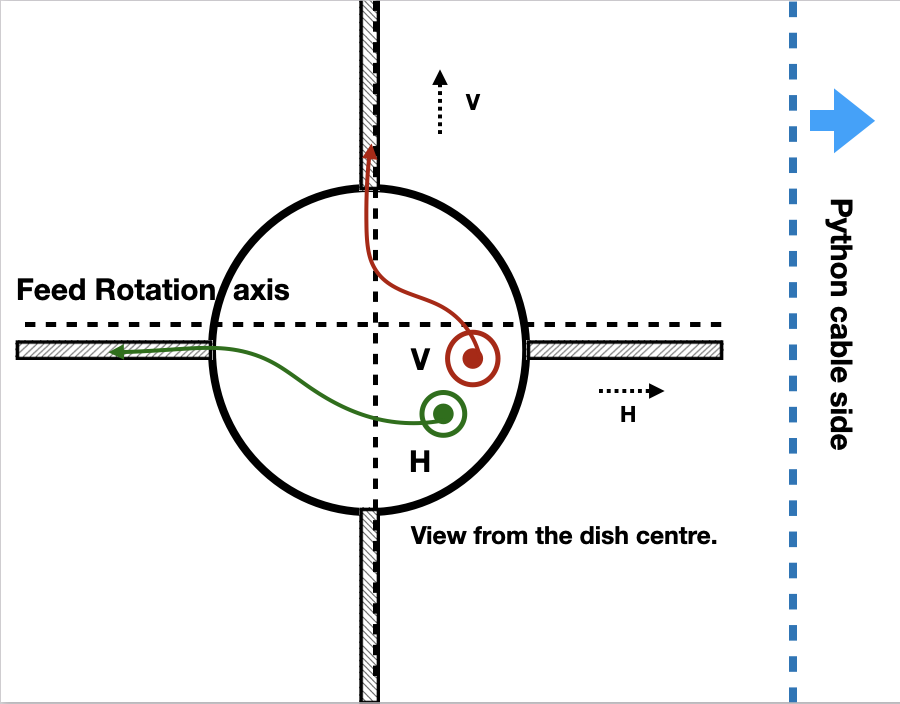}
\includegraphics[width=0.43\textwidth]{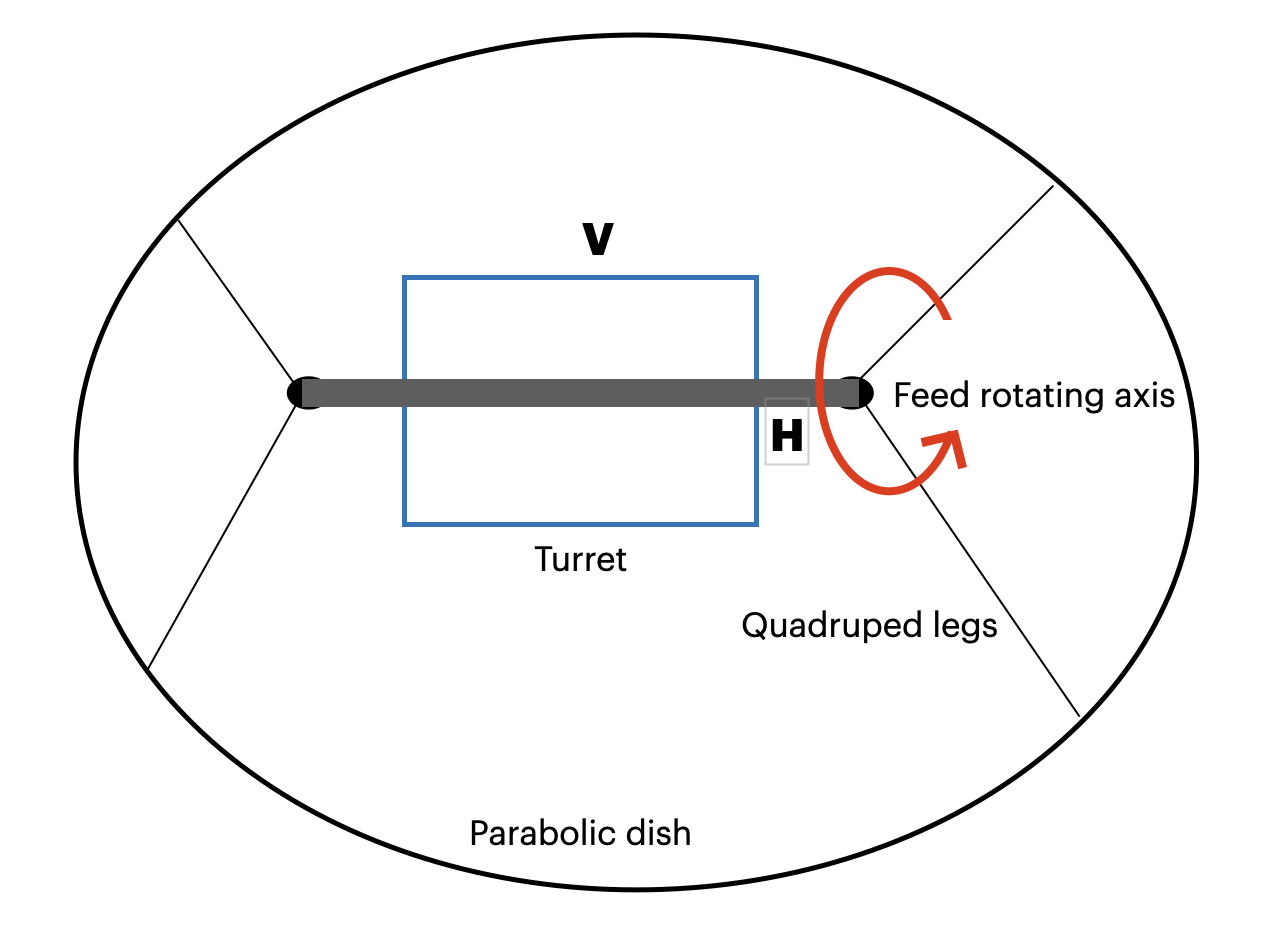}
\caption{{\textit Left:} This figure shows the cross dipole. There are two connectors V and H, corresponding to vertical and horizontal component, respectively. The connector and element marked as H goes parallel to the rotating axis of FPS system. The marking on the feed is aligned to the python side which is H input.  {\textit Right:} The feed connection,  if viewed from top of the dish.}  \label{fig:dipole}
\end{figure*}

The figures \ref{fig:balun} and \ref{fig:dipole} show the cross dipole feed with the dipole output connectors marked as V and H. The V is connected to the Vertical element and H to the horizontal element. The V connector centre pin is connected to the lower arm of the vertical dipole. The H connector centre pin is connected to the horizontal dipole arm on the opposite side of the connector. The connector side is placed near the python cable. The  dipole arms for Horizontal polarisation is mounted parallel to the Feed Position System (FPS) rotating axis and the Vertical dipole arms is orthogonal to the FPS rotating axis at prime focus of the parabolic dish. This convention is followed in all bands and at all antennas of GMRT array both for legacy and upgraded systems of GMRT. 

In summary the H dipole element is parallel to FPS rotating axis and V dipole element is perpendicular to FPS rotation axis. The feed to hood connecting cable is of equal length and could be used to swap the cables if needed, which can give polarisation swap (Figure
\ref{fig:dipole}).
 
\begin{figure*}
\centering
\includegraphics[width=0.43\textwidth]{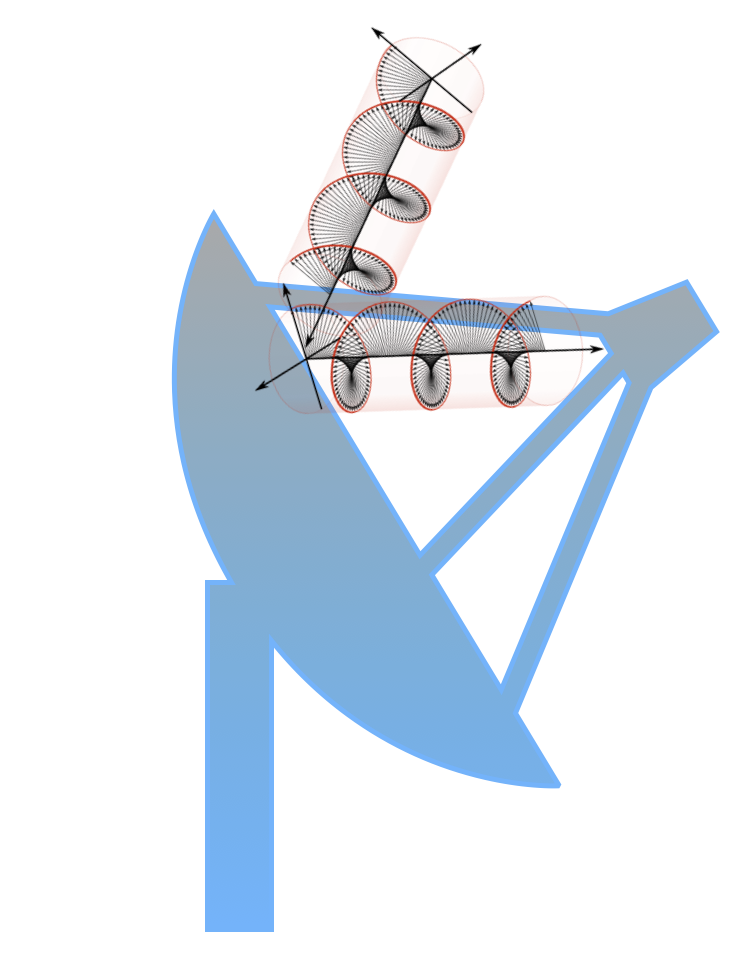}
\caption{This figure shows the configuration of the GMRT feed. The radiation from a celestial source first falls on the dish  and then reaches the feed after the reflection from the antenna dish. This will   cause a reversal of the circular polarisation } \label{fig:primefocus}
\end{figure*}

{\bf GMRT antennas are prime focus antennas, i.e. the radiation falling on the feed reverses its circular polarisation. We show this effect in Fig \ref{fig:primefocus}. If this reflection is not taken into account, it will result in a reversal of circular polarisation. We need to keep this in mind.}

\section{Initial Test Observations}\label{sec:initial}
\subsection{Target properties}
The aim of  these test observations is to investigate the polarisation convention of the u/GMRT and understand whether these follow the standard IAU/IEEE
convention   mentioned above. The results of these test observations has already been reported as NCRA technical report (REF. Barnali's report) and re-described here for completeness purpose. To  test the convention we choose a test source and carry out uGMRT observations in  both the interferometric and the phase-array mode. The chosen source is pulsar B1702--19 (RA: $17^h05^m36^s.099$, Dec: $-19^d06^m38^s.60$, B1950), with a relatively higher degree of circular polarisation ($\approx 40\%$) as per the pulse profiles from EPN database \footnote{ \url{http://www.epta.eu.org/epndb/\#gl98/J1705-1906/gl98_610.epn}}. 
%{\sout{and the ATNF Pulsar catalogue\footnote{ \url{https://www.atnf.csiro.au/research/pulsar/psrcat/}}.}
% ATNF catelogue does not provide polarisation information, hence I remove it.

We also look at the published 1369 MHz Stokes V profile of the pulsar with the Parkes radio telescope \citet{johnston2017} and 1408 MHz profile with the Lovell telescope \citet{gould1998}. Parkes Telescope explicitly mentions that the telescope follows the PSR/IEEE convention (According to PSR/IEEE convention Stokes V is the difference of the Left circular polarisation (LCP) and right circular polarisation (RCP) and the sign is positive if LCP is higher than RCP, i.e. V$=$LCP -- RCP \footnote{See Section \ref{sec:intro} for more details of the definitions}.). Since the signs match in the published data with both the  Parkes and Lovell telescopes, it indicates that the convention adopted by Lovell telescope is  same as Parkes telescope, i.e. the PSR/IEEE convention. As the EPN database uses Lovell telescope data for this pulsar, we infer that the convention used for Stokes $V$ in the EPN database is also V$=$LCP -- RCP. 

\begin{figure}
\centering
\includegraphics[width=0.47\textwidth]{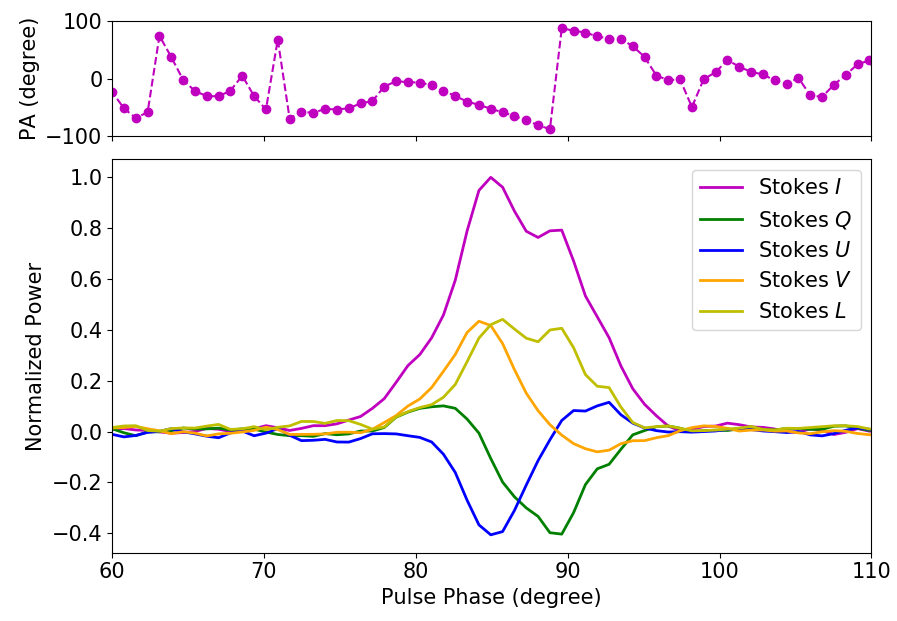}
\includegraphics[width=0.49\textwidth]{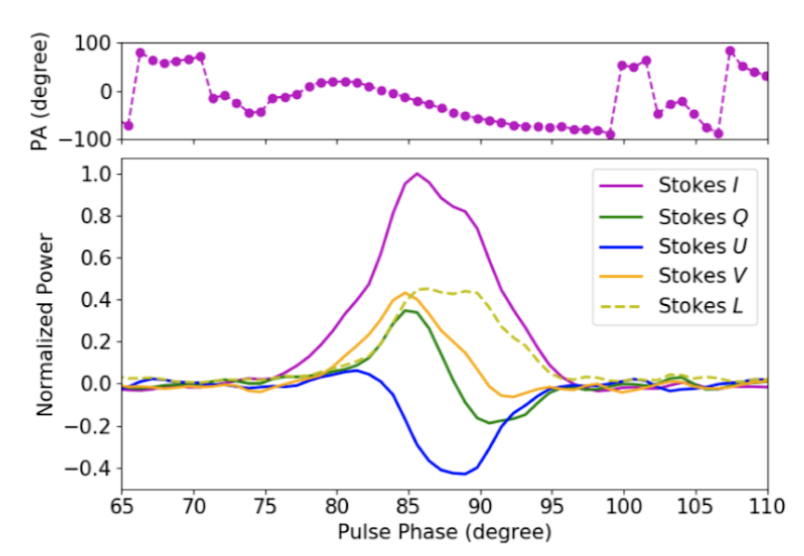}
\caption{{{\bf Left:} The 610 MHz profiles for the pulsar B1702--19 for Stokes $I$ (magenta curve), $Q$ (green curve), $U$ (blue curve) and $V$ (orange curve) obtained by \citet{gould1998} using the Lovell telescope. Also shown are the angle of polarisation (PA) on top panel and linear polarisation $L=\sqrt{Q^2+U^2}$ in yellow in the bottom panel. {\bf Right:} Same as the left, but at 408 MHz  for PSR B1702-19. Note that the convention for Stokes $V$, here, is $V=RR^*-LL^*$ even though the convention used in the original paper \citep{gould1998} was $V=LL^*-RR^*$. This is done to avoid confusion since GMRT convention for defining $V$ is $RR^*-LL^*$. }\label{fig:epn_profile}}
\end{figure}

If we convert the Stokes profiles from PSR/IEEE convention into IAU/IEEE convention (V$=$RCP -- LCP, see Section \ref{sec:intro} for details), then the pulsar is $\approx 40\%$ right circularly polarised. In Figure \ref{fig:epn_profile}, we show the Stokes  $I$, $Q$, $U$, $V$ published profiles at 610 MHz  (left panel) and 400 MHz (right panel) for this pulsar, generated from the data available in the EPN database \citep{gould1998}. To make it consistent with the rest of the report, we follow the uniform IAU/IEEE convention, and Stokes V profile is plotted as,  V$=$RCP -- LCP.

\subsection{uGMRT observations}
We observed PSR B1702--19 in band 4 (550--750 MHz) of the uGMRT on March 19, 2020 with 200 MHz bandwidth, divided into 2048 channels. 3C286 was used as the flux density and bandpass calibrator, and J1822-096 (RA: $18^h22^m28.71^s$, Dec: $-09^d38^m56^s.84$, J2000) was used as the phase calibrator. The on-source time of the pulsar was around 40 minutes. The data were recorded both in the interferometric and the  pulsar phased array mode. 
The same pulsar with the same set of calibrators were also observed in uGMRT band 3 with 200 MHz bandwidth with default settings (first LO 500 MHz, 500--300 MHz) on May 20, 2020.

In the pulsar mode, the array was first phased using the phase calibrator J1822-096. The data were then recorded on the target pulsar in the standard format followed at uGMRT (co- and cross-polar voltage products, viz. $RR^*$, $LL^*$, and real and imaginary parts of $RL^*$ \footnote{For u/GMRT channel 1 is labelled as  $RR^*$ and channel 2 $LL^*$. For details see next section.}) with a time resolution of 327.68 microseconds. 
Since the rotation period of the target pulsar is about 280 ms, time resolution of 0.33 ms is appropriate. These data were analysed using the full polarisation data analysis pipeline of the uGMRT described in \citet{kudale2008}. $R$ and $L$ powers were separately scaled by its average bandshape to take out the bandpass effect. The final outputs were Stokes $I$, $Q$, $U$ and $V$ ($=RR^*-LL^*$).
%\footnote{as per GMRT conventions, channel 1 is designated as $RR^*$ and channel 2 is $LL^*$}
{\bf Note that no polarisation calibration was applied for the phased array data.} 

The interferometric data were flagged and calibrated using standard tasks in $\textsc{casa}$ \citep[e.g.][]{das2019}. The $RR^*$ and $LL^*$ data of the target were separately self-calibrated to make the final images. We present the results obtained from these analysis in the following two subsections.

\begin{figure*}
\centering
\includegraphics[width=0.44\textwidth]{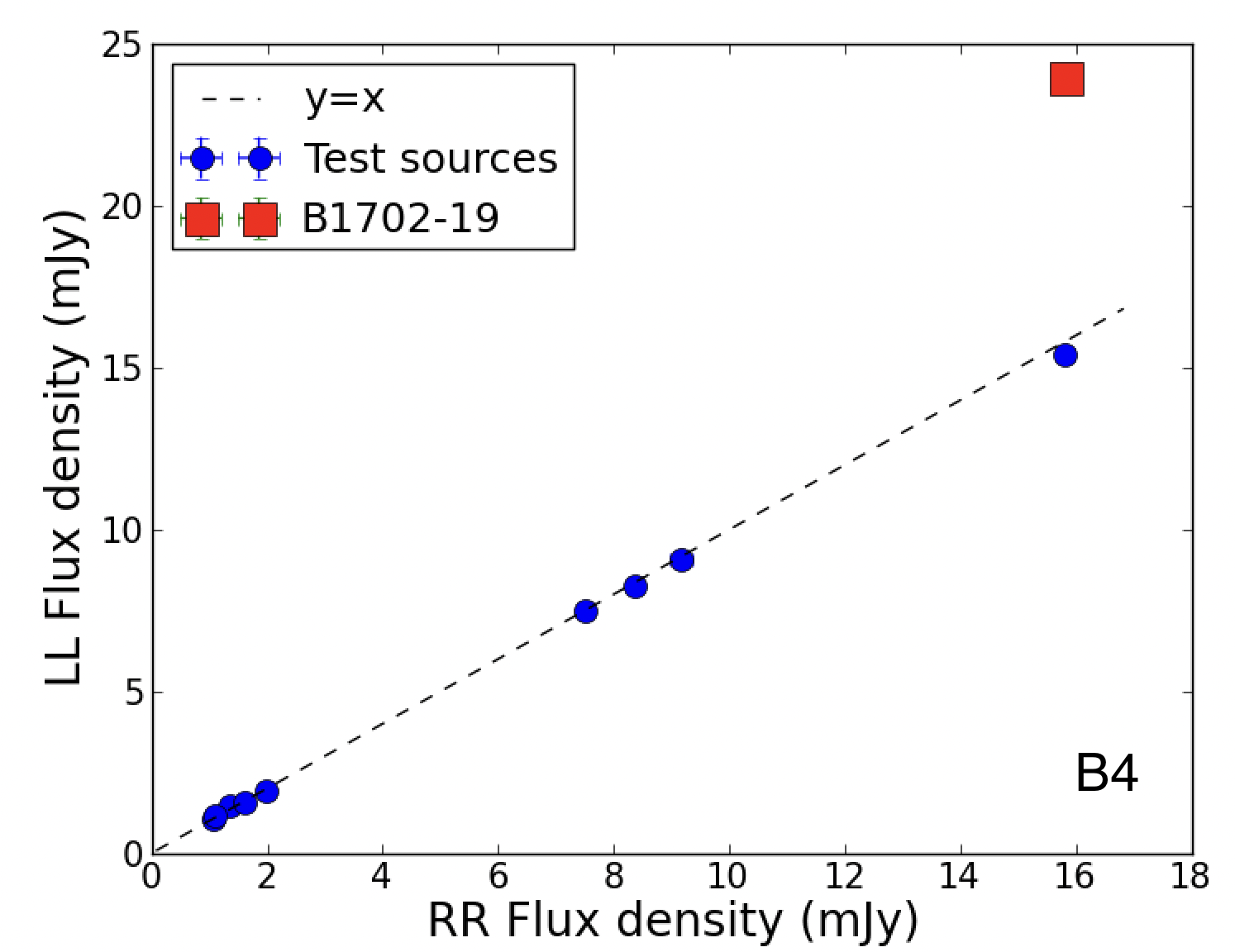}
\includegraphics[width=0.44\textwidth]{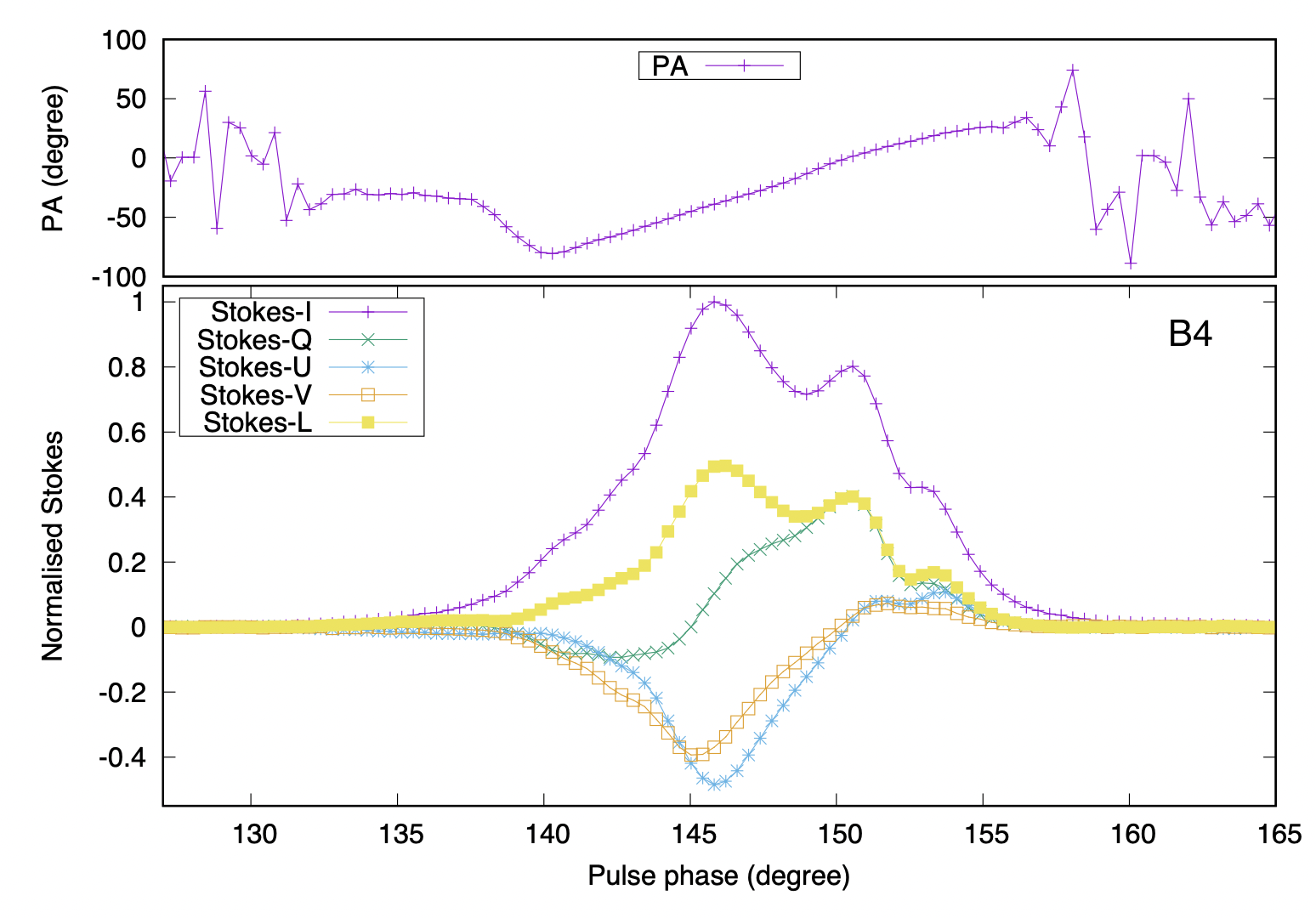}
\includegraphics[width=0.45\textwidth]{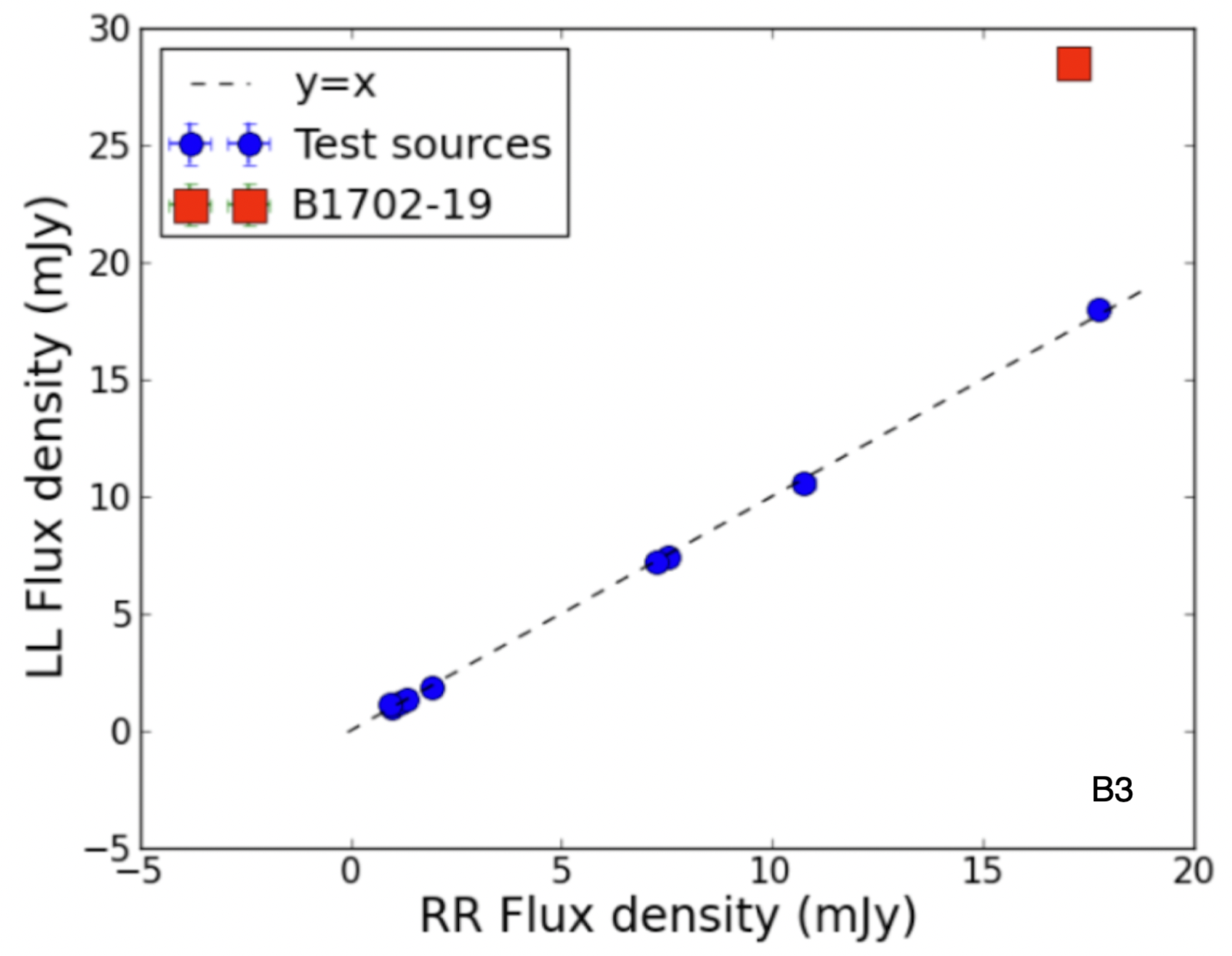}
\includegraphics[width=0.45\textwidth]{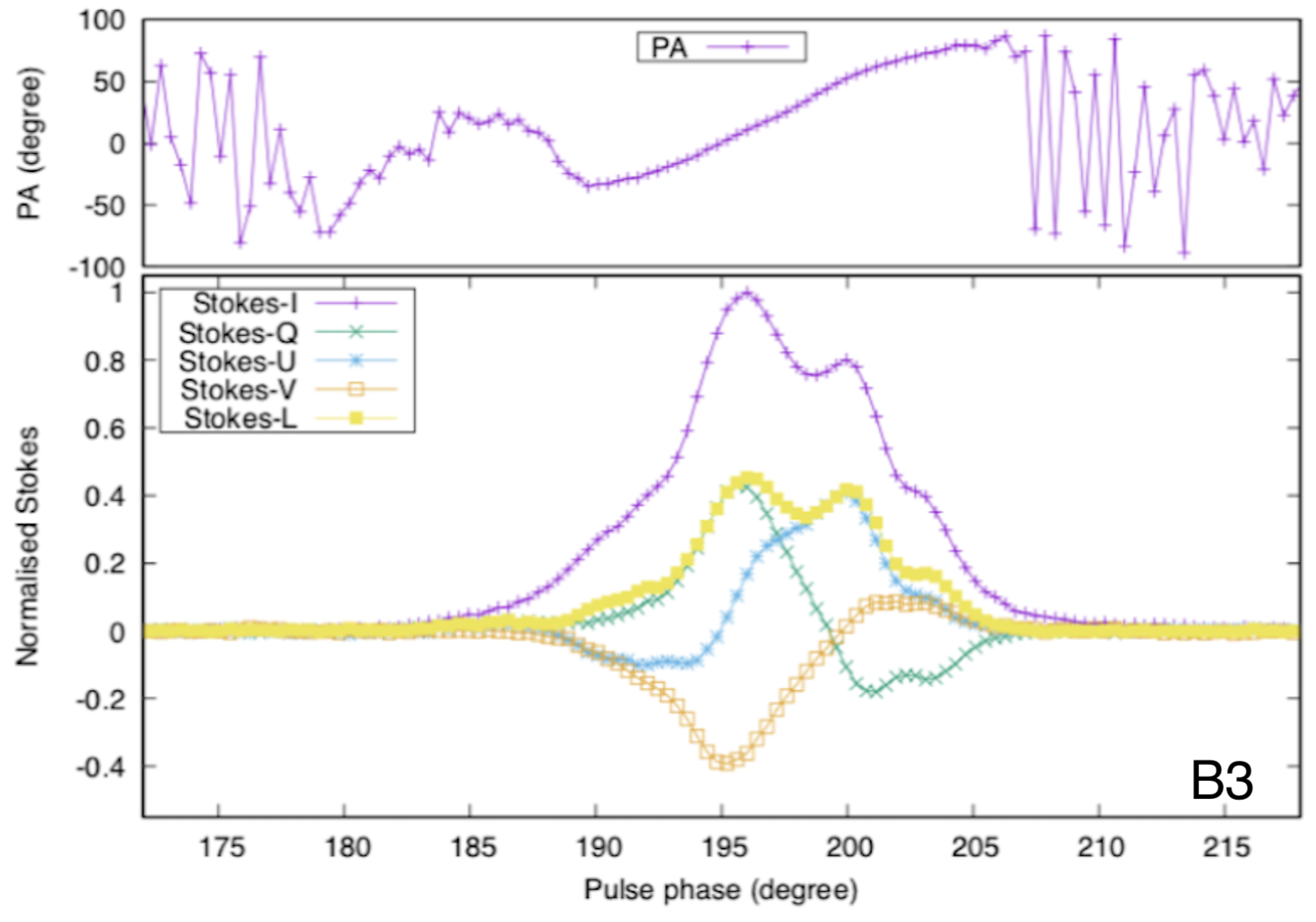}
\caption{{{\bf Top Panel:} Left: Comparison of $RR^*$ and $LL^*$ flux densities for the pulsar B1702-19, obtained by imaging the interferometric data, and that for a few test sources in the field of view (FoV)  in band 4 (550--750 MHz) of the uGMRT. Right: The Stokes $I$ (magenta), $Q$ (green curve), $U$ (blue curve) and $V$ (orange curve) profiles for the pulsar B1702--19 at 650 MHz as measured by the uGMRT. Also shown is the linear polarisation $L=\sqrt{Q^2+U^2}$ in yellow. {\bf Bottom Panel:} Left:  A comparison of $LL^*$ and $RR^*$ flux densities of the pulsar B1702-19 (red square) in band 3 of the uGMRT, along with a few other test sources in the FoV, obtained from imaging of interferometric data. Right:  The I, Q, U, V profiles obtained from uGMRT band 3 data. The colour scheme is same as that for band 4. Note that no polarisation calibration was carried out in the phased array data, and Stokes $V$ is defined as
$V = RR^* - LL^*$. }
\label{fig:band4_ll_rr}}
\end{figure*}

\subsubsection{Results from the interferometric mode observations}\label{subsec:interferometry_mode}
At uGMRT, \textbf{channels 1 and 2 are designated as  $R$ and $L$}, respectively. In the left panels of Figure \ref{fig:band4_ll_rr}, we show the results obtained from the interferometric data in bands 3  (bottom panel) and band 4 (top panel). We  plot the flux densities in the $LL^*$  against the flux densities in the $RR^*$ for the target (marked by red square) along with a few other test sources in the field of view. The flux densities of the test sources were measured in $RR^*$ and $LL^*$ to make sure that the inferred difference in the target flux density is not due to any systematic. As seen in the Figure \ref{fig:band4_ll_rr}, these test sources have equal flux densities at the $RR^*$ and $LL^*$, and hence the flux density difference in the target in the $RR^*$ and $LL^*$ is real.

In band 4,  $RR^*$ flux density of the pulsar is $15.8\pm 0.1$ mJy, whereas the $LL^*$ flux density comes out to be $23.9\pm 0.2$ mJy. This gives percentage circular polarisation $V/I=-20\%$, while using  IAU/IEEE convention, i.e. $V=(RR^*-LL^*)/2$. This uGMRT observation suggests the pulsar is left handed circularly polarised at band 4, whereas it is right handed circularly polarised (Stokes V positive) under IAU/IEEE convention as per EPN database. Therefore, it appears that the uGMRT channel 1, which is
currently designated as $R$) is LCP ($L$), and opposite for channel 2, if one follows IAU/IEEE convention in band 4.

In band 3, the $RR^*$ flux density of the pulsar is $\sim 17$\, mJy and $LL^*$ flux density is $\sim 28$\,mJy. This gives circular polarisation fraction, $V/I=-24\%$. Thus even in this case, the  observation using uGMRT suggests that the pulsar is  left circularly polarised. Therefore, on the face of it, the uGMRT channel 1 is also LCP ($L$) and vice-versa for channel 2 in band 3 as per IAU/IEEE convention.

\subsubsection{Result from the pulsar mode observations}\label{subsec:pulsar_mode}
The phased array observations provide pulse profiles of Stokes parameters, which can not be obtained from interferometric observations. These pulse profiles can be compared with the EPN pulse profiles directly. In the right panels of Figure \ref{fig:band4_ll_rr}, we show the pulse profiles for Stokes $I$, $Q$, $U$ and $V\,(=RR^*-LL^*)$. The summary of the results obtained for both bands are presented below.

{\bf Band 4:}\\ 
The maximum circular polarisation observed  from uGMRT observation is $-40\%$. By comparing these profiles (top right panel of Figure \ref{fig:band4_ll_rr}) with those in the left panel of Figure \ref{fig:epn_profile} using the EPN database, we find that:
\begin{enumerate}
    \item The signs of the Stokes $Q$ profiles in the two datasets do not match, but the Stokes $U$ profiles agree with each other.
    \item The sign of the uGMRT Stokes $V$ profile does not match with that of the EPN Stokes $V$ profile.
    \item The sweeps of the polarisation angle (PA) are opposite in the uGMRT from the EPN data.
\end{enumerate}

{\bf Band 3:}\\ 
The maximum circular polarisation observed  from uGMRT observation  is $-28\%$. By comparing these profiles (bottom right panel of Figure \ref{fig:band4_ll_rr}) with those  in the right panel of Figure \ref{fig:epn_profile} using the EPN data base, we find that:
\begin{enumerate}
    \item The signs of Stokes $Q$ profiles match between the uGMRT and EPN data, however, the signs of Stokes $U$ do not match. This is opposite to that of the result in band 4.
    \item The sign of the uGMRT Stokes $V$ profile does not match with that of the EPN Stokes $V$ profile.
    \item The sweeps of the polarisation angle (PA) are opposite in the two datasets.
\end{enumerate}

\subsection{Legacy GMRT archival data}\label{sec:legacy_GMRT_analysis}
\subsubsection{Interferometric data}
\begin{figure*}
\centering
\includegraphics[width=0.42\textwidth]{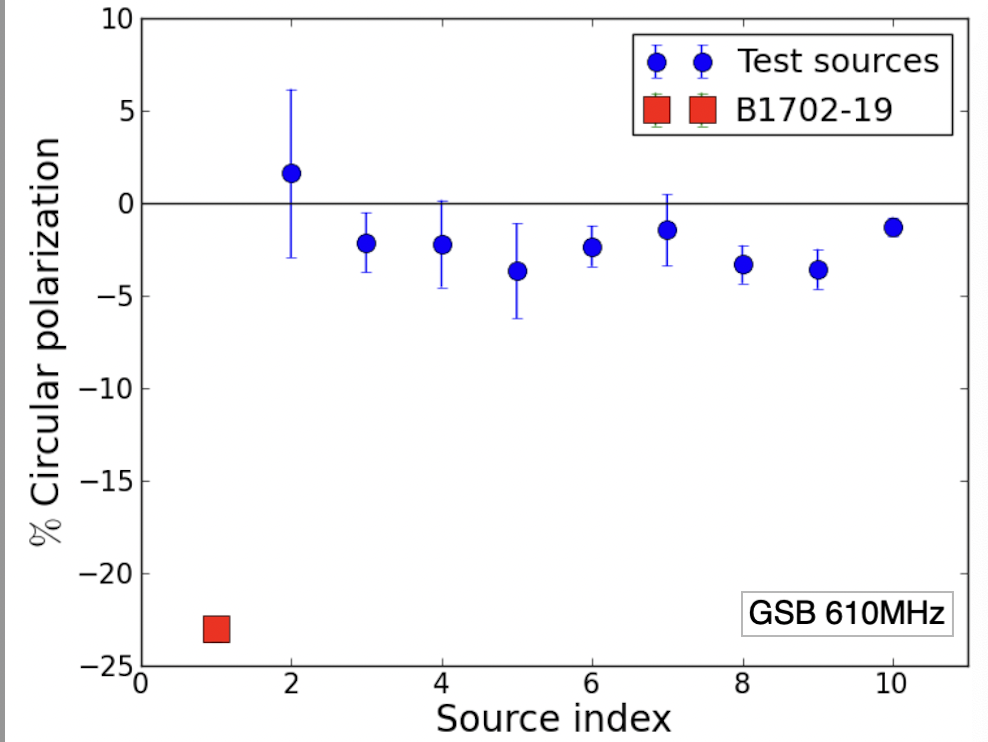}
\includegraphics[width=0.43\textwidth]{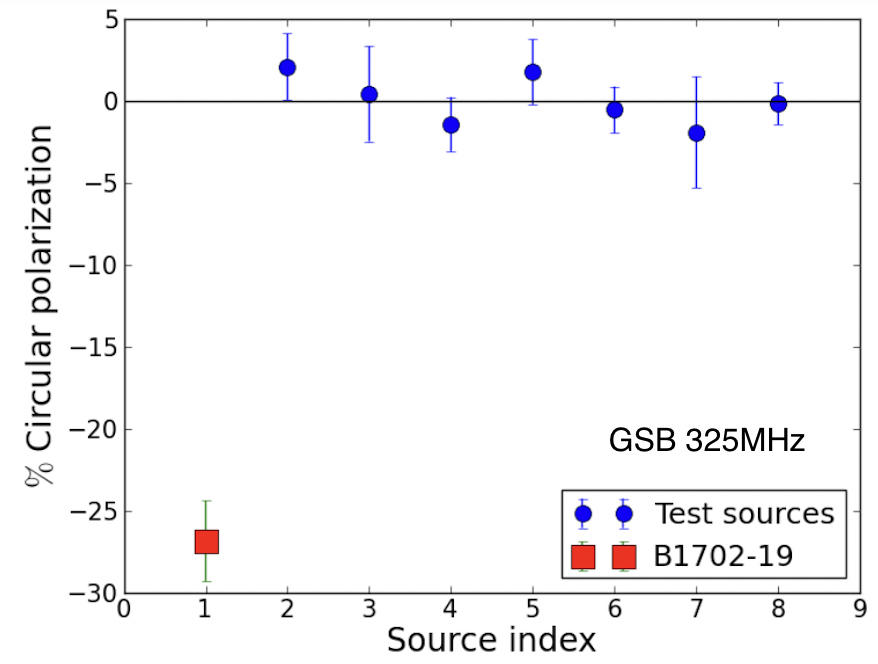}
\caption{{\it Left panel: The percentage circular polarisation for the pulsar B1702-19 (red square) and a few other sources in FoV at legacy GMRT 604 MHz frequency for data taken in 2007. 
Right panel: The percentage circular polarisation for the pulsar B1702-19 (red square) and a few other sources in FoV at legacy GMRT  331 MHz frequency for data taken in 2007. Note that Stokes V is defined as per IAU/IEEE convention, $RR^* -LL^*$.} 
\label{fig:gsbpol}}
\end{figure*}

We also searched for the legacy GMRT interferometric data for pulsar B1702-19. We found several datasets and downloaded observations taken on 15 May 2007 (P.I.: Bhaswati Bhattacharyya) at 610 MHz with 16 MHz bandwidth, and data taken on 2 Sep 2007 (PI: Bhaswati Bhattacharyya) at 330 MHz. There were no flux calibrators in both the observations, and hence the flux densities cannot be believed in terms of their absolute values, though the relative flux density in channels 1 and 2 can give significant information on Stokes V. We found Stokes V ($RR^*-LL^*$) to be negative, and Stokes $V/I=-22$\% at 610 MHz as shown in the left panel of Figure \ref{fig:gsbpol}, which is similar to what we reported for uGMRT band 4 data in Section \ref{subsec:interferometry_mode}. For 330 MHz data, we found the $LL^*$ flux density to be greater than $RR^*$ flux density as well. Here circular polarisation fraction, $V/I$ was found to be $-25$\% as shown in the right panel of Figure \ref{fig:gsbpol}. This is close to the value found in uGMRT band 3 observation mentioned in Section \ref{subsec:interferometry_mode}.

These results have an important implication. It means that the sense of polarisation remains unchanged between legacy GMRT and uGMRT interferometric data. Hence, we conclude that the legacy GMRT data also followed convention opposite to that of IAU/IEEE. 

\subsubsection{Pulsar mode data}
In Sanjay Kudale’s Master's thesis, a detailed comparison has been made for many pulsars with the EPN database. The thesis concluded that the
 sense of $V$ polarisation matches the two data sets. The thesis uses the Stokes V definition as $RR^*-LL^*$, whereas the EPN database uses $LL^*-RR^*$ as Stokes V.  
 This difference was not considered in the thesis while comparing the profiles with the EPN database. Hence, in principle, the thesis establishes that the phased array data 
 in  legacy GMRT followed convention opposite to that of IAU/IEEE. Since many of  the  datasets used in this thesis were with the GMRT hardware backend (GHB), it means 
 the sense of polarisation has remained opposite to that of the IAU/IEEE from the beginning of GMRT, i.e. GHB, GMRT Software Backend (GSB) as well as current uGMRT GWB correlator. 

\subsection{Implications of the initial test results}
The above exercise gives the following results. {\bf The sense of 
circular polarisation is the same for interferometric data as well as beam data. This also remains unchanged in the GHB, GSB, and GWB data.}

Thus, there seems to be a reversal in the sense of Stokes $V$ in the GMRT convention with that of the IAU/IEEE convention. If taken on the face value, the uGMRT LCP should be RCP and the RCP  should be LCP, to follow the IAU/IEEE convention. This is valid for both band 4  and band 3 as well.

\subsection{Issue with Stokes Q and U}\label{subsec:stokesQU_issue}
However, there is a bigger issue, which is the discrepancy of Stokes parameters $Q$ and $U$. In band 4 the signs of the Stokes $Q$ profiles do not match between uGMRT and EPN data, but the Stokes $U$ profiles agree with each other. In addition, the sweeps of the polarisation angle (PA) are opposite in the uGMRT vs EPN data. The Stokes V results suggest that $R$ and $L$ are swapped, however, simply swapping $R$ and $L$ can not explain the behavior of $Q$ and $U$. From Equations \ref{eq:Q} and \ref{eq:U},
\begin{equation}
    \begin{split}
       Q=RL^*+R^*L,\quad U=-i(RL^*-R^*L), \quad PA=1/2 \arctan(U/Q) 
    \end{split}
\end{equation}
It is evident from these equations that swapping $R$ and $L$ will give an opposite sign of $U$ and not that of Stokes $Q$. 

We next consider the possibility that $X$ and $Y$ are swapped (\S\ref{sec:intro}). This is because the GMRT feeds are linear feeds that are converted into circular feeds for bands 2, 3, and 4. Thus it is important to explore the possibility of a swap between X and Y itself. For clarity, we denote uGMRT $R$ and $L$ as $\tilde{R}$ and $\tilde{L}$ respectively. In addition polarisation angle (PA) for GMRT can be written as $\tilde{PA}$. If $X$ and $Y$ are swapped, we will get:
\begin{equation}
    \begin{split}
        \tilde{R}&=\frac{Y+iX}{\sqrt{2}}=i\frac{X-iY}{\sqrt{2}}=iL\\
         \tilde{L}&=\frac{Y-iX}{\sqrt{2}}=-i\frac{X+iY}{\sqrt{2}}=-iR 
    \end{split}
    \label{eq:26}
\end{equation}
 where, $R,\ L$ are IAU/IEEE convention.

In that case, we will get $\tilde{V}=\tilde{R}\tilde{R^*}-\tilde{L}\tilde{L^*}=LL^*-RR^*$, which is opposite to the EPN database values converted into IAU/IEEE convention.
 For Stokes $Q$ and $U$, we will get:

\begin{equation}
    \begin{split}
        \tilde{Q}&=\tilde{R}\tilde{L}^*+\tilde{R}^*\tilde{L}=(iL)(iR^*)+(-iL^*)(-iR)=-(RL^*+R^*L)=-Q\\
    \tilde{U}&=-i(\tilde{R}\tilde{L}^*-\tilde{R}^*\tilde{L})=-i\{(iL)(iR^*)-(-iL^*)(-iR)\}=-i(RL^*-R^*L)=U \\
    \tilde{PA}&=1/2\arctan(\tilde{U}/\tilde{Q})=1/2\arctan(U/-Q)=-1/2\arctan(U/Q)=-PA
    \end{split}
    \label{eq:27}
\end{equation}
 This is similar to what was observed in uGMRT band 4 observation. Thus,  this initial test at band 4 infers that the $X$ and $Y$ dipoles need to be swapped to explain the test results of band 4 uGMRT data.

In band 3, the signs of Stokes $Q$ profiles match between the uGMRT and EPN data, however, the signs of Stokes $U$ do not match. This is opposite to that of the result in band 4. These results can be explained by {\sout{a simple}} $R$ and $L$ swap,  but not using $X$ and $Y$ swapping as mentioned for band 4. Again, if we denote uGMRT $R$ and $L$ as $\tilde{R}$ and $\tilde{L}$ respectively, then $\tilde{R}=L$ and $\tilde{L}=R$.
In this case
\begin{equation}
    \begin{split}
      \tilde{Q}&=\tilde{R}\tilde{L}^*+\tilde{R}^*\tilde{L}=LR^*+L^*R =Q\\
   \tilde{U}&=-i(\tilde{R}\tilde{L}^*-\tilde{R}^*\tilde{L})=-i(LR^*-L^*R)=-U\\
   \tilde{PA}&=1/2\arctan(\tilde{U}/\tilde{Q})=1/2\arctan(-U/Q)=-1/2 \arctan(U/Q)=-PA  
    \end{split}
    \label{eq:28}
\end{equation}
Thus, this initial test at band 3 infers that the $R$ and $L$ dipoles need to be swapped to explain the test results of band 3 uGMRT data.
\vspace*{0.4cm}
\noindent{\bf\large  Goal of current tests}:
In the above section, requiring  $X$ and $Y$ to swap in band 4 and $R$ and $L$ to swap for band 3 to explain the observed results are quite incomprehensible in the first place.  Since polarisation calibration is not done in the previous experiment, it is not obvious whether the discrepancy between band 3 and band 4 results is due to polarisation calibration issues or intrinsic. In this report, we aim to find out the polarisation convention adopted by the uGMRT  \citep{gupta2017} and legacy GMRT in various bands \citep{swarup+91}.  

%\newpage

\section{Understanding the GMRT polarisation}
With this background, we can start to investigate the results seen in bands 3 and 4 and understand the polarisation convention of the GMRT. 

\begin{figure*}
\centering
\includegraphics[width=0.43\textwidth]{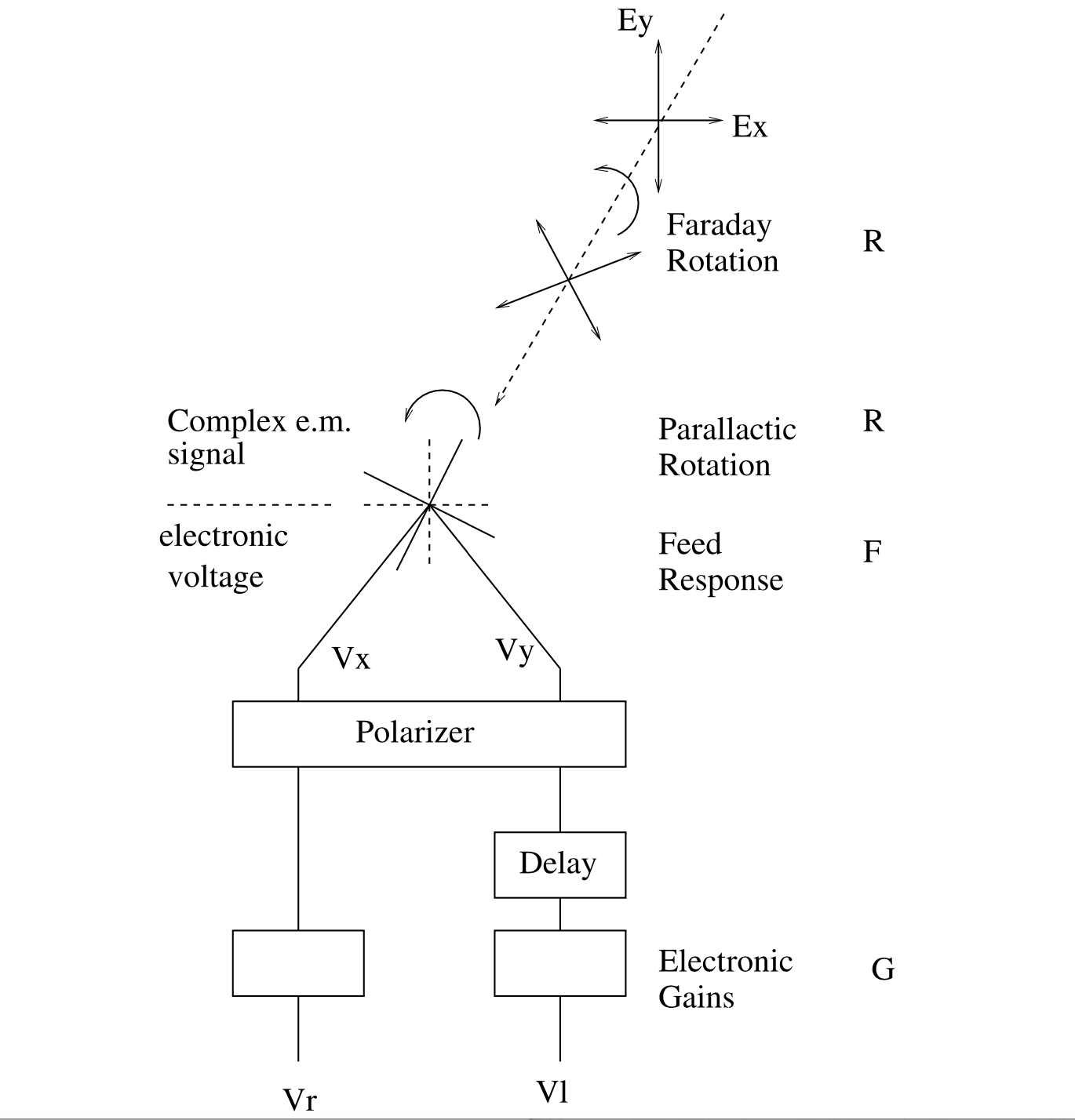}
\caption{Schematic of polarisation signal propagation in space, and through electronics which affects the state of polarisation (MSc thesis: Sanjay Kudale  adapted from \cite{hamaker1996}).} \label{fig:polbad}
\end{figure*}

Multiple factors could affect true source polarisation of a radio wave. Once radio waves emanate from the source, they travel through the interstellar medium (ISM). 
The magnetic field of the ionized ISM can cause Faraday rotation, causing the intrinsic polarisation of the source to change even before it reaches the telescope. Radio waves then interact with the antenna dish. Here they are reflected and brought to a focus at a feed. This reflection will cause a reversal  of polarisation and needs to be taken into account. The signals received at the feeds are amplified and they encounter a large number of electronic components before being recorded as final data products. Thus various external and internal factors can affect the polarisation (Figure \ref{fig:polbad}). From the propagation of the radiation from the astrophysical source of interest to the output of the correlator, there are various stages where one needs to correct for the above effects. 
We check for the following (in no particular order):
\begin{itemize}
\item Faraday rotation
\item  Parallactic rotation due to alt-az mount
\item Signal propagation from the front end (when it reaches the dish) to the optical fibre
\item Optical fibre to back end (user-friendly data) 
\item Calibration issues
\item Dependence on observational settings
\end{itemize}

\subsection{Faraday effects}
If a linearly polarised light is propagating through a magnetized plasma, It is found that the transmitted light is still linearly polarised, but that the plane of polarisation is rotated by an angle proportional to the projection of the magnetic field along the direction of the propagation of light. This is called the Faraday rotation (or the Faraday effect). ISM generally has some magnetic field. Magnetic fields in the ISM added to the presence of free electrons, can change the polarisation properties of an electromagnetic wave traveling through it. 

\begin{figure}[h]
\centering
\includegraphics[width=0.65\textwidth]{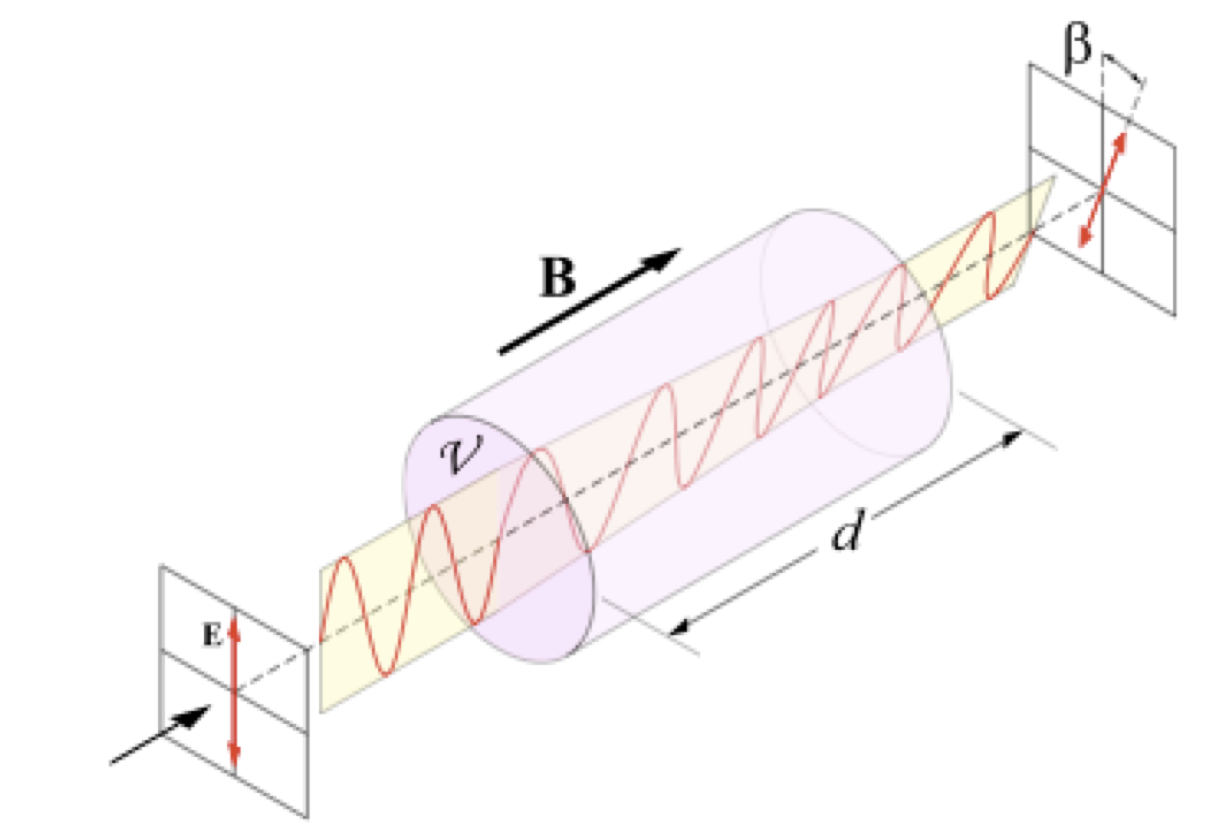}
\caption{The Faraday rotation effect. One can see the rotation of the plane of polarisation when passing through a medium with a magnetic field $B$. 
Faraday rotation of the plane of polarisation of an electromagnetic wave traveling through a region with a non-zero magnetic field and free electron density.
 \label{fig:faraday}}
\end{figure}

In circularly polarised light the direction of the electric field rotates at the frequency of the light, either clockwise or counter-clockwise. This electric field causes a force on the ISM electrons. The motion thus affected will be circular, and these circularly moving electrons create their magnetic field in addition to the ISM magnetic field. The created field will be parallel to the external field for one (circular) polarisation, and in the opposing direction for the other polarisation direction. The result of this is that the net magnetic field is enhanced in one direction and diminished in the opposite direction. This results in
left and right circularly polarised waves propagating at slightly different speeds,  known as circular birefringence. Since a linear polarisation can be decomposed into the superposition of two circularly polarised components of opposite handedness and equal amplitude, the effect of a relative phase shift, induced by the Faraday effect, is to rotate the orientation of a wave's linear polarisation (Figure \ref{fig:faraday}).

In contrast to the Faraday effect in solids or liquids, interstellar Faraday rotation has a simple dependence on the wavelength of light $\lambda$. The observed frequency-dependent polarisation angle,  $\chi(\lambda^2)$ is  given as  $\chi=\chi_0+RM\lambda^2$,  where $\chi_o$ is intrinsic source polarisation angle, and the rotation measure (RM, in units of rad m$^{-2}$) is defined as,
\begin{equation}
    RM =0.81 \int_{l}^{0}  n_e B_\parallel.dl
    \label{eq:29}
\end{equation}
where $l$ is the distance travelled by the polarised emission (in parsecs), $n_e$ is the number density of free electrons (in cm$^{-3}$) and $B_\parallel$ is the magnetic field strength component along the line-of-sight (LOS) (in $\mu$G). The accepted sign convention is that positive RM indicates a magnetic field oriented
toward the observer, and negative RM indicates a magnetic field oriented away from the observer. 

While the effect of Faraday rotation is to rotate only the phases and ideally should not result in depolarisation. However, since the radio instruments have a certain
 spectral resolution and the Faraday rotation is slightly different at different  frequencies,  the finite spectral resolution of the instrument causes a depolarisation of the
  polarised signal. However, this will not result in the reversal of sign of the polarisation. Thus Faraday rotation can be ruled out as a possible cause of reversal of Stokes Q or U in GMRT 
  bands 3 and 4. Faraday rotation rotates the plane of polarisation of the linearly polarised emission. Hence, it can not cause the sign change of Stokes V.

\iffalse
{\color{green} This paragraph seems unnecessary and does not relate to the rest of the report. We may remove this.} One needs to note that the rotation of the plane of polarisation is larger at lower frequencies, and depends on the magnetic field along the line of sight, denoted by $B_\parallel$. Thus, when observing a linearly polarised radio source with large bandwidths, uncorrected {\color{red} the effect of} Faraday rotation can cancel some amount of linear polarisation in the signal due to rotation of the plane of polarisation across the bandwidth of the receiver. In some cases, where the electron distribution to the source is well known (using the DM), a measurement of the RM can provide a useful estimate of the average magnetic field along the line of sight to the source.

\fi

\subsection{Role of parallactic angle}\label{subsec:parallactic_angle}
\begin{figure*}
\centering
\includegraphics[width=0.45\textwidth]{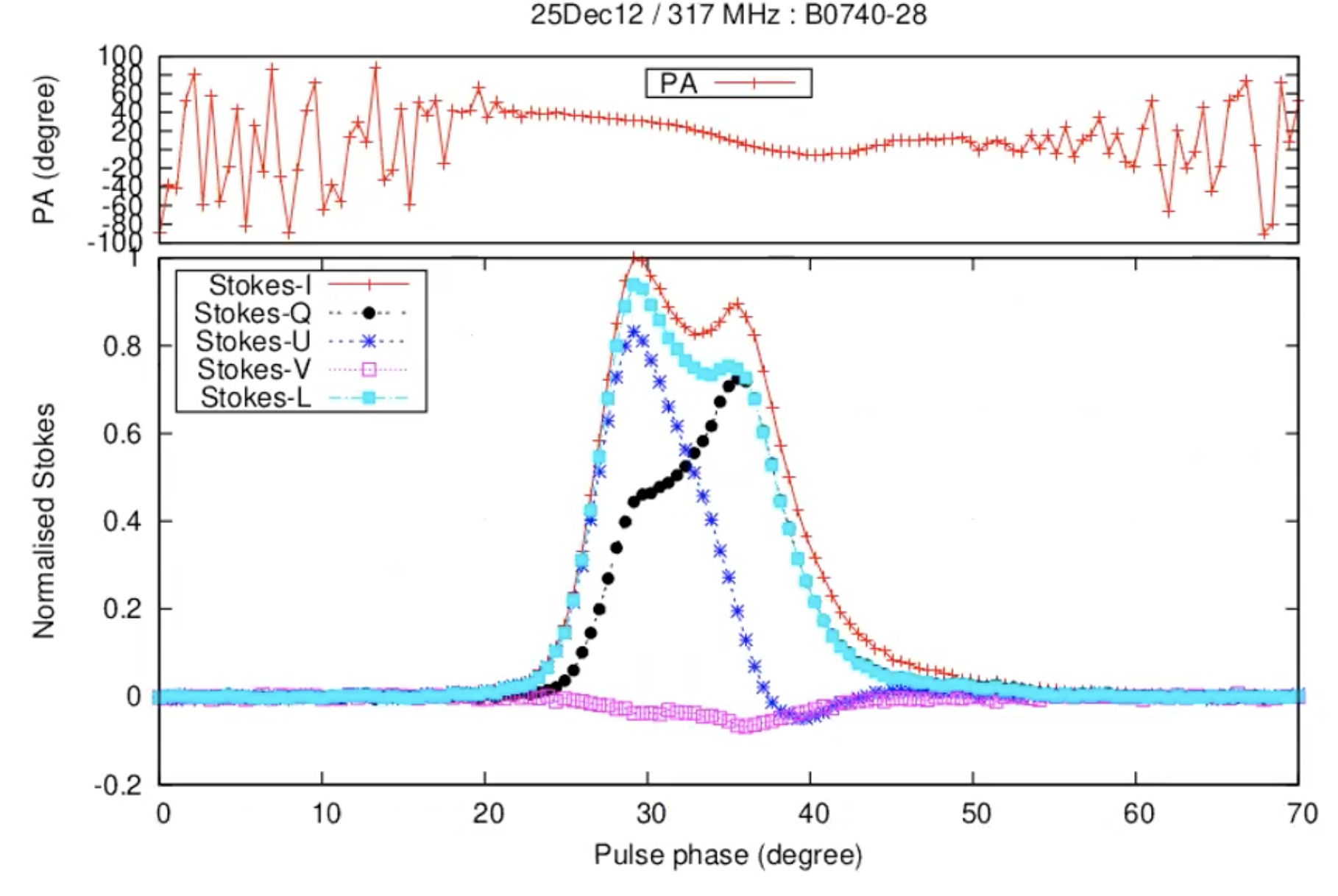}
\includegraphics[width=0.45\textwidth]{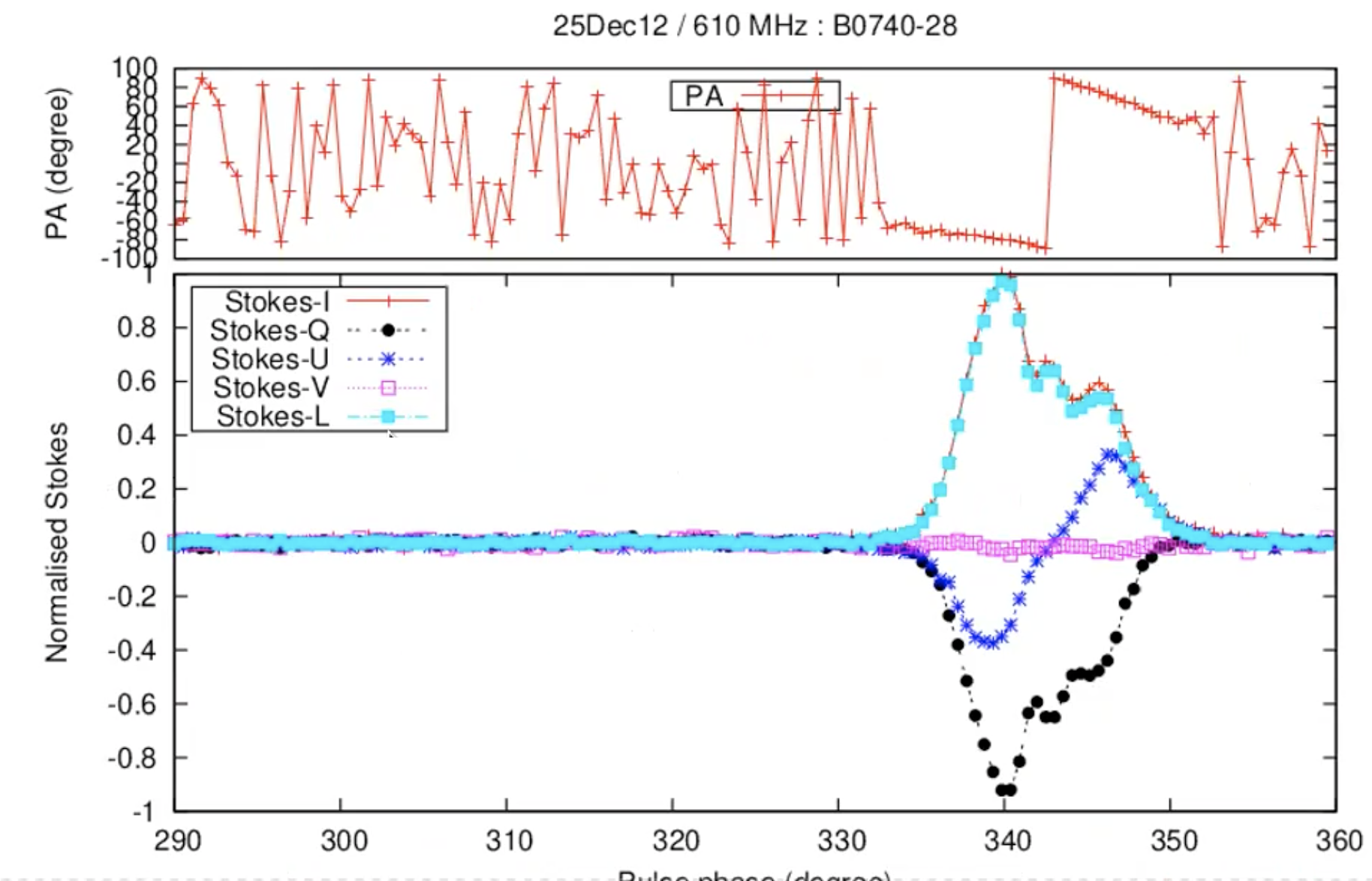}
\includegraphics[width=0.45\textwidth]{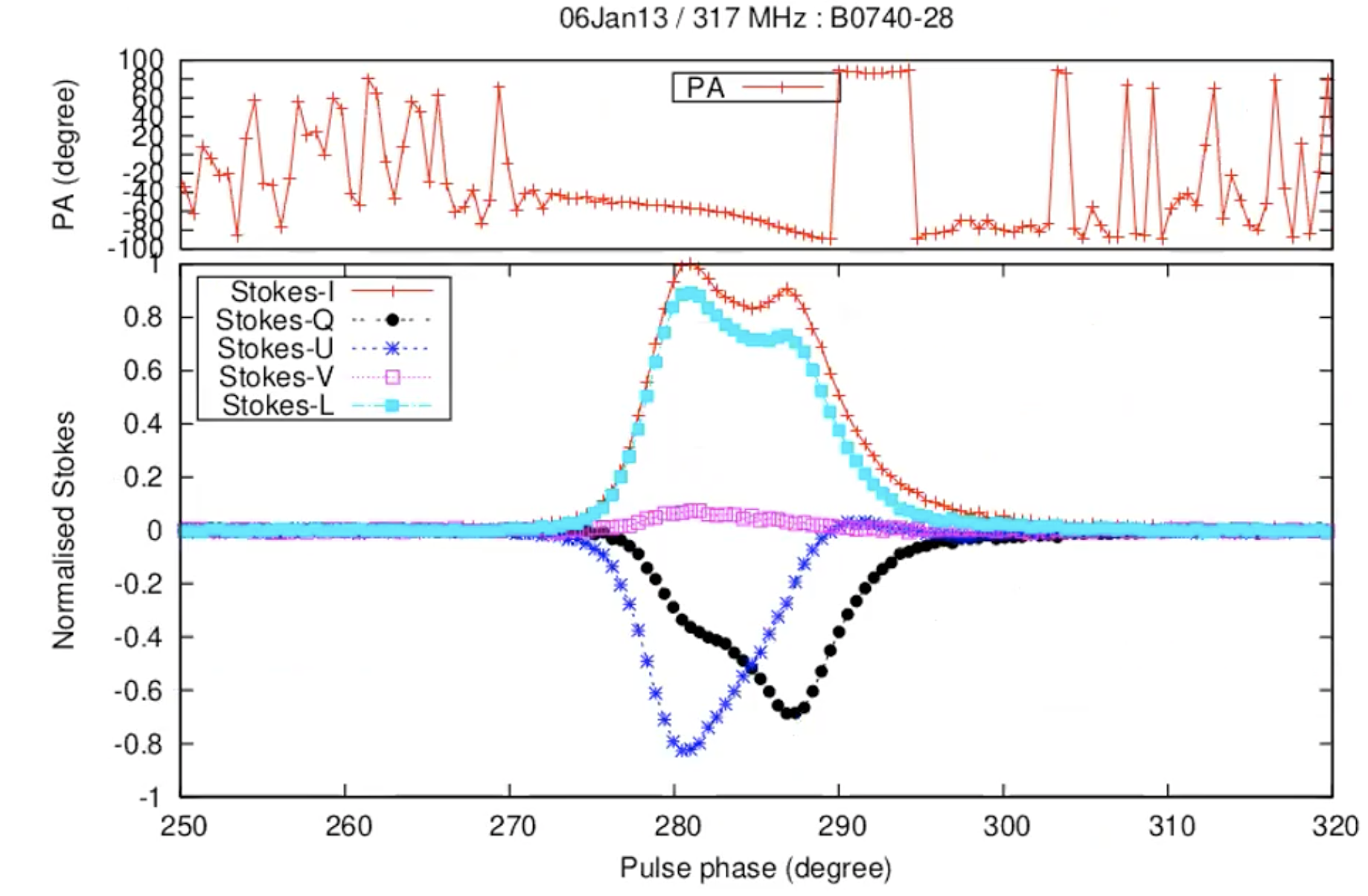}
\includegraphics[width=0.45\textwidth]{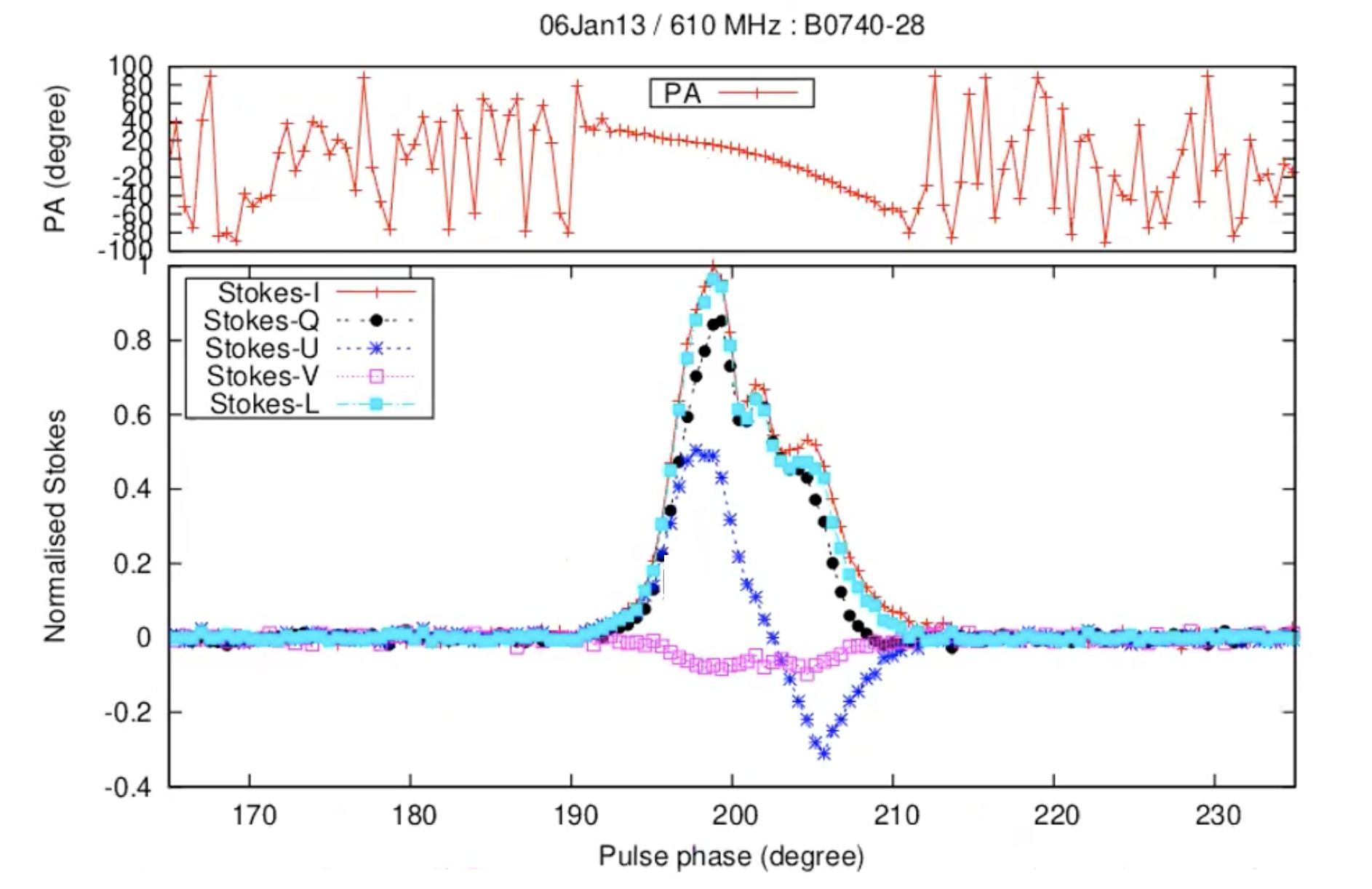}
\includegraphics[width=0.45\textwidth]{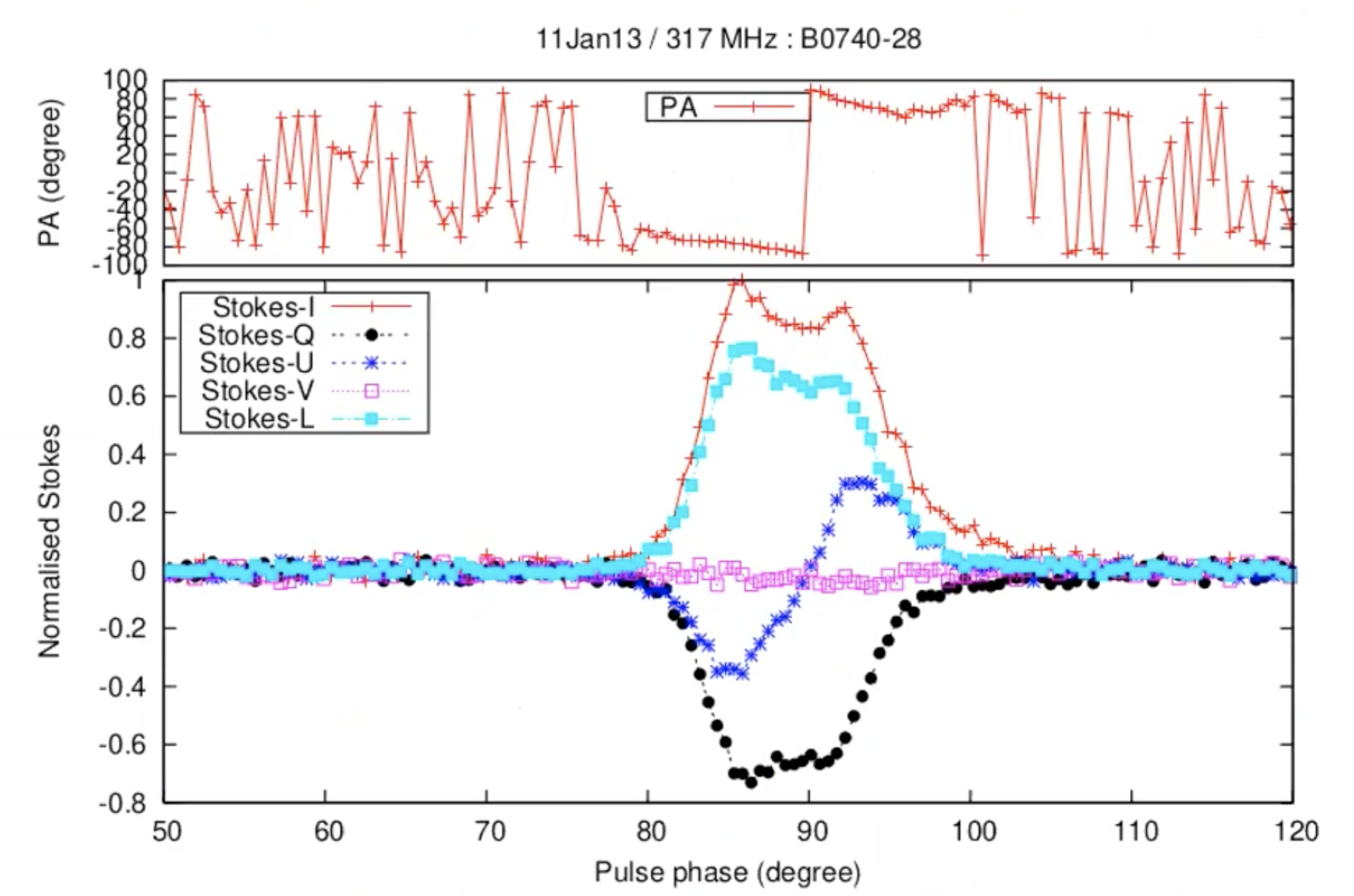}
\includegraphics[width=0.45\textwidth]{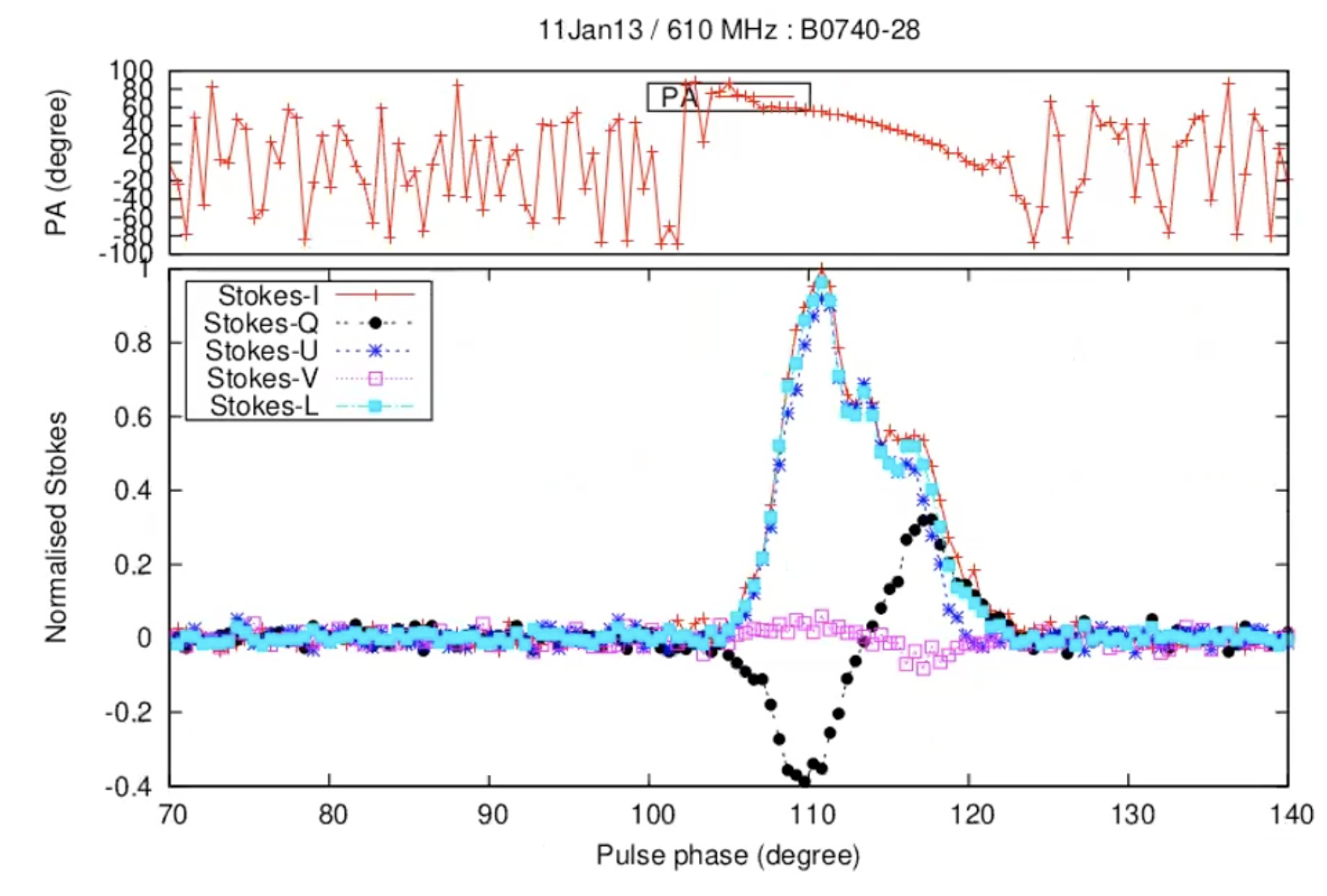}
\includegraphics[width=0.45\textwidth]{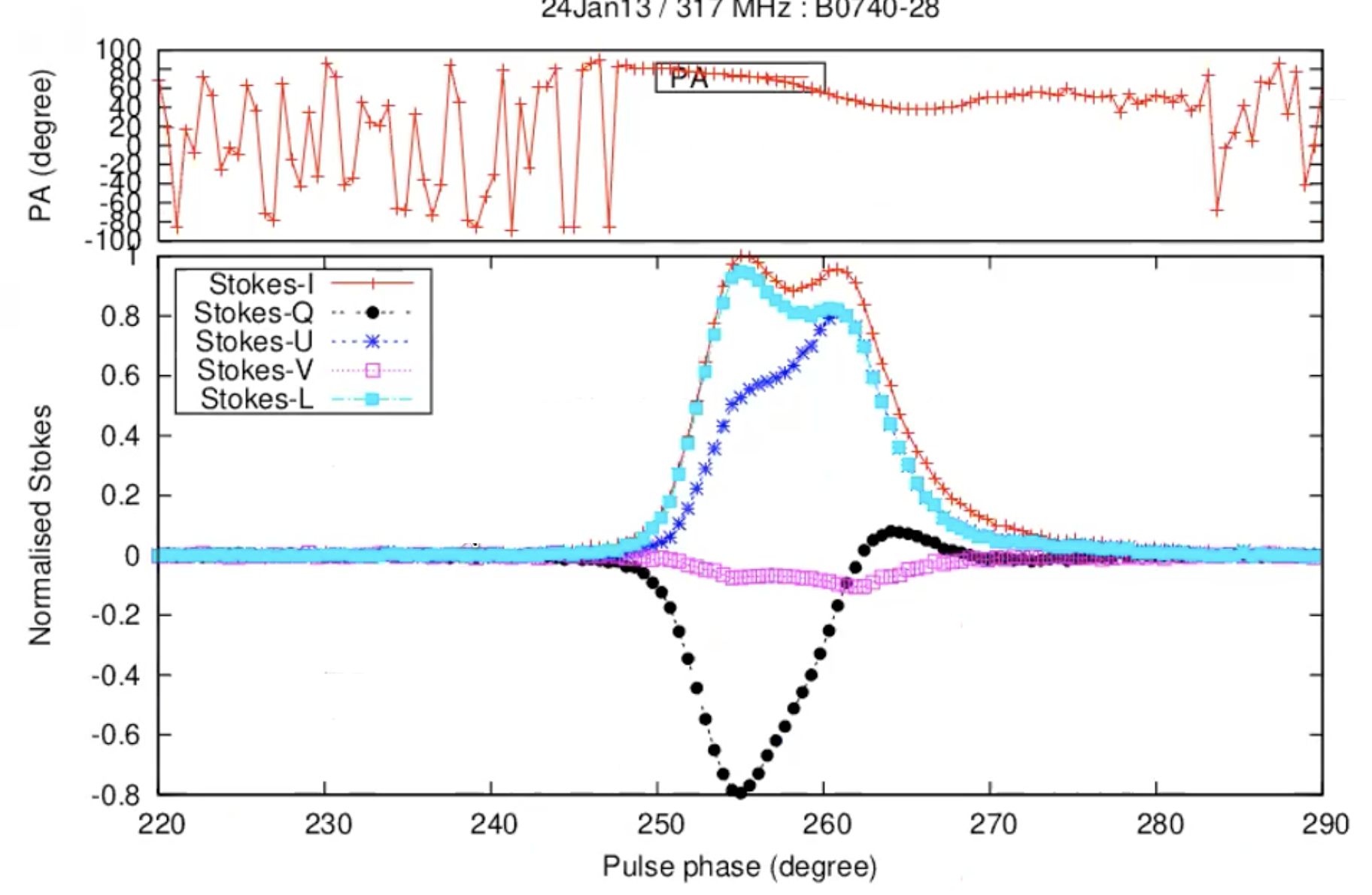}
\includegraphics[width=0.45\textwidth]{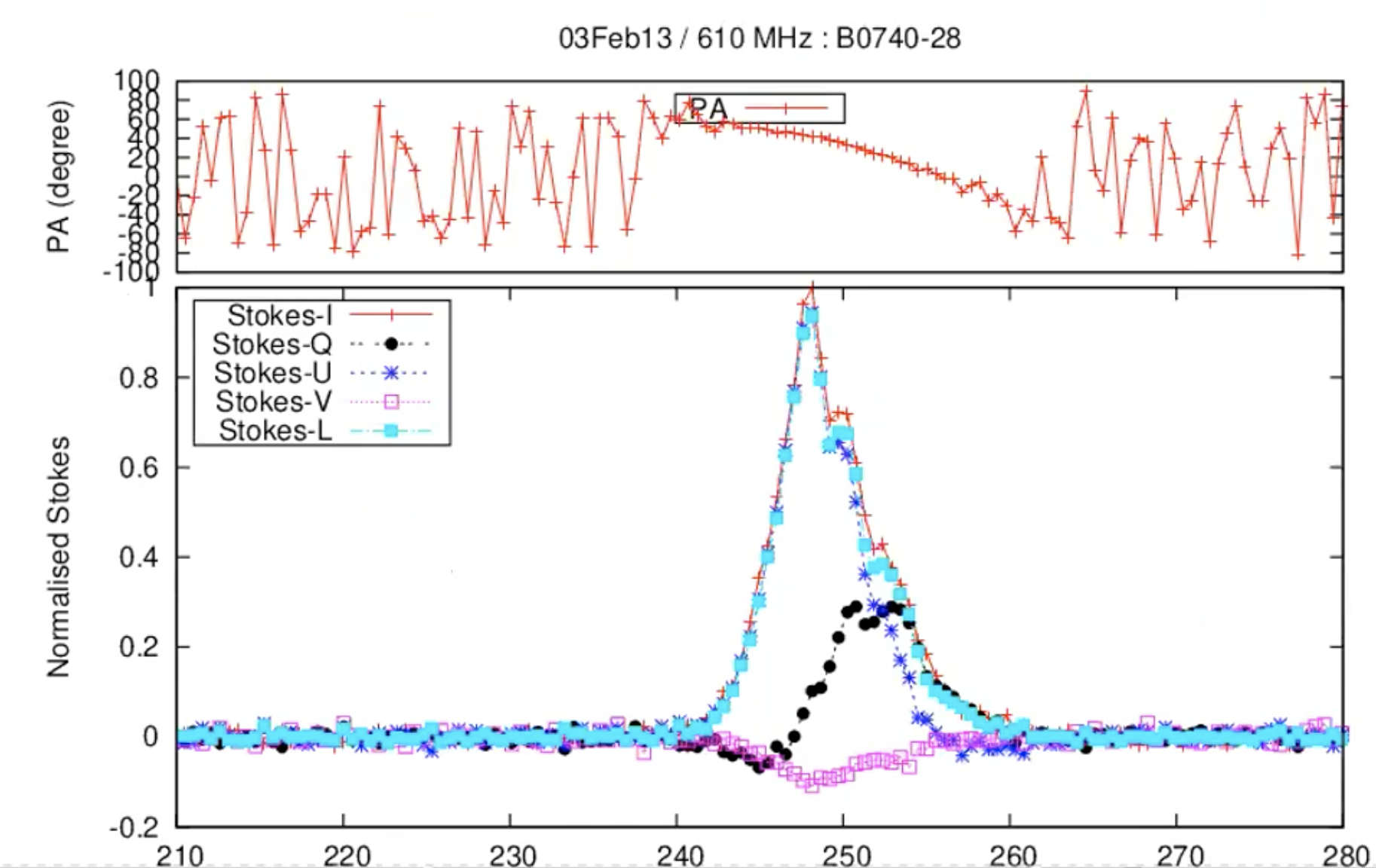}
\caption{GMRT phased array observations of pulsar B0740-28 at different epochs in 2013, at different parallactic angles. The left panels show band 3 data and the right panels show band 4 data (courtesy Yashwant Gupta). Please note that these data are taken in USB mode and no
correction or calibration has been applied to the data.
 \label{fig:pa}}
\end{figure*}

For the case of an alt-azimuth mounted telescope, the feed dipoles rotate in the plane of the sky, and hence to the plane of polarisation of the incoming radiation. This rotation of the feed to sky coordinate is characterized by parallactic angle. Correction of parallactic angle is necessary to decipher the correct polarisation of radiation.  In figure \ref{fig:pa}, we plot the raw beam data taken for a pulsar B0740-28 on multiple days. The data plotted are without correcting for parallactic angle. One can see without parallactic angle correction, Stokes profiles are not to be believed. Parallactic angles need to be calculated for each time and corrected for.%To correct for parallactic angle one does not need to observe over 60 degrees, it is needed when one does not have a polarised calibrator with a known polarization angle.

\subsection{Necessity of polarisation calibration}\label{subsec:calibration}
% This exercise aims to check whether, under any circumstances, polarization can be swapped
The signals received in the two polarisation channels go through the receiver chain which may have unequal gains. Gain calibration is required to correct for  these effects. We aim to determine if, under certain circumstances, lack of  polarisation calibration can cause the results seen in bands 3 and 4, as mentioned in the Section \ref{subsec:stokesQU_issue}.

In full polarimetric calibration, the EM signal is {\sout{fully}} treated as a vector. For a single baseline of two antennas `j' and `k', the incoming signal in an $XYZ$ coordinate system 
can be written as a column vector
\begin{equation}
e_+= \begin{pmatrix}
e_x\\
e_y
\end{pmatrix}
\label{eq:30}
\end{equation}

The $e_+$ vector can be written on the circular basis
\begin{equation}
e_\odot=  Ce_+ =\frac{1}{\sqrt{2}}
\begin{pmatrix}
i  & i\\
1 & -i
\end{pmatrix}
\begin{pmatrix}
e_x\\
e_y
\end{pmatrix}=
\begin{pmatrix}
e_R\\
e_L
\end{pmatrix}
\label{eq:31}
\end{equation}

Then the correlation between these two signals can be written in the form of a {\sout{fully}} coherence matrix
\begin{equation}
\mathbf {E_{jk}}=  \mathbf  {e_j e_k^\dagger} =
\begin{pmatrix}
e_{jR}\\
e_{jL}
\end{pmatrix}
\begin{pmatrix}
e^*_{kR} & e^*_{kL}
\end{pmatrix}=
\begin{pmatrix}
e_{jR}e^*_{kR}  & e_{jR}e^*_{kL} \\
e_{jL}e^*_{kR}  & e_{jL}e^*_{kL} 
\end{pmatrix}
\label{eq:32}
\end{equation}

If the coordinate system is chosen according to IAU/IEEE convention the coherency matrix 
can be written in terms of Stokes parameters
\begin{equation}
\mathbf {E_{jk}}= \frac{1}{2}\begin{pmatrix}
I+V & Q+iU\\
Q-iU & I-V
\end{pmatrix}
\label{eq:33}
\end{equation}

But the true source signal is always modified by the atmospheric and instrumental effects. The transformation of the signal by the instrument can be represented by $2 \times 2$ Jones matrices as, $\mathbf{w_j=J_je_j}$, where $\mathbf{w_j}$ is the output signal from the single antenna chain. Then the output coherency matrix will be 
\begin{equation}
\mathbf{W_{jk}=w_j w_k^\dagger  = J_j E_{jk} J_k^\dagger}
\label{eq:34}
\end{equation}
In polarimetric calibration, one wants to find out these Jones matrices for each antenna.

Now Jones matrix can be decomposed in the following form
$$\mathbf{J_i=G_iB_iD_i}$$
where $\mathbf{G_i,\,\,B_i}$ are the time dependent antenna gains and bandpass, respectively,  and 
$$\mathbf{D_i}=\begin{pmatrix}
1 & d_R\\
d_L & 1
\end{pmatrix}
$$ is the leakage matrix representing the polarisation leakage from one polarisation channel to another polarisation channel.

Using unpolarised calibrators these factors can easily be solved within an unknown unitary 
transformation between the corrected coherency matrix and true source coherency matrix: $\mathbf{E'_{jk}}$ as
$\mathbf{E'_{jk}=X E_{jk} X^\dagger}$ where $\mathbf{X}$ is the unitary matrix \citep{hamaker2000}. To calibrate this unitary ambiguity we need polarised calibrator sources.

The unitary transformation can represent two things:
\begin{itemize}
\item Reflection of the vector 
\item Rotation of the vector
\end{itemize}
In all interferometric analysis software only the rotational property is considered and not the reflection property.

The rotation matrix represents the rotation of a polarisation vector $\mathbf{P}=(Q, U, V)$ in the 3D polvector space and can be decomposed into rotations about three perpendicular axes.

In circular basis it is decomposed as,
\begin{equation}
\mathbf{X=X_q(\phi)X_u(\epsilon) X_v(\theta)}=
\begin{pmatrix}
e^{i\phi} & 0\\
0 & e^{-i\phi} 
\end{pmatrix}
\begin{pmatrix}
\cos \epsilon & i \sin \epsilon \\
i \sin \epsilon & \cos \epsilon
\end{pmatrix}
\begin{pmatrix}
e^{i\theta} & 0\\
0 & e^{-i\theta} 
\end{pmatrix}
\end{equation}

\begin{itemize}
\item $\mathbf{X_q(\phi)}$ represents the phase difference between the R and L polarisation channel.
%and will cause a rotation of Stokes Q and U and polarisation angle
%will change.
\item $\mathbf{X_u(\epsilon)}$ represents the ellipticity matrix, which will cause leakage from stokes Q to stokes U.
\item $\mathbf{X_u(\epsilon)}$ represents the geometrical rotational matrix, which includes Faraday rotation, parallactic rotation, and any absolute physical rotation of the feed itself. In circular basis it takes the form $\left(\begin{smallmatrix}   
e^{i\theta} & 0\\
0 & e^{-i\theta} 
\end{smallmatrix}\right)$. This will also cause a change in the polarisation angle.
\item An important property of rotation matrices to note here is that the determinant of a rotation matrix is  $ +1$.
\end{itemize}

The important property of the reflection matrix is that the determinant of a reflection matrix is $-1$. Although both rotation and reflection matrices are unitary, this property distinguishes the reflection matrix from the rotation matrix.

These properties can be used to understand the possible reason causing  an incomprehensible result from the initial test described in Section \ref{subsec:stokesQU_issue}. We will attempt this in both band 3 and band 4.

\subsubsection{Possible reason of uGMRT band 3 and 4 initial test results}
In uGMRT band 3, the observations showed that the Stokes V and Stokes U change sign, but stokes Q remained the same. This means that a true source coherence matrix 
$\mathbf{B_3}=\left(\begin{smallmatrix}   
I+V & Q+iU\\
Q-iU & I-V
\end{smallmatrix}\right)$ is transformed to
\begin{equation}
\mathbf{B'_3}=\begin{pmatrix}   
I-V & Q-iU\\
Q+iU & I+V
\end{pmatrix}=\mathbf{X} \begin{pmatrix}   
I+V & Q+iU\\
Q-iU & I-V
\end{pmatrix} \mathbf{X^\dagger}
\end{equation}

The required unitary matrix to carry out the above operation is 
$$\mathbf{X}=
\begin{pmatrix} 
0 & 1\\
1 & 0
\end{pmatrix}
$$
The determinant of this unitary matrix is $-1$. This means it is an effect of reflection, not rotation. Thus the calibration software package cannot calibrate this effect.

It was suggested above that such a flip in band 3 could be due to the swapping of R and L.
This can also be seen if we write the coherency matrix in R and L form.
\begin{equation}
\mathbf{XB_3X^\dagger}=
\begin{pmatrix} 
0 & 1\\
1 & 0
\end{pmatrix}
\begin{pmatrix} 
RR^* & RL^*& \\
LR^* & LL^*
\end{pmatrix}
\begin{pmatrix} 
0 & 1\\
1 & 0
\end{pmatrix}=
\begin{pmatrix} 
LL^* & LR^*& \\
RL^* & RR^*
\end{pmatrix}
\end{equation}

There is a possibility that other rotational effects are present. If we do not calibrate these rotational effects,  it may also change the values of the Stokes parameters, but cannot explain the simultaneous sign change of stokes U and V.

In band 4, the signs of Stokes V and Q were found to be swapped but the Stokes U sign remained the same. It can be easily shown that there exists no unitary matrix on a circular basis to explain this. But on a linear basis, there exists a unitary transformation, which
can be caused by $X$ and $Y$ channel swapping. 

\subsubsection{A coherent framework explaining bands 3 and 4 results}\label{subsec:coherent_swap}
An important question is whether both band 3 and band 4 polarisation swaps can be described in a single coherent framework or not. There are rotational effects, which we did not calibrate for the initial tests. We need to first take care of those before justifying the problem.

Let us first assume that R and L are swapped in both bands 3 and 4. Then in this swapped coordinate system, the coherency matrices for bands 3 and 4 can be
related as
\begin{equation}
\mathbf{B'_4=X' B'_3 X'^\dagger}
\end{equation}
which becomes
\begin{equation}
\mathbf{\begin{pmatrix}
I-V & -Q+iU\\
-Q-iU & I+V
\end{pmatrix}=X' \begin{pmatrix}
I-V & Q-iU\\
Q+iU & I+V
\end{pmatrix}
X'^\dagger}
\end{equation}

From this expression the corresponding unitary matrix is 
$$\mathbf{X'}=\begin{pmatrix} e^{i\pi/2} & 0 \\ 0 & e^{-i\pi/2} \end{pmatrix}$$. The determinant of this matrix is $+1$, thus it is a rotational matrix. This rotational effect can be calibrated using the calibration software. Then both band 3 and band 4 conventions will be the same and can be explained by only the R and L  swap.

In this exercise above, an assumption has been made that the polarisation properties of the source do not change between bands 3 and 4. But in reality the source polarisation can be different in different bands. In that case the true rotation angle could be different than $\pi/2$. While polarisation fraction may vary, there is no physical mechanism that can cause reflection in the polarisation vector 3D polvector space. 

The conclusion of the above mathematical analysis is as follows:
\begin{itemize}
\item The sign changes between bands 3 and 4 cannot be explained by any kind of rotation, like Faraday rotation and parallactic rotation, absolute geometric rotation of the feed, and cross-phase between two polarisation channels.
\item The change in the Stokes sign can only be explained by reflection, which cannot be calibrated by any software package because the reflection effect is never considered in the calibration framework. This reflection effect can simply arise from the swapping of the dipoles with respect to the IAU/IEEE convention.
\item The difference of the Stokes Q and U signs between band 3 and band 4 can be explained by rotation. If we calibrate the rotational effects then both band 3 and band 4 Stokes signs can be explained in a single coherent  framework, that is R and L polarisation channels are swapped. However, these are theoretical results and they need to be confirmed with observations, as we show in later sections.
\end{itemize}
To summarise, both band 3 and band 4 results can be explained in a single coherent  scenario involving R \& L swap (reflection effect), and uncorrected rotational effect in the case of band 4.

\section{New Sets of Experiments}
\subsection{Tracking polarisation from the feed to fibre optic system}
The main aim here is to determine how H and V are positioned as this will determine which channel is RCP and LCP. The first step was to track the polarisation from the feeds to the fibre optic system.

\subsubsection{Determine the GMRT Azimuth 0 direction}
\begin{figure}
\centering
\includegraphics[width=0.48\textwidth]{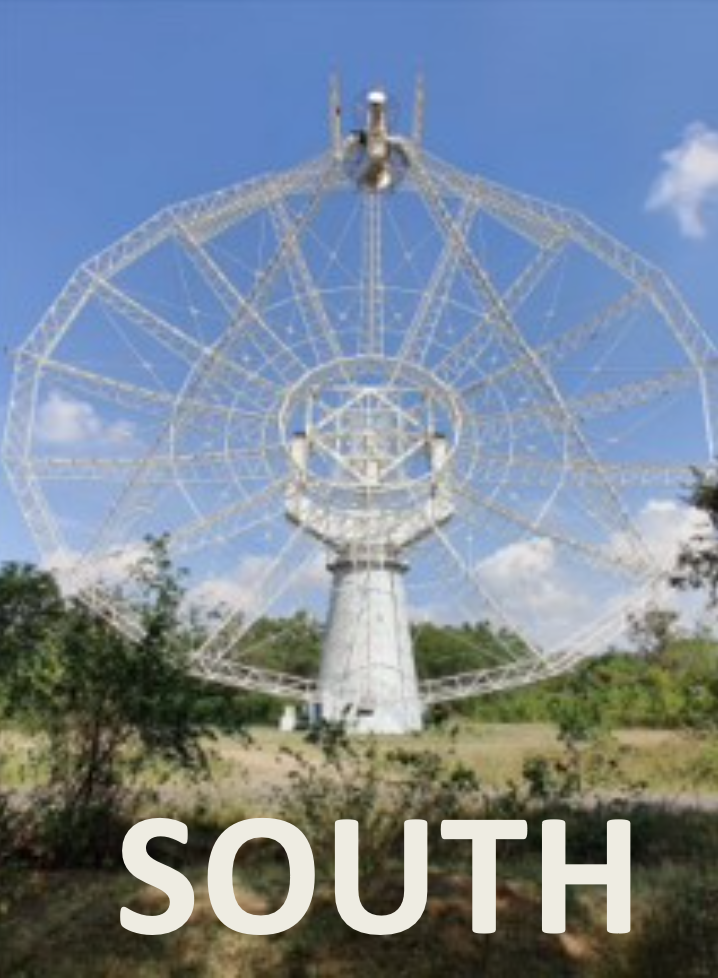}
\includegraphics[width=0.30\textwidth]{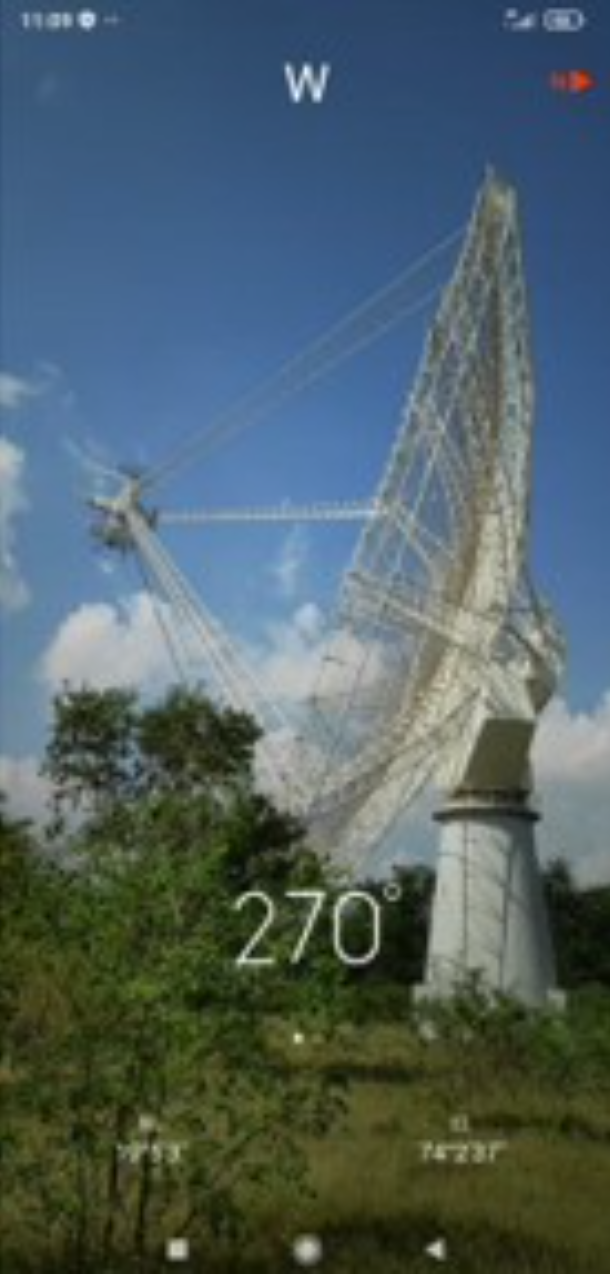}
\caption{GMRT antenna  positioned at Azimuth $0^o$ and Elevation at $18 ^o$ from  the control room. The  Az $0^o$ of the antenna points in the South direction.  \label{fig:az}}
\end{figure}

As a first step, we aimed to see whether the Azimuth $0^o$ of a GMRT antenna is towards the North or South? For this exercise, a GMRT antenna was positioned at Azimuth $0^o$ and Elevation at $18 ^o$ from the control room. The servo reading was checked from the antenna base and confirmed the antenna position. We checked the direction to which the antenna is pointing using a GPS receiver (Figure \ref{fig:az}). {\bf We confirmed that Az $0^o$ of the antenna points towards the South direction. This is different than the regular convention which is always towards the North, which will cause in swap in the angle which is corrected in the GMRT.} This was also verified at the antenna site and obtained the same result.

\subsubsection{Identify feed rotating axis and the direction of feed rotation}
\begin{figure}
\centering
\includegraphics[width=0.31\textwidth]{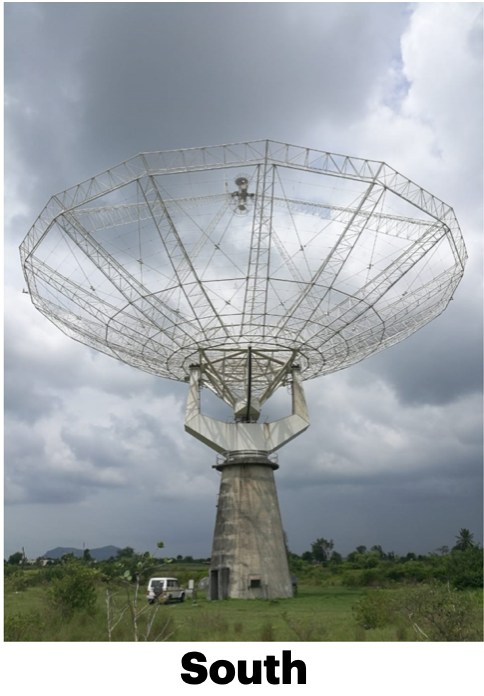}
\includegraphics[width=0.31\textwidth]{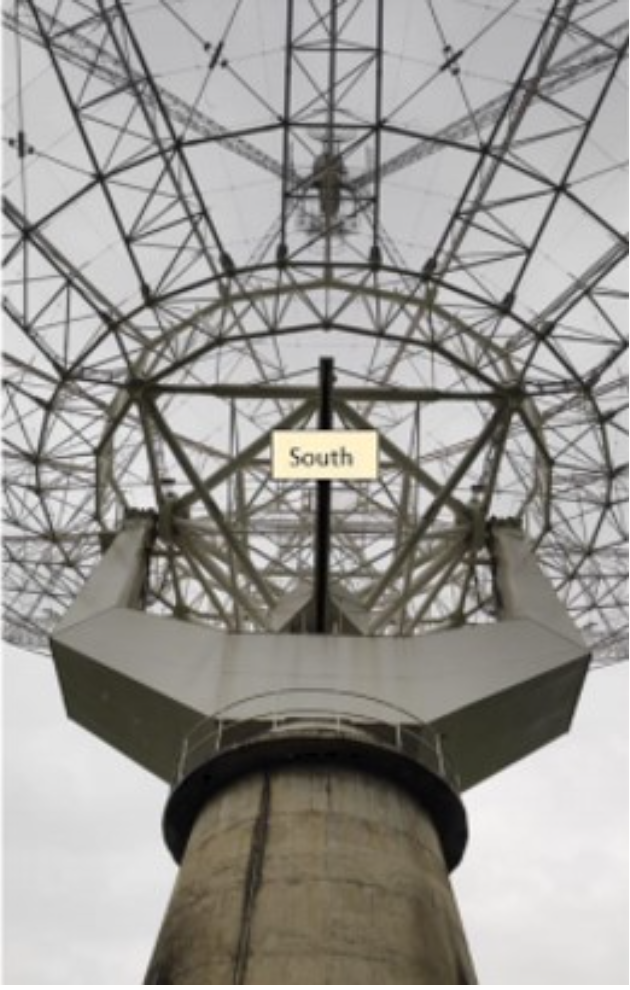}
\includegraphics[width=0.31\textwidth]{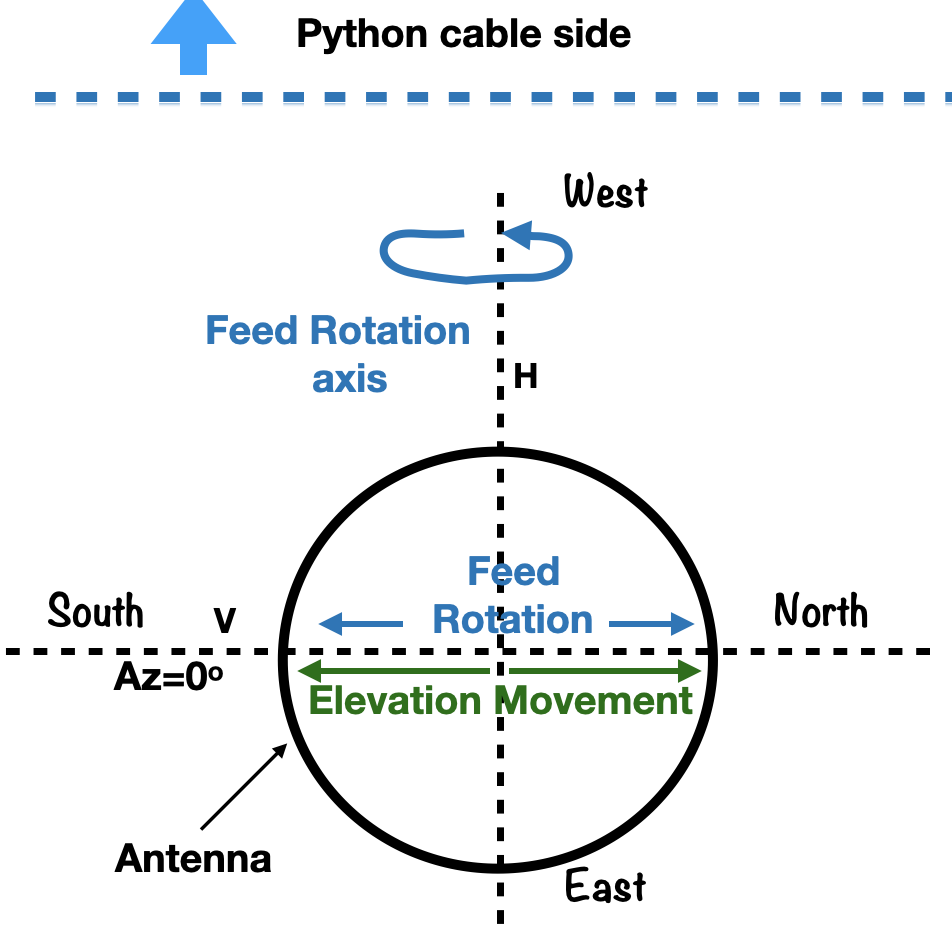}
\caption{ Experiment to identify the feed rotating axis and the direction of feed rotation. The GMRT antenna is positioned at $Az=0^o$ and $EL=90^o$. The Horizontal dipole element is  along East-West Direction and 
the Vertical dipole element is mounted in a North-South direction.
\label{fig:az2}}
\end{figure}

The next step was to determine the direction of the feed rotating axis and feed rotation. For this test, the antenna was positioned at $Az=0^o$ and $EL=90^o$  (Figure \ref{fig:az2}). We find that the feed moves along North-South or South-North direction
i.e along the elevation gear of the antenna. The feed rotating axis is positioned along the East-West direction i.e perpendicular to the elevation gear of the antenna

The Horizontal dipole element (H) is placed along the East-West Direction (Python assembly (West) is along this axis which is taken as a reference to mount the Horizontal dipole element). The Vertical dipole element (V) is mounted along the North-South direction (Fig \ref{fig:az2}).

\begin{figure}
\centering
\includegraphics[width=0.91\textwidth]{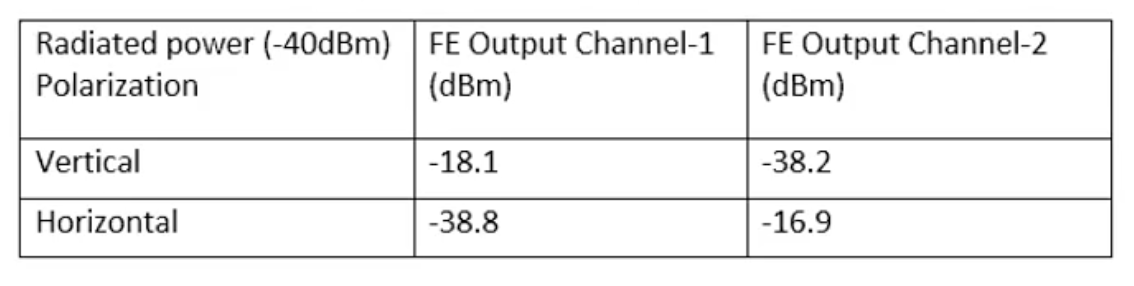}
\caption{label \label{fig:lband}}
\end{figure}

To test these further, we also created a dipole element and put it on the ground, and directly let the feed get it. The horizontal component goes to V and the vertical goes to H (Figure \ref{fig:lband}).

Since IAU/IEEE convention defines North to be $X$ and East to be $Y$, we can conclude that 
the Vertical element receives $X$ component of the electric field $E_x$. The Horizontal polarisation receives $Y$ component of the electric field $E_y$. The Same convention is followed to orient the dipole
elements for band 2, band 3, and band 4.

\subsubsection{Lab test to determine H and V mapping with Channels 1 and 2}
So far we have determined that the H represents $Y$ and V represents $X$. We know that as per IAU/IEEE convention, radiation is RCP if $X$ leads $Y$ and LCP if vice versa (Figure \ref{fig:iau}). Now we have to identify the characteristics of GMRT
channels 1 and 2 at the receiver end. 

\begin{figure*}
\centering
\includegraphics[width=0.46\textwidth]{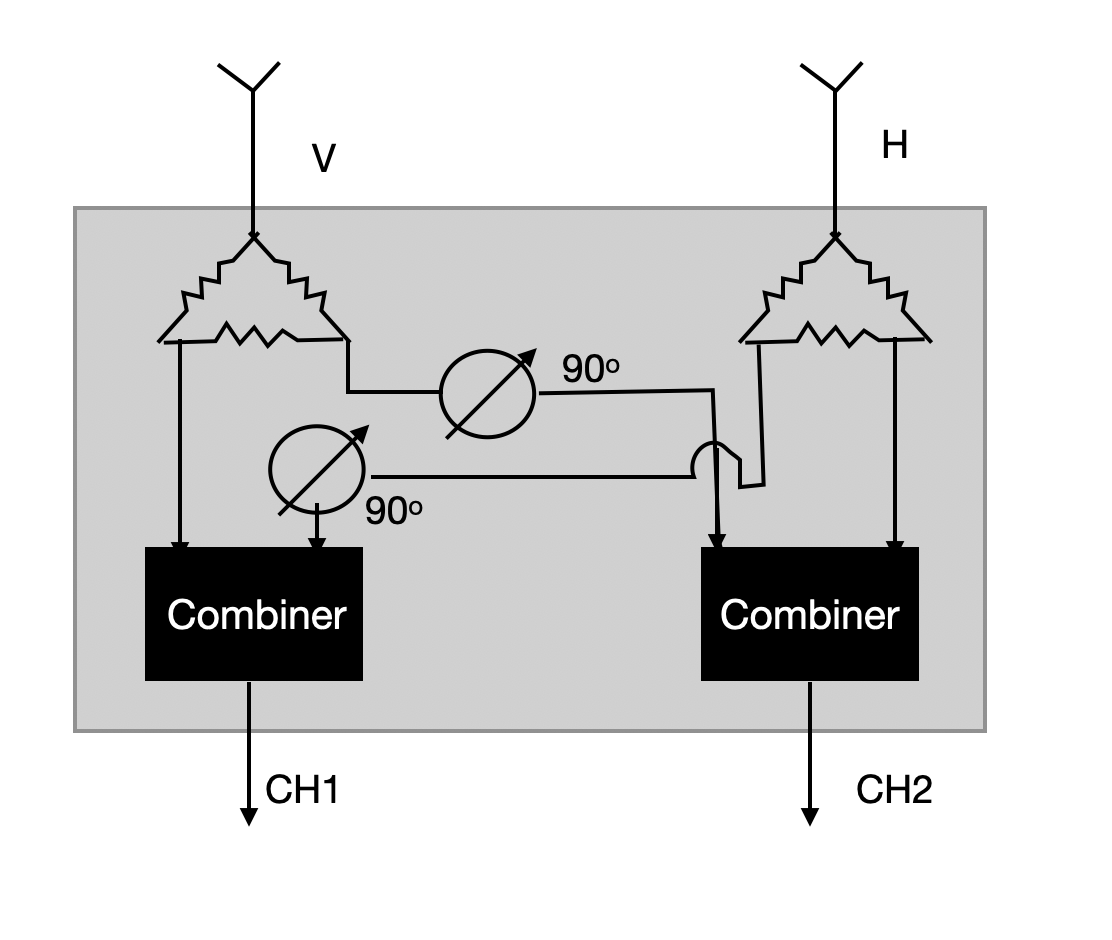}
\caption{The sketch diagram shows the functional diagram of the polariser of all 
bands, which converts the linear polarisation to circular polarisation i.e as Channel 1 and Channel 2.} \label{fig:polariser}
\end{figure*}

As we saw, each GMRT antenna has V and H polarisers, 90 degrees in phase. The polariser at bands 2, 3, and 4 converts linear polarisation appearing in H and V, to circular polarisation appearing in Channel 1 and 2 (Fig \ref{fig:polariser}).

To determine this mapping, first, the signal was fed at V input of the polariser, and H input was terminated with a 50 $\Omega$ resister. V is then split into two beams and the output is recorded in channels 1 and 2, both amplitude and phase. The result shows that while the signal directly comes to Channel 1, it goes to channel 2 with a 90 degrees phase shift. When the same experiment is done with H, channel 2 receives 0 phase shift and channel 1 receives 90 degree phase shift. This has been tracked throughout the electronic chain till the fibre output. Our results from these experiments are summarised in Table \ref{tab:expt}.

Hence we can conclude that,
\begin{equation}
   \begin{split}
       \mathrm{Channel\ 1}= V(0 \,\rm deg)+ H (90 \,\rm deg) =V+iH \\
        \mathrm{Channel\ 2} = H(0 \,\rm deg) + V(90\, \rm deg) = H+iV\\
   \end{split} 
   \label{eq:40}
\end{equation}

Since previous experiment showed $V\equiv X$ and $H\equiv Y$, hence,
\begin{equation}
    \begin{split}
        \mathrm{Channel\ 1}=X+iY \\
        \mathrm{Channel\ 2}=Y+iX \\
    \end{split}
    \label{eq:41}
\end{equation}

Thus channel 1 is RCP and channel 2 is LCP. {\bf This is a very important result, which shows that the GMRT convention of channel 1 being $RR^*$ and channel 2 being $LL^*$ is correct,  and there is no swap from the feed to} at least till the optical fibre.

\begin{table}
\begin{center}
\caption{Experiment to determine mapping of V and H to channels 1 and 2 \label{tab:expt}}
\begin{tabular}{lcc}
\hline
Signal input & Channel 1 & Channel 2\\
\hline
V & 0 deg & 90 deg\\
H & 90 deg & 0 deg\\
\hline
\end{tabular}
\end{center}
\end{table}

\subsection{Tracking the polarisation from optical fibre to user end}
In the GMRT, the signal flow from the front end to the correlator through many electronic blocks, parallel in two separate electronic chains. There are chances of two polarisation signals to interchange their electronic path at a couple of places. These two chains are called channel 1 (130 MHz polarisation channel in the legacy GMRT) and channel 2 (175 MHz polarisation in the legacy GMRT) and further, the autocorrelation of these polarisations are named RR$^*$ and
LL$^*$ respectively. It may be possible that the incoming R \& L signals have flipped in the receiver chain in which case definitions of Stokes V would have changed its meaning. We performed the following tests to check the conventions of Stokes parameters.

\begin{figure*}
\includegraphics[width=0.6\textwidth]{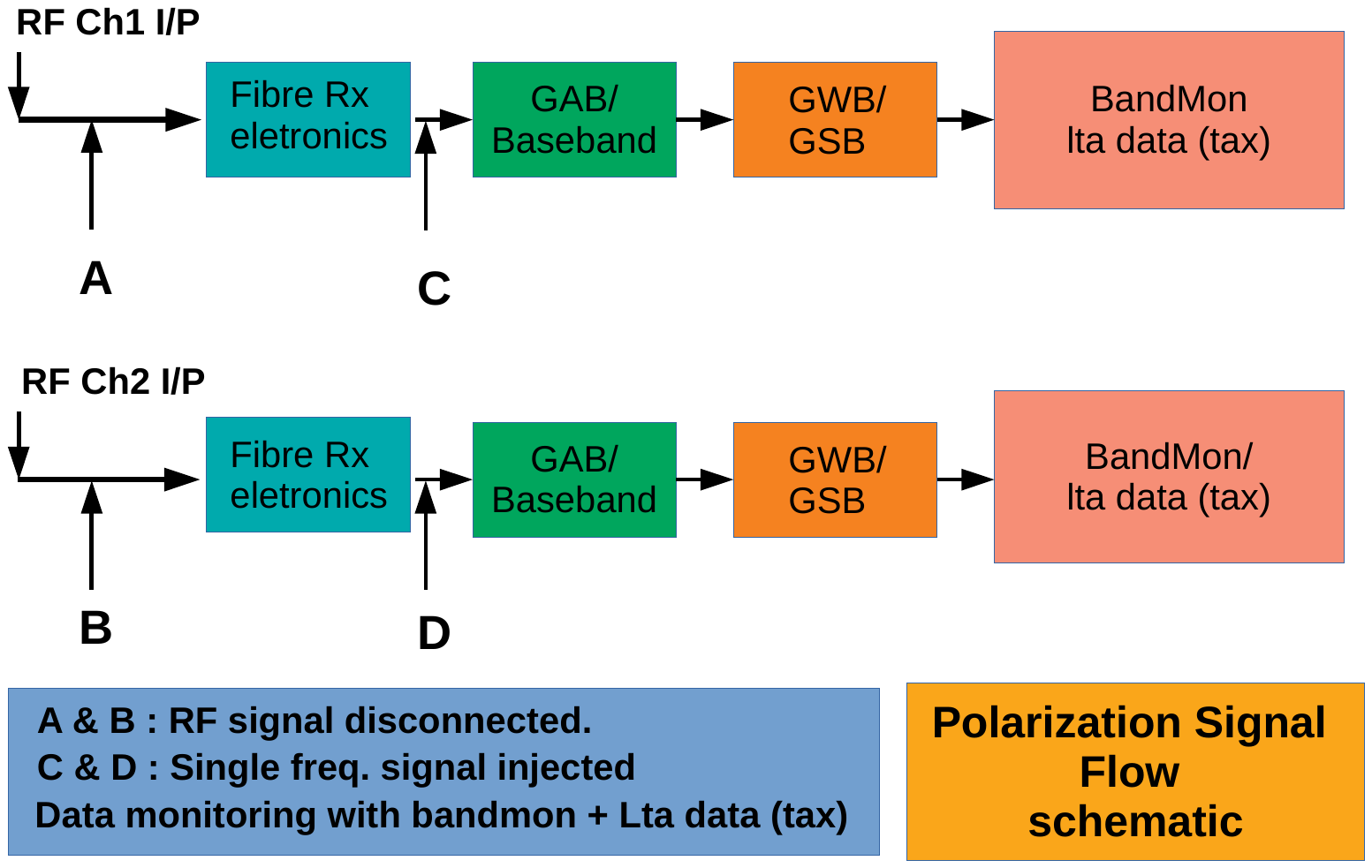}
\caption{u/GMRT signal flow from the optical fibre to the user end. Cable lines come from optical fibre to GWB input. From GWB input, the electric cable goes to the hardware. One signal goes to the interferometer, and another to the beamformer. In correlator room, Channel 1 is on the right side of the correlator facing the wall in the correlator room.}
\label{fig:flow}
\end{figure*}

\vspace*{0.4cm}
{\large{ \bf Test 1}}:\\
In the first test, the signals from the noise generator were fed to V and H dipole elements with a 90 degree phase shift. The signal for the injection is generated in the lab using a noise generator and can be split into two by a power splitter and quadrature hybrid generating two polarisations, say Pol 1 and Pol 2 (90-degree phase shift with respect to Pol 1). The test will not determine the RCP versus LCP nature of the two polarisations but will determine if there is a swap or not  from optical fibre to the user end. We inject this signal at the analog backend input as well as input to the optical fibre at the antenna base. No swap of the signal was found.

\vspace*{0.4cm}
{\large{ \bf Test 2}}:\\
The second test was a mixed setup test where one channel was fed with the output from the antenna and the second from the power generated by the noise generator. Here also, the aim was to look for any kind of swap of cables. For the antenna signal, we pointed the antenna to a calibrator and used the received signal. Several combinations were checked. Data were recorded in visibility and beam modes simultaneously. Below we show various combinations which were used. As mentioned above Pol 1 and Pol 2 refer to noise-generated signals.
\begin{enumerate}
\item  GWB CH-1 was fed with Pol 1 input and GWB CH-2 was connected to the antenna.
\item  GWB CH-1 was fed with Pol 2 input and GWB CH-2 was connected to the antenna.
\item  GWB CH-2 was fed with Pol 1 input and GWB CH-1 was connected to the antenna.
\item  GWB CH-2 was fed with Pol 2 input and GWB CH-1 was connected to the antenna.
\item  GWB CH-1 was fed with Pol 1 input and GWB CH-2 was fed with Pol 2 input.
\item  GWB CH-1 was fed with Pol 2 input and GWB CH-2 was fed with Pol 1 input.
\end{enumerate}

The output was checked at various monitoring levels and using different tools, which are:
\begin{enumerate}
\item {\it  bandmon:} Realtime bandshape monitoring of visibility data.
\item {\it tax:}  Offline data checking of visibility data.
\item {\it CASA} Offline data processing package to monitor visibility (in FITS) + imaging.
\item {\it gptool:} Full polar beam data analysis package.
\end{enumerate}

Following are our findings:
\begin{itemize}
\item We found no swap of polarisation anywhere between the optical fibre to the back end in test 1.
\item When Pol 1  frequency signal is injected, CH-1 is showing higher signals at all four levels listed above.
\item With Pol 2 frequency signal injected, CH-2 is showing signals at all four levels listed above.
\item There is no swap of the signal path from  optical fibre output to GWB output, to beam data, to visibility data to FITS data.
\item Pol 1 represents channel 1, Pol 2 represents channel 2.
\end{itemize}
 
\vspace*{0.4cm}
{\large{ \bf  Test 3}}:\\
Another test was done by Analog Backend team\footnote{courtesy Mr. Abhijit Dhende \& Ms. Sweta Gupta} to see if two polarisation signals are exchanged in their path
using GWB  \& GSB.

{\bf Experiment 1: }
\begin{enumerate}
    \item Signal coming from RF chan 1 input was disconnected (at point A in Figure \ref{fig:flow}).\\
    Findings: GWB Bandmon Pol1 showed power drop, and the fringe vanished.
    \item Signal coming from RF chan 2 input disconnected (at point B in Figure \ref{fig:flow})\\
    Findings: GWB Bandmon Pol2 showed power drop, and the fringe vanished.
\end{enumerate}

{\bf Experiment 2: }
\begin{enumerate}
    \item RF Terminated \& Single frequency injected at input GAB/baseband (at point C in Figure \ref{fig:flow})\\
    Findings: GWB showed in bandmon a single line in the band at the desired location for Pol 1.
    \item RF Terminated \& Single frequency injected at input GAB/baseband (at point D in Figure \ref{fig:flow})\\
   Findings: GWB showed in bandmon a single line in the band at the desired location for Pol 2.
\end{enumerate}
Hence this proves that there is no swap in the signal path.

\subsection{Observational test results using astronomical source}\label{sec:astro_test}
Now we explore the effect of observational settings and calibration in the actual astronomical data. Several test observations were carried out to understand polarisation convention by analyzing beam as well as interferometric data (Table \ref{tab:obs}). The target pulsars were chosen such that their polarisation properties are well known with significant amount of both linear and circular polarisation. Our chosen targets are PSR B0740- 28, PSR B0138+57 and PSR  B1702-19. 
 
Band 3 \& band 4 observations were carried in default sideband mode for basic tests, i.e. band 4 in USB (550 -- 750 MHz), band 3 in LSB (500 -- 300 MHz). Additionally, we carried observations in the reversed side-band mode in a few observations. Data were recorded in full polar mode with 2048 channel in all observations. Band 3 \& band 4 observations are recorded in 200 MHz bandwidth mode some band 4 observations were carried out in 100 MHz bandwidth mode). Time resolution (sampling interval)  differ from one observation to another, which was chosen based on pulsar period, available disk space and on-pulse duty cycle. {\bf We now discuss beam data results in band 4 and all  interferometric data analysis and results. The details of beamformer data and other bands' results will be discussed by Sanjay Kudale in a 
separate technical report.}

\subsubsection{Beamformer observation (pulsar mode observation)}\label{sec:new_pulsarmode_obs}
Beamformer observation has an advantage over imaging analysis that it produce high time resolution Stokes profile of the pulsar and swing of polarisation position angle of linear polarisation across pulse (PPA),  which can be directly compared with the pulsar profiles obtained from EPN database. Beam data were recorded with the GWB which offers the beam products as $RR^*$, $Im(RL^*)$, $LL^*$ and $Re(RL^*)$. To check data quality first analysis was done with gptool. After data was found to be good, it was further processed with analysis process described in Kudale \& Gupta, 2008 (NCRA : internal technical report). {\bf The pipeline is modified and details of the modifications are subject to another GMRT technical report by Sanjay Kudale.}

\begin{table}[]
\caption {Details of all the test observations} \label{tab:obs} 
\begin{tabular}{lllllllll}
\hline
Sr & Date of     & Target  & Freq & Side  & Starting  & BW & Samp     & Stokes Q, U, V      \\
   & Observation &             & Band      &     Band      & Freq, MHz           & MHz        & $\mu$sec & IAU/IEEE            \\
   \hline
1  & 19 Mar 2020 & B1702-19    & B4        & USB       & 550           & 200        & 327.68   & $-$23\%, $-$10\%, +15\% \\
2  & 20 May 2020 & B1702-19    & B3        & LSB       & 500           & 200        & 327.68   & $+$8\%, $-$19\%, +26\%  \\
3  & 13 Feb 2021 & B1702-19    & B4        & USB       & 550           & 200        & 163.84   & $-$23\%, $-$10\%, +15\% \\
$\**$  4  & 04 Dec 2021 & B1702-19    & B3        & LSB       & 500           & 200        & 327.68   & $+$8\%, $-$19\%, +26\%  \\
   \hline
4  & 12 Feb 2021 & B0136+57    & B4        & USB       & 550           & 200        & 163.84   & $-$27\%, $-$64\%, $-$16\% \\
5  & 09 Mar 2021 & B0136+57    & B3        & LSB       & 500           & 200        & 163.84   & $-$62\%, $-$26\%, $-$5\%  \\
6  & 19 Mar 2021 & B0136+57    & B4        & USB       & 550           & 100        & 327.68   & $-$27\%, $-$64\%, $-$16\% \\
7  & 19 Mar 2021 & B0136+57    & B4        & LSB       & 650           & 100        & 327.68   & $-$27\%, $-$64\%, $-$16\% \\
8  & 31 May 2021 & B0136+57    & B3        & LSB       & 500           & 200        & 163.84   & $-$62\%, $-$26\%, $-$5\%  \\
   \hline
9  & 24 Mar 2021 & B0740-28    & B4        & USB       & 550           & 100        & 163.84   &       +59\%, +88\%, +5\%              \\
10 & 24 Mar 2021 & B0740-28    & B4        & LSB       & 650           & 100        & 163.84   &     +59\%, +88\%, +5\%                  \\
11 & 24 Mar 2021 & B0740-28    & B3        & USB       & 400           & 100        & 163.84   &       $-$63\%, +34\%, +6\%                \\
12 & 24 Mar 2021 & B0740-28    & B3        & LSB       & 500           & 100        & 163.84   &            $-$63\%, +34\%, +6\%           \\
$\**$ 13 & 2 May 2021  & B0740-28    & B4        & LSB       & 650           & 100        & 327.68   &      +59\%, +88\%, +5\%                 \\
$\**$ 14 & 4 May 2021  & B0740-28    & B4        & USB       & 550           & 100        & 163.84   &     +59\%, +88\%, +5\%                  \\
$\**$ 15 & 4 May 2021  & B0740-28    & B4        & LSB       & 650           & 100        & 163.84   &    +59\%, +88\%, +5\%     \\
   \hline             
\end{tabular}
$\**$ These observations were done with RF SWAP switch.
\end{table}

\vspace*{0.4cm}
\noindent {\bf \large{Band 4 observations:}}\\
%4 Band 4 Observation :
%{\color{green} Comment: Dk, the result of this section actually right thing to present, because no such arbitrary PPA offset is not added here.}
Band 4 observations were carried out  in many modes, including USB, LSB as well as RF swap model. To record in LSB and USB mode of uGMRT, we could record only 100 MHz bandwidth. LSB \& USB observations were  separated by very short time (about an hour). A few  observations 
were also  carried with RF swap applied to check the effect of exchanging R-voltage and L-voltage on final profiles.

\vspace*{0.4cm}
\noindent {\bf  LSB vs USB observations:}\\

\begin{figure}
\centering
\includegraphics[width=0.44\textwidth]{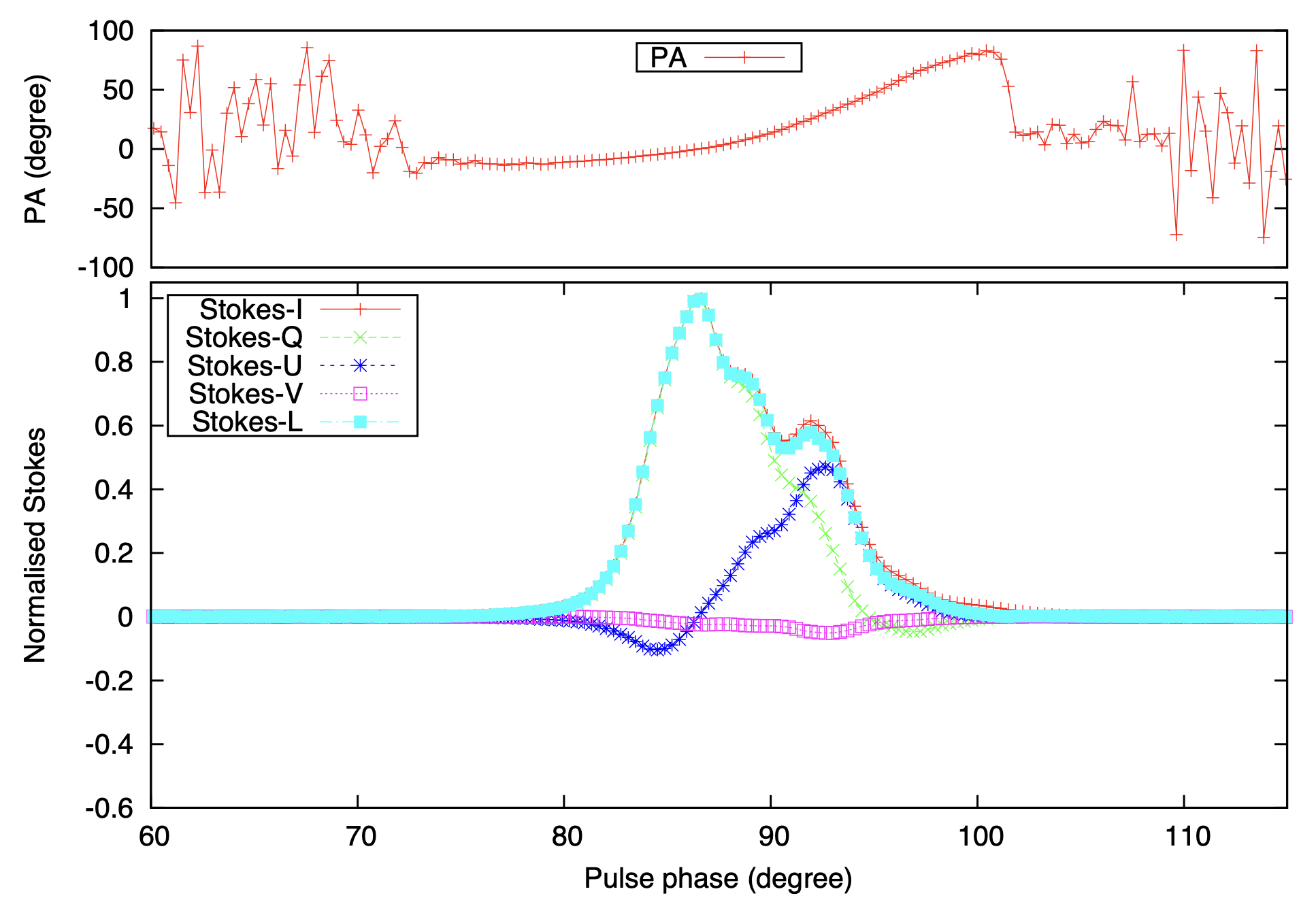}
\includegraphics[width=0.44\textwidth]{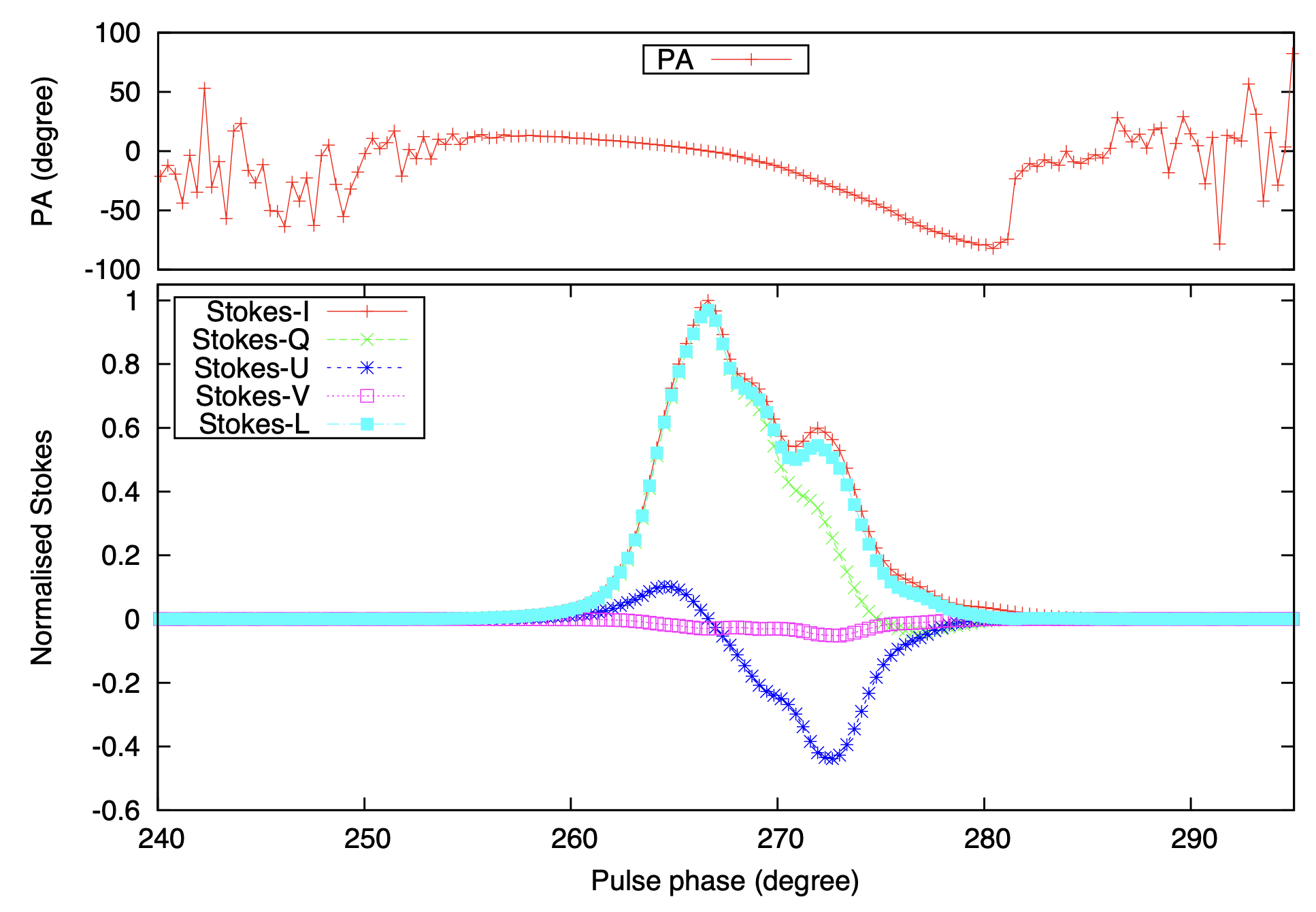}
\includegraphics[width=0.44\textwidth]{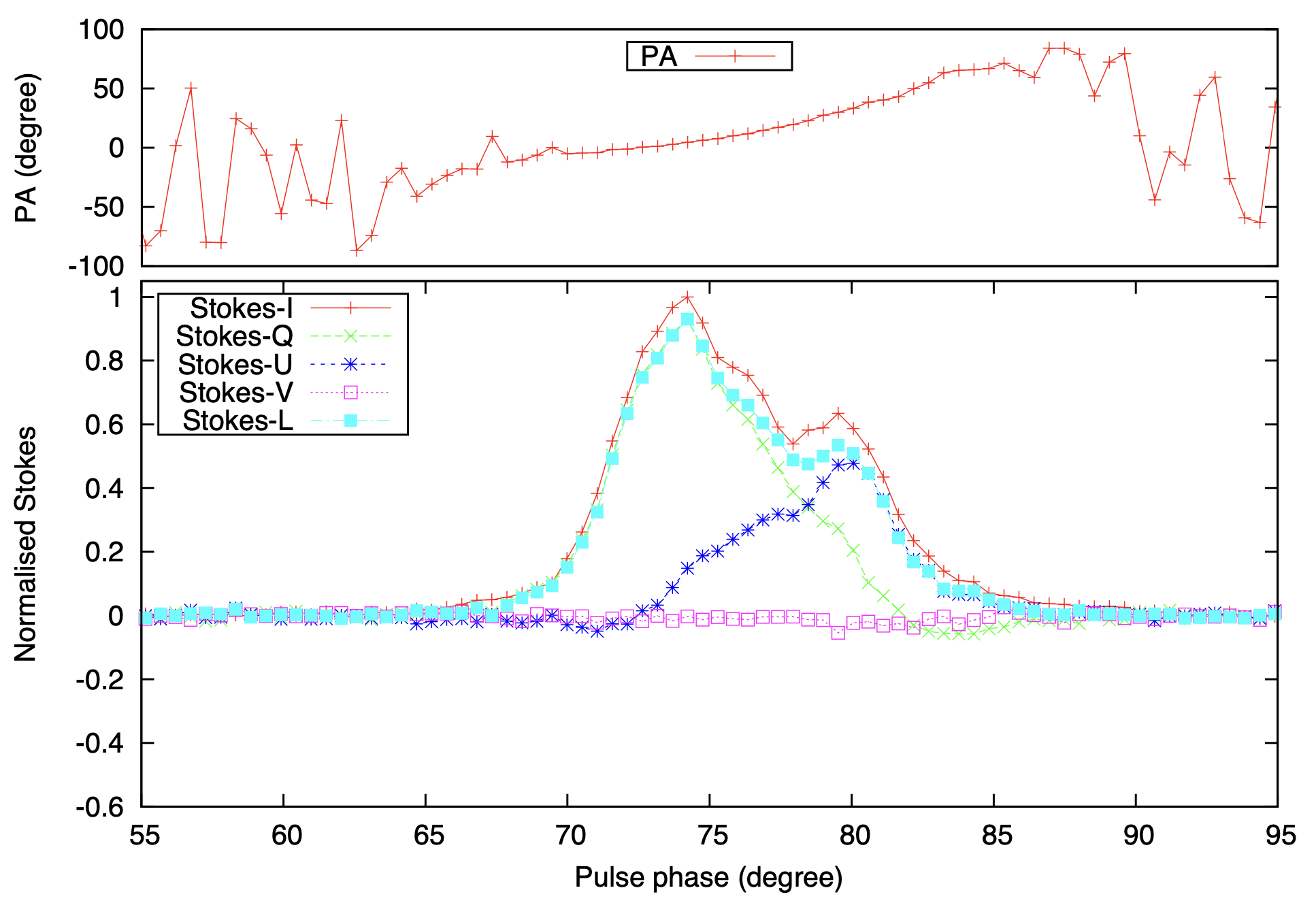}
\includegraphics[width=0.44\textwidth]{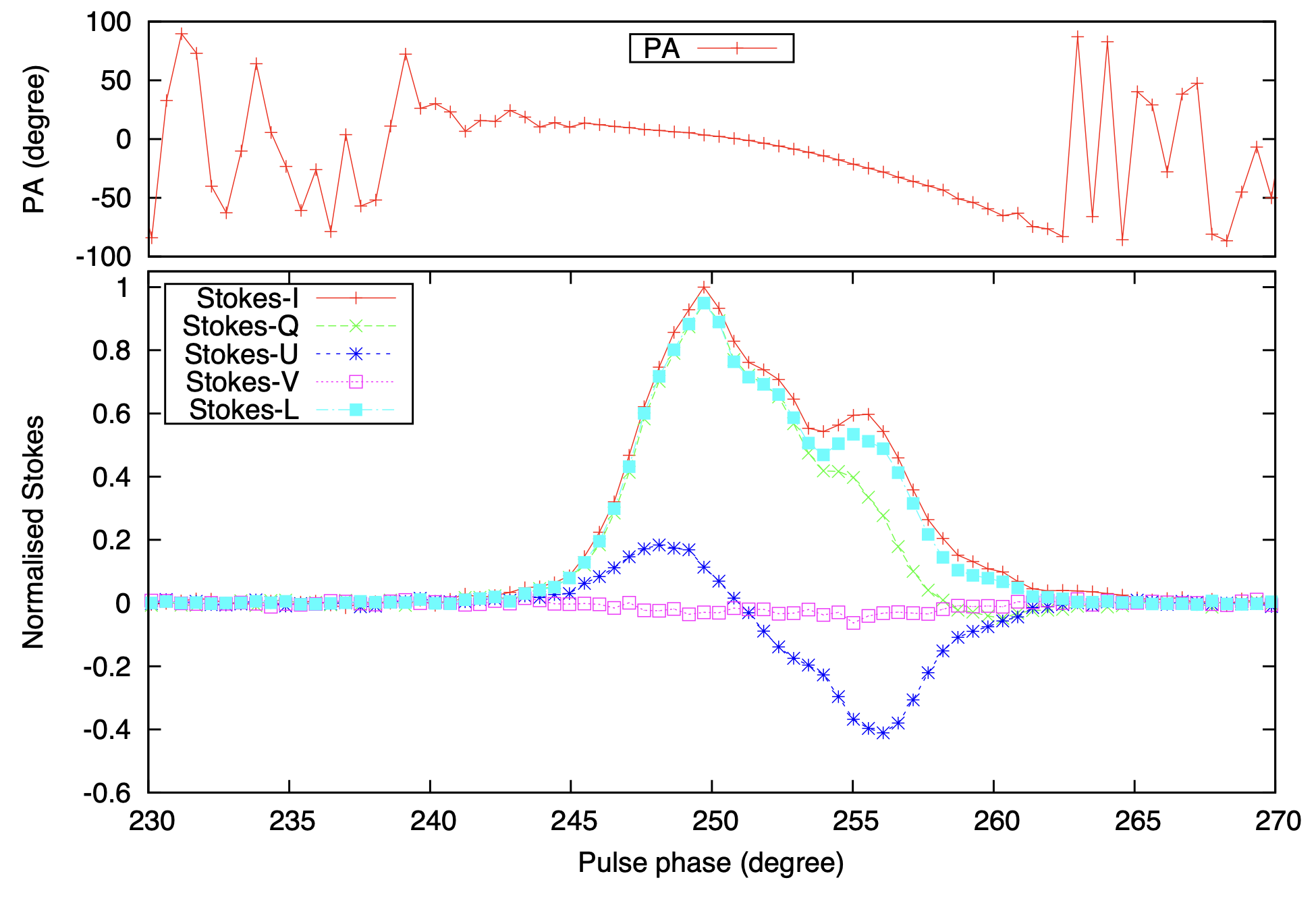}
\caption{Top panel: The uGMRT Stokes profile of PSR B0740-28 observed at band 4 in USB mode (left) and LSB mode (right).
%along with EPN profile on the right.
Bottom panel: Same as top left two panels but for GSB 610 MHz data.
 \label{fig:b4lsbusb}}
\end{figure}

To check how  the Stokes profile vary from one frequency to other and from lower sideband to upper sideband we observed pulsar B0740-28  with both sidebands with recording bandwidth to be 100 MHz. All these observations were  carried over just 3.5 hours so that there is no major  change any environment which can cause profile change. During analysis PPA is made zero at the peak in total intensity (and linear polarisation) pulse profile.  We show band 4 profiles in (Figure \ref{fig:b4lsbusb}), along with
610 MHz EPN profile corrected for IAU/IEEE convention.

For given frequency band profiles of LSB \& USB are exactly opposite for Stokes-U, Stokes-V  and for the PPA. 
 These results are consistent for GWB as well as GSB. %(Fig \ref{fig:b4gsblsbusb}).
The profiles obtained by these are showing match band3 \& band4 for LSB and band3 \& band4 for USB separately (Figure \ref{fig:b4lsbusb} and Figure \ref{fig:b3lsbusb}).  

\begin{figure}
\centering
\includegraphics[width=0.44\textwidth]{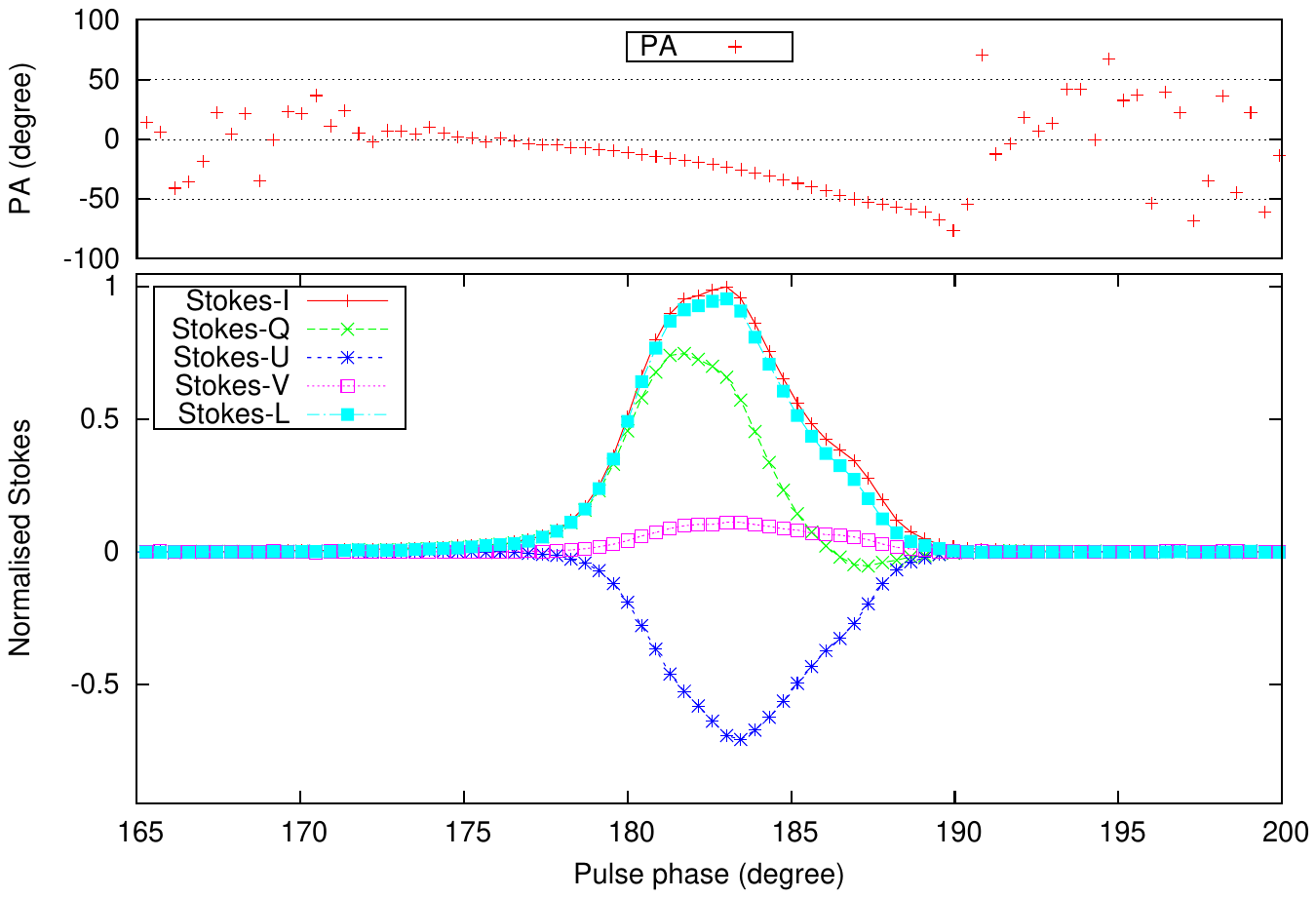}
\includegraphics[width=0.44\textwidth]{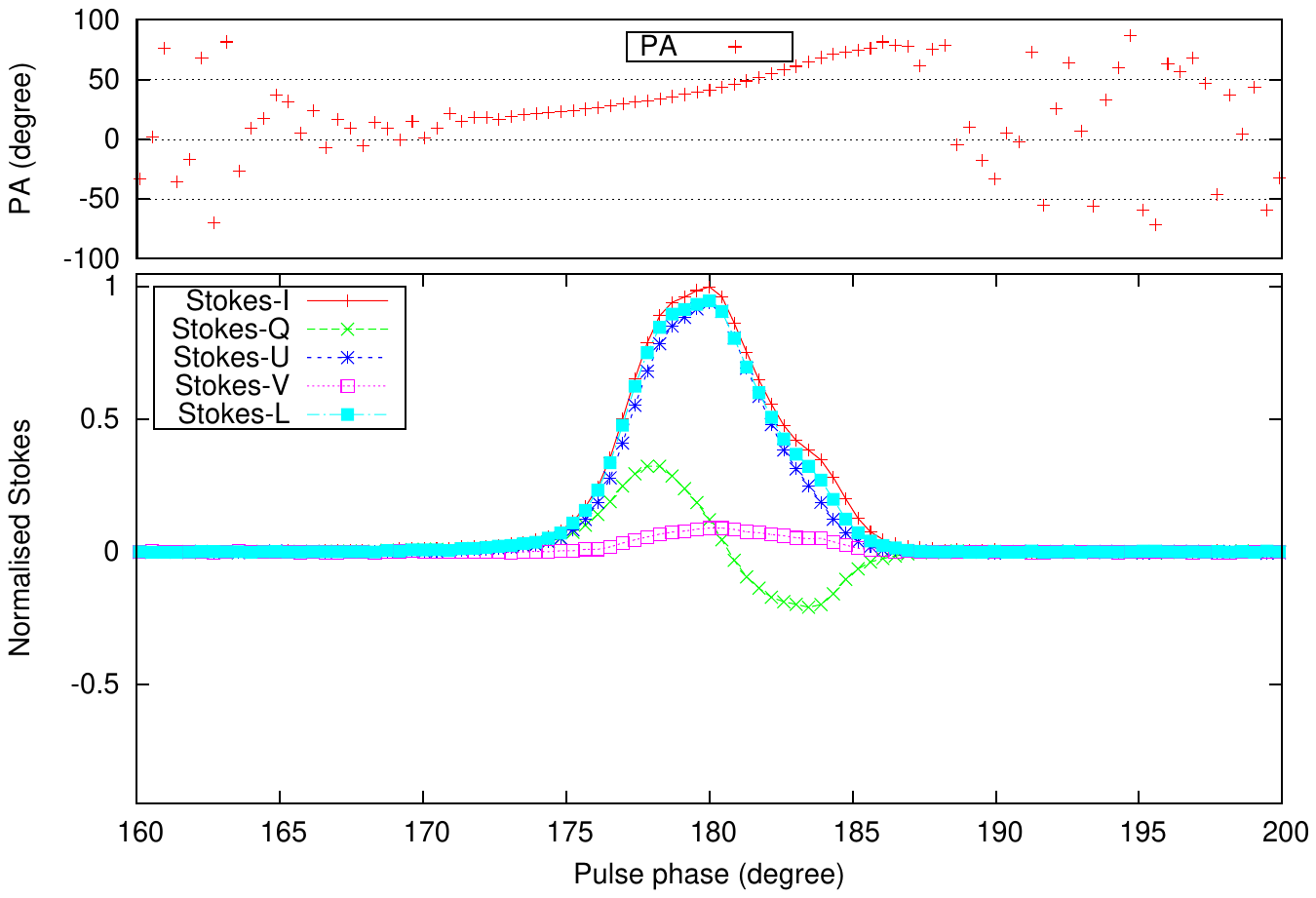}
\caption{The  uGMRT Stokes profile of PSR B0136+57 observed at band 4 in USB mode (left) and LSB mode (middle), 
along with EPN profile on the right. Stokes U indeed seems to invert in USB vs LSB.
%Bottom panel: Same as top left two panels but for GSB 610 MHz data.
 \label{fig:b4lsbusb1}}
\end{figure}

The same trend can also be seen in the pulsar B0136+57 (Fig \ref{fig:b4lsbusb1}).

%To check if software is generating
%inversion of Stokes-U in analysis pipeline, we carried analysis of same data in two ways. The recorded USB data was converted to LSB by offline software by inverting channel sequence. 
%With the required changes in parameters, e.g. start frequency, sideband etc in psr analyse.in and giving appropriate sign for Faraday rotation calculation in calibrate we generated Stokes profiles for these cases. 
%For comparison purpose we made PPA zero in peak power in total intensity of the pulsar. With PPA setting to zero, the slope of cross-phase across channels (linear curve) is seen opposite to each other. The profiles produced from recorded USB data and converted LSB data are similar to each other confirming that inversion of Stokes-U is not generated in software, instead it is carried in with data depending upon whether the observations are taken in LSB or USB.

\begin{figure}
\centering
\includegraphics[width=0.34\textwidth]{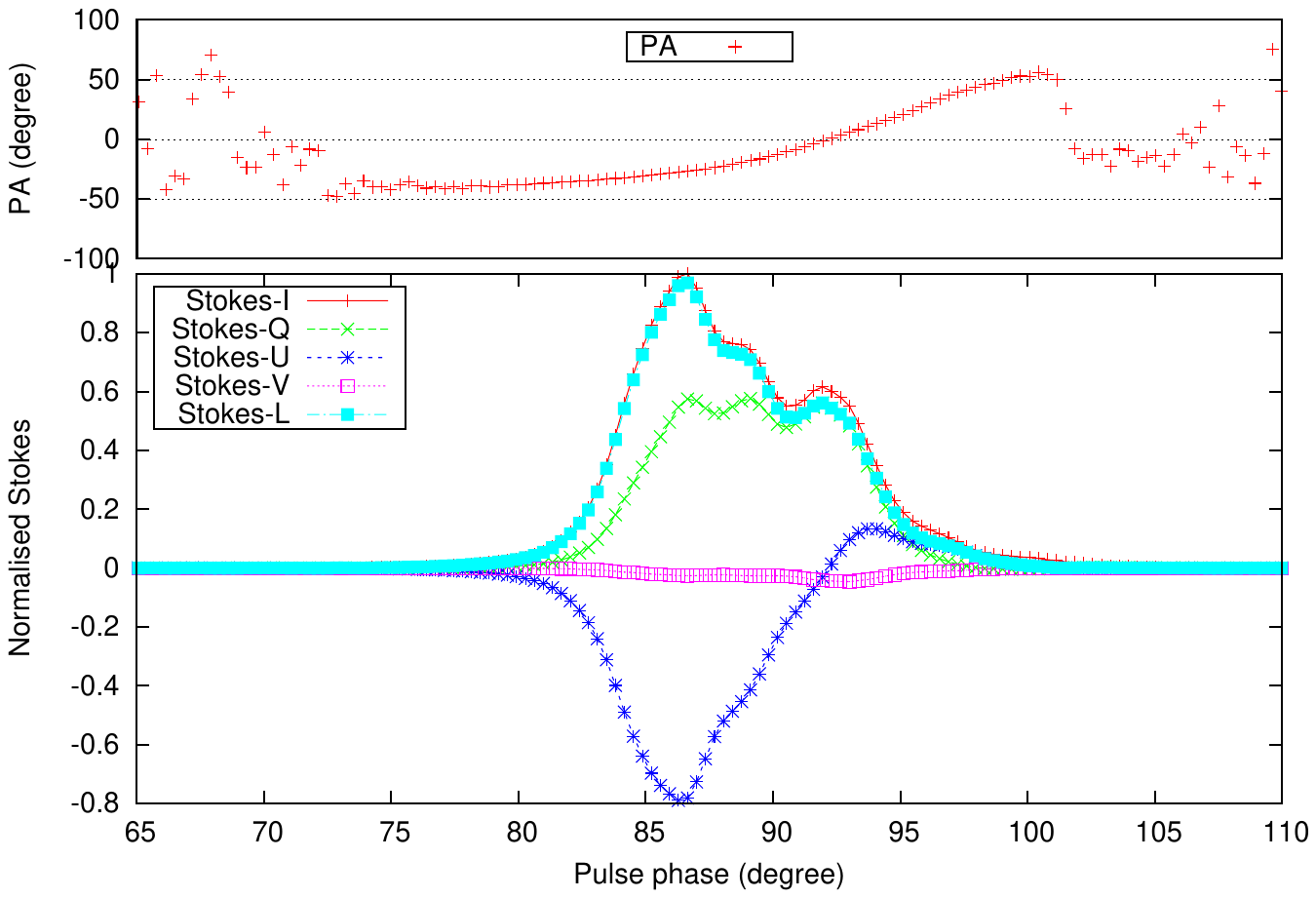}
\includegraphics[width=0.34\textwidth]{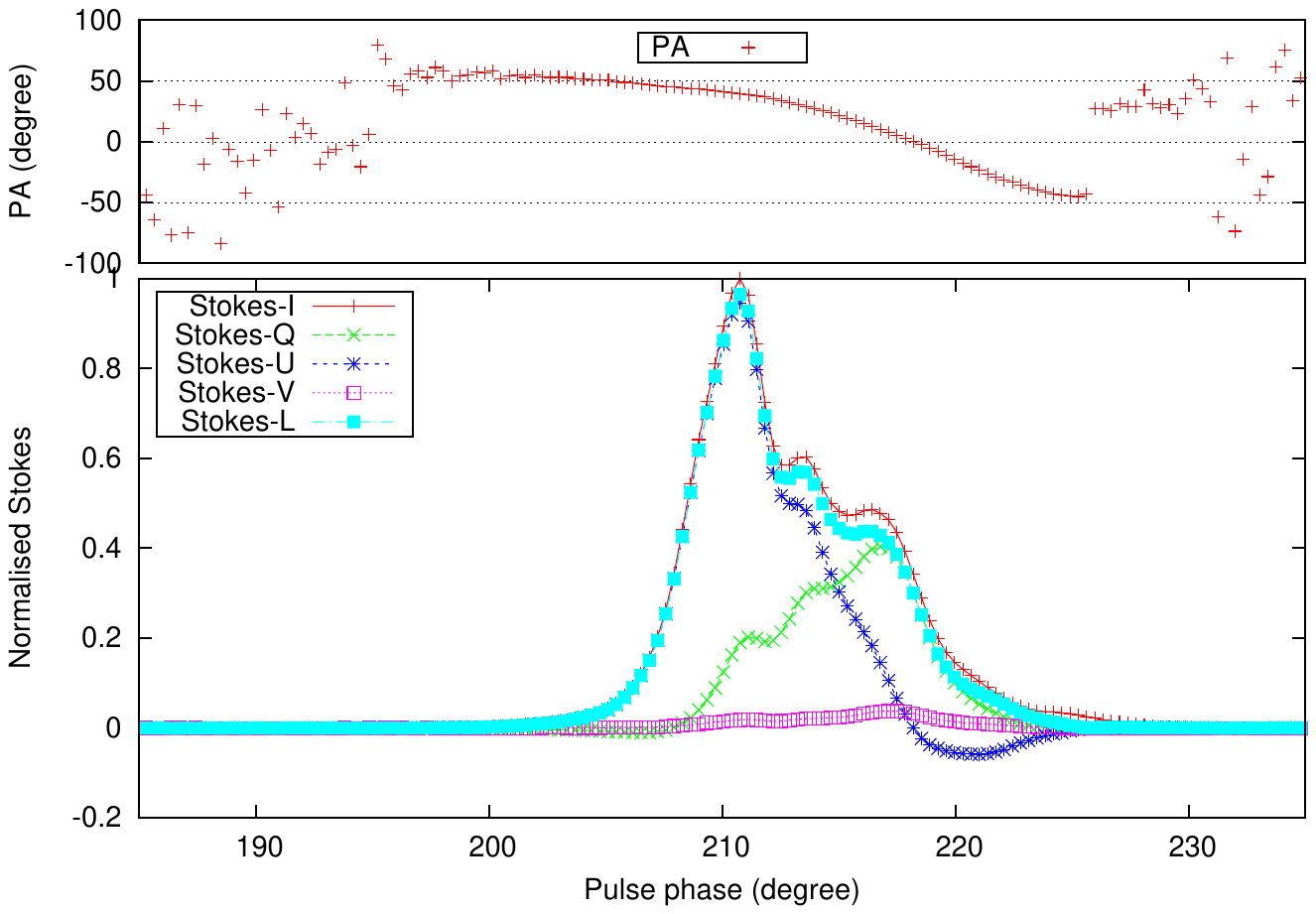}
\includegraphics[width=0.3\textwidth]{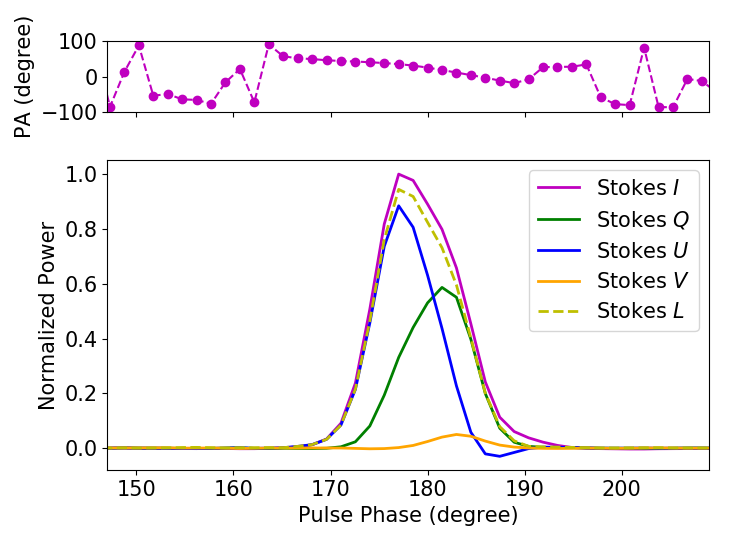}
\caption{Left panel shows USB band 4 data for PSR B0740-28 . Middle panel shows the RF SWAP data for the same pulsar and right are the EPN profile. 
%Bottom row: Top row: Left panel shows LSB band 4 data for B0740-28. Middle panel shows the RF SWAP data for the same pulsar for LSB (here Stokes $U$ is inverted manually) and right are the EPN profile. 
One can see swapping 'R' and 'L' in the USB mode data, the results match with the EPN profile.
 \label{fig:swap}}
\end{figure}

\vspace*{0.4cm}
\noindent {\bf  RF swap observation:}\\
As we mentioned in section 2, the GMRT antennas have RF swap switch at the end of the common box, which when 
set in swap mode,  interchange the signal path and interchange R and L.
We carried two observations of PSR B0740-28  to check if R \& L channels are exchanged how it would affect the final full Stokes profiles. In figure \ref{fig:swap}, we show USB and LSB profiles for regular data, data with RF swap and EPN data.

The result of above beam observations in different settings are the following:
\begin{itemize}
\item In USB observations, after PPA correction, Stokes $U$ and $V$ remain mismatch from that of EPN profiles.
\item In LSB mode, Stokes $U$ and PA are inverted than that of USB data.
\item For observations made in RF swap mode, in USB mode, all Stokes profiles match with that of EPN data.
\item For observations made in RF swap mode, in LSB mode, all Stokes profiles except Stokes $U$ and PA match with that of EPN data.
%\item Above results are true for both bands 3 and 4. {\color{green} Comment :DK, If this is true, then band 3 USB and LSB, should shown different profiles before RF swap, which is not true if we look at row 11 and 12 (24 Mar obs on B0740-28) in Table 2. Please confirm from Sanjay. If the result is similar as band 4, then we should show some plots similar to Figure 23.}
\end{itemize}

\subsubsection{Interferometric data}
We now focus on the results of the interferometric data. For interferometric observations, we have used the pulse phase averaged values of the flux densities and polarisation fractions obtained from EPN database for each of the target pulsars. EPN database followed PSR/IEEE convention. We have converted those values in IAU/IEEE convention.

Polarisation calibration has been performed before deriving the Stokes information. Calibration has been performed in \textsc{CASA}. We have calibrated the following effects as a part of instrumental polarisation calibration: 
\begin{enumerate}
\item {\bf Instrumental bandpass and flux density scale:} Using the flux density calibrator.
\item {\bf  Time variable instrumental gain:}  Phase calibrator close to the target source.
\item {\bf Cross-hand delay:}  Cross-hand delay causes a phase ramp with frequency between R and L polarisation channels. Cross-hand delay can be calibrated using a polarised calibrator. We have used the strong polarisation angle calibrator for this purpose.
\item {\bf  Polarisation leakage:}  Polarisation leakage has been calibrated using the unpolarised leakage calibrator. In one epoch, 21 May 2020, we have used the phase calibrator 1822-096 for leakage calibration, because no unpolarised calibrator observation was available for this epoch.  We have verified that 1822-096 does not have strong polarisation, hence can be used as a unpolarised calibrator at uGMRT frequencies.
\item {\bf  Cross-hand phase and absolute polarisation angle:} In circular basis, the cross-hand phase (phase difference between R-L polarisation channels) and absolute polarisation angle calibration are degenerate. Both of them are calibrated together using the polarisation angle calibrator 3C 138 or 3C 286.
\end{enumerate}

Below we discuss the data analysis procedure.
\vspace*{0.4cm}

{\bf Flagging :}\\
We have performed standard flagging of the bad antennas and bad channels. At the initial stage, strong RFIs are flagged using CASA task {\em flagdata} using its {\em tfcrop} mode. 
We also performed flagging in between calibration processes using {\em rflag} mode of {\em flagdata} task. Since central square baselines are more prone to RFI and some of our 
data is also affected by the issue with central square baselines\footnote{\href{http://www.gmrt.ncra.tifr.res.in/gmrt\_users/help/csq\_baselines.html}{Central Square Issues}}, we have 
flagged all central square baselines from the beginning of the data analysis.

\vspace*{0.4cm}
{\bf Model import :}
For polarisation calibration, one needs a full Stokes model of the calibrator. The standard models available in \textsc{CASA} are the Stokes I model only. Thus those models have not been used for the polarisation angle calibrators; 3C 286 and 3C 138. For phase calibrators and instrumental leakage calibrators, the standard Stokes I model has been used. The polarisation model of 3C 286 and 3C 138 available in \cite{perley2013}, have been used here. The \textsc{CASA} task {\em setjy} has been used in $standard=manual$ mode to import the polarisation models.

3C 286 is a very good polarisation angle calibrator and has polarisation fraction $\sim 9$\%, and the polarisation angle is almost constant over our observing frequency. 3C 138 on the other hand {\color{red}has} $\sim$6\% linear polarisation, and the polarisation angle varies with frequency. Thus two avoid bandwidth depolarisation (particularly for 3C138), calibration has been performed at the high-frequency resolution ($\sim 97$ kHz). The flux density of 3C 138 is a little lower compared to 3C 138, thus large integration time has been used for 3C 138.

\vspace*{0.4cm}
{\bf Calibration steps:}
\begin{itemize}
\item Step 1: First we performed calibration of the flux density calibrator. We choose a small chunk of good channels ($\sim 5-10$ MHz) and perform a phase-only gain calibration to calculate the phase variation over the entire scan of the flux calibrator. We used {\em gaincal} task only over the selected channels.
\item Step 2: We then used this phase solutions and performed the geometric delay calibration using {\em gaincal} task with {\em gaintype=`k'} and bandpass calibration over the entire band using bandpass task. We obtained a normalized bandpass solution.
\item Step 3: Using the delay and bandpass solution, we have now estimated amplitude and phase variations of the flux density calibrator and phase calibrator. After these three steps, we performed some level of flagging on the residual visibilities and repeat these steps until we found that there are no significant RFI present.
\item Step 4: When we converged on the previous steps, we first apply the gain and bandpass solutions on the unpolarised leakage calibrator and on the polarisation angle calibrators. After applying the solutions, we performed a flagging using {\em tfcrop} mode of {\em flagdata}.
\item Step 5: Then cross-hand delay is estimated from the polarised calibrator sources. In this case, we used the polarisation angle calibrator source (3C 286 or 3C 138). The cross-hand delays are estimated using the entire observing band using \textsc{CASA} task {\em gaincal} with {\em gaintype=`KCROSS'}.
Step 6: After applying the cross-hand phase solutions, instrumental leakages are estimated using the un-polarised calibrator source. Leakages are estimated for every frequency channel separately. We have used \textsc{CASA} task {\em polcal} with the {\em poltype=`Df'} and with {\em solint=`ınf'}. This calibrates the leakage using un-polarised source per frequency channels.  
\item Step 7: After applying the correction for the instrumental leakages, the cross-hand phases are estimated for every frequency channel using the polarisation angle calibrator. This step performs the absolute polarisation angle calibration. \textsc{CASA} task {\em polcal} has been used for this purpose with {\em poltype=`Xf'}, which denotes the polarisation angle calibration per frequency channels.

Once all the calibrations are done, the calibration solutions are applied to the target data. The corrected target data is then flagged for RFIs. Since the calibration method used in \textsc{CASA} is only valid for the low-level instrumental leakages (< 10\%), the calibration solutions are flagged when the instrumental leakage for an antenna at a certain frequency is more than 10\%. After flagging these high leakage frequency channels per antenna, Steps 1 to Step 7 are repeated from scratch. When this is done, several self-calibration iterations are performed to further improve the image dynamic range.

\item Step 8: Self-calibration is performed using separate sky models for both the co-polar visibilities (RR and LL). This is done to make sure that if there is a strong circularly polarised source present in the field, the circular polarisation fraction remains the same with the self-calibration iteration. Ideally, one should perform self-calibration until the dynamic range is conserved. To reduce the computational burden, three rounds of phase-only and four rounds of amplitude-phase self-calibration have been performed. A single flagging is performed after the first amplitude-phase self-calibration round.
\end{itemize}

\vspace*{0.4cm}
{\bf  Final Imaging:}
Once the self-calibration is done, the final spectral cube of Stokes parameters is made. At different epochs, depending on the integration time, spectral cubes are made at smaller or larger frequency resolutions. Only Stokes Q and U spectral cubes are made and  for  Stokes V (circular polarisation) a continuum image is made over the entire band. The spectral slices where the source is not detected either in Stokes Q or U are removed from further analysis.

\vspace*{0.4cm}
{\bf RM fit:}
RM fit was performed on each separately using Stokes Q and Stokes U spectral cube. PSR B1702-19 has low RM ($RM = -19$ rad\,m$^{-2}$) with sufficient circular polarisation, however, the linear polarisation fraction is low. PSR B0136+57 has high RM ($RM = -94$ rad\,m$^{-2}$), with large fraction linear polarisation but low circular polarisation. Thus it is a good source for the sign check of Stokes Q and Stokes U.
%Polarisation properties pulse phased average values, which we will detect in the image plane.

\begin{figure}
\centering
\includegraphics[width=0.45\textwidth]{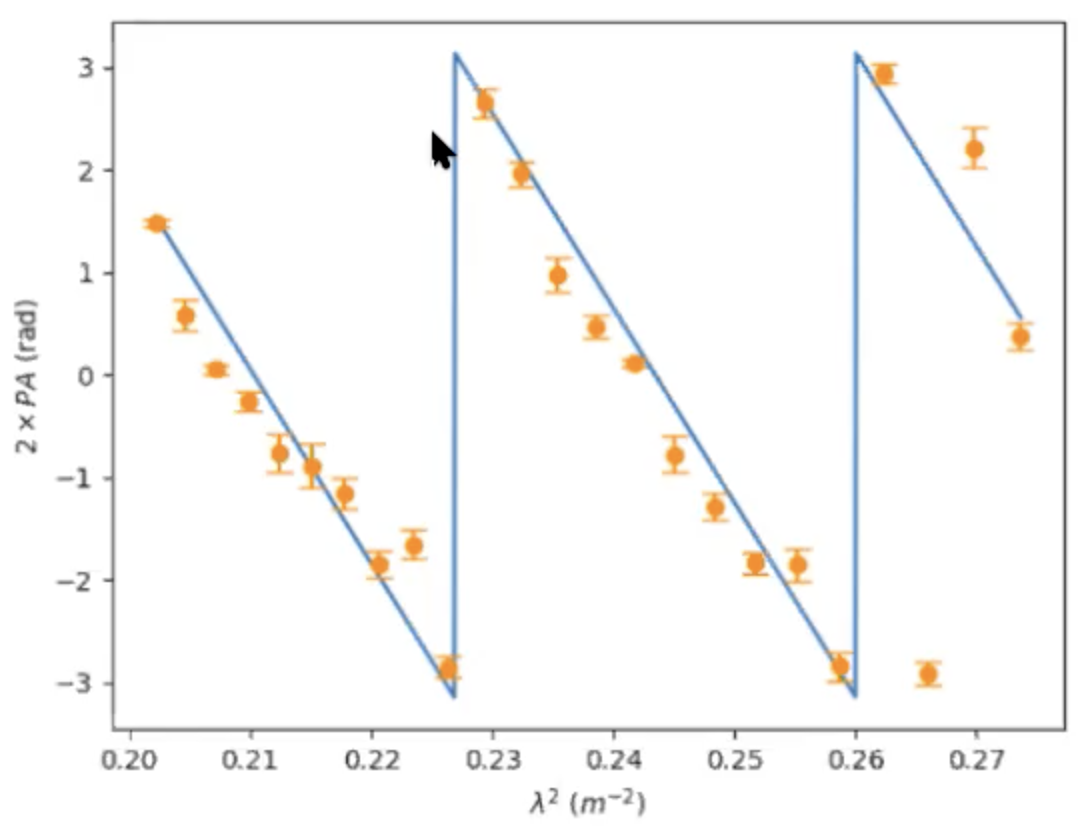}
\caption{RM fit to PSR B0136+57, band 4 USB data on 12 Feb 2021. \label{fig:exp1}}
\end{figure}

Below are our results:

\vspace*{0.5cm}
\noindent{\bf \large{PSR B0136+57, band 4 USB}}\\
\begin{itemize}
\item Pulse averaged source polarisation properties (IAU/IEEE):  RM $= -94$ rad/m$^2$, Stokes Q = -27\%, U = -64\%, V = -16\%
\item Observations date and settings: 12 Feb 2021; Band 4, LO I: 550 MHz, USB
\item Polarisation leakage and position angle calibration was done using 3C147 and 3C138 respectively.
\item Post calibration and post RM fit results:
\begin{enumerate}
\item RMfit = $-90$ rad/m$^2$
\item  $2\chi_0$ ($\chi_0$ = Intrinsic position angle) = 123 degree (Figure \ref{fig:exp1}).
\item Stokes Q $= -$ve, Stokes $U = +$ve, Stokes $V = +$ve.
\item Result: Stokes U and V flipped, and Stokes Q remained the same.
\end{enumerate}
\end{itemize}

\vspace*{0.5cm}
\noindent{\bf \large{PSR B1702-19,   band 4 USB}}\\

\begin{figure}[h]
\centering
\includegraphics[width=0.45\textwidth]{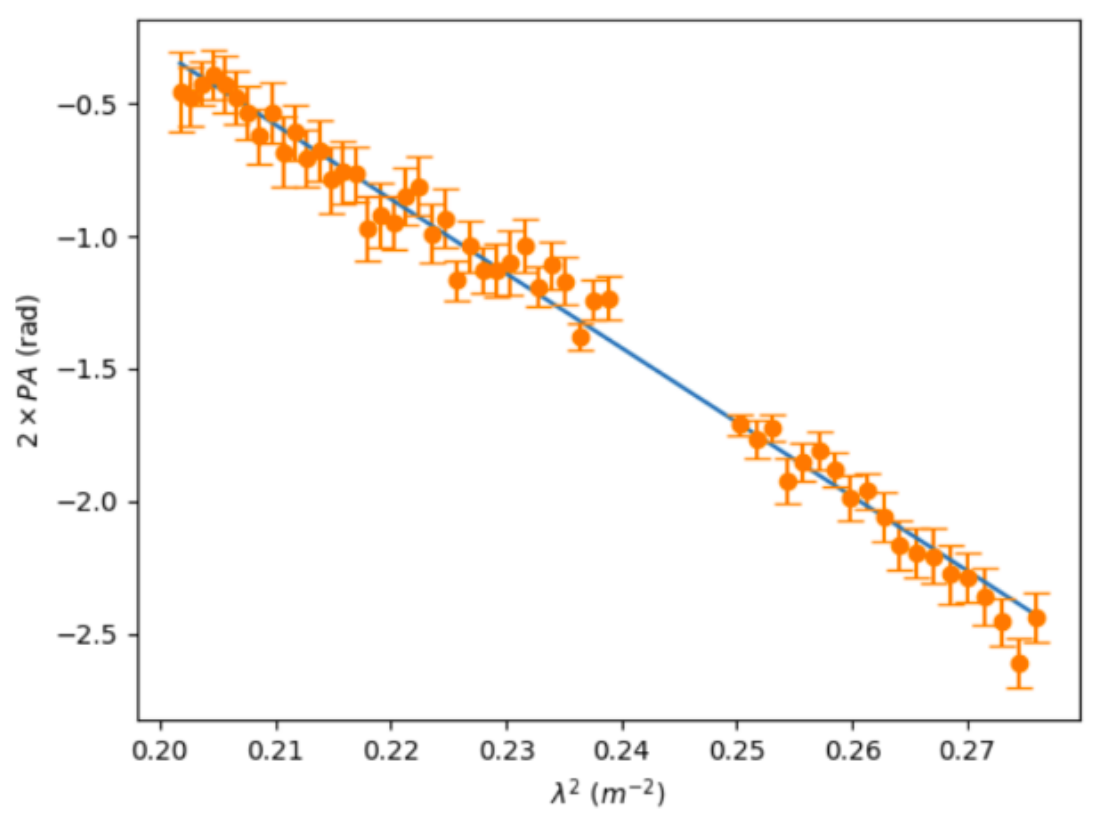}
\caption{RM fit to PSR B1702-19,   band 4 USB data on 13 Feb 2021. \label{fig:exp2}}
\end{figure}

\begin{itemize}
\item Pulse averaged source polarisation properties (IAU/IEEE): RM $= -19$ rad/m$^2$, Q = -23\%, U = -10\%, V = +15\%
\item Observations date and settings: 13 Feb 2021; Band 4, LO I : 550 MHz LO, USB
\item Polarisation leakage and position angle calibration was done using 2355+498 and 3C 286 respectively.
\item Post calibration and post RM fit results:
\begin{enumerate}
\item RMfit = $-14$ rad/m$^2$
\item  $2\chi_0$ ($\chi_0$ = Intrinsic position angle) = 124 degree (Figure \ref{fig:exp2}).
\item Stokes Q $= -$ve, Stokes $U = +$ve, Stokes $V = -$ve.
\item Result: Stokes U and V flipped, and Stokes Q remained the same.
\end{enumerate}
\end{itemize}

\vspace*{0.5cm}
\noindent{\bf \large{PSR B0136+57,   band 4 LSB}}\\

\begin{figure}
\centering
\includegraphics[width=0.45\textwidth]{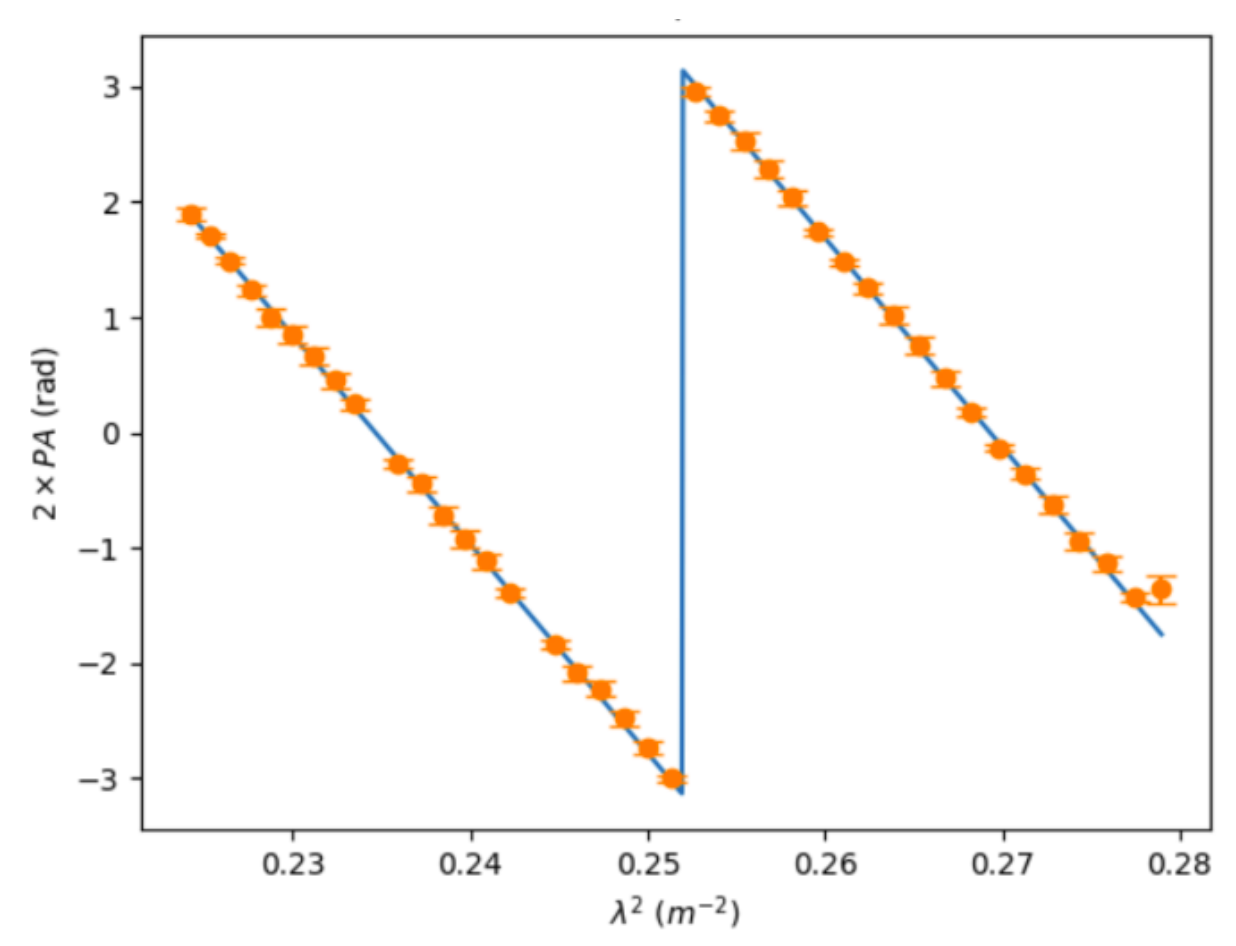}
\caption{RM fit to PSR B0136+57, band 4 LSB on 19 Mar 2021. \label{fig:exp3}}
\end{figure}

\begin{itemize}
\item Pulse averaged source polarisation properties (IAU/IEEE):  RM $= -94$ rad/m$^2$, Stokes Q = -27\%, U = -64\%, V = -16\%
\item Observations date and settings: 19 Mar 2021; Band 4, LO I : 650 MHz LO, LSB
\item Polarisation leakage and position angle calibration was done using 3C48 and 3C138 respectively.
\item Post calibration and post RM fit results:
\begin{enumerate}
\item RMfit = $-90$ rad/m$^2$
\item  $2\chi_0$ ($\chi_0$ = Intrinsic position angle) = 102 degree (Figure \ref{fig:exp3}).
\item Stokes Q $= -$ve, Stokes $U = +$ve, Stokes $V = +$ve.
\item Result: Stokes U and V flipped, and Stokes Q remained the same.
\end{enumerate}
\end{itemize}

\vspace*{0.5cm}
\noindent{\bf \large{PSR B0136+57,   band 3 LSB}}\\

\begin{figure}[h]
\centering
\includegraphics[width=0.45\textwidth]{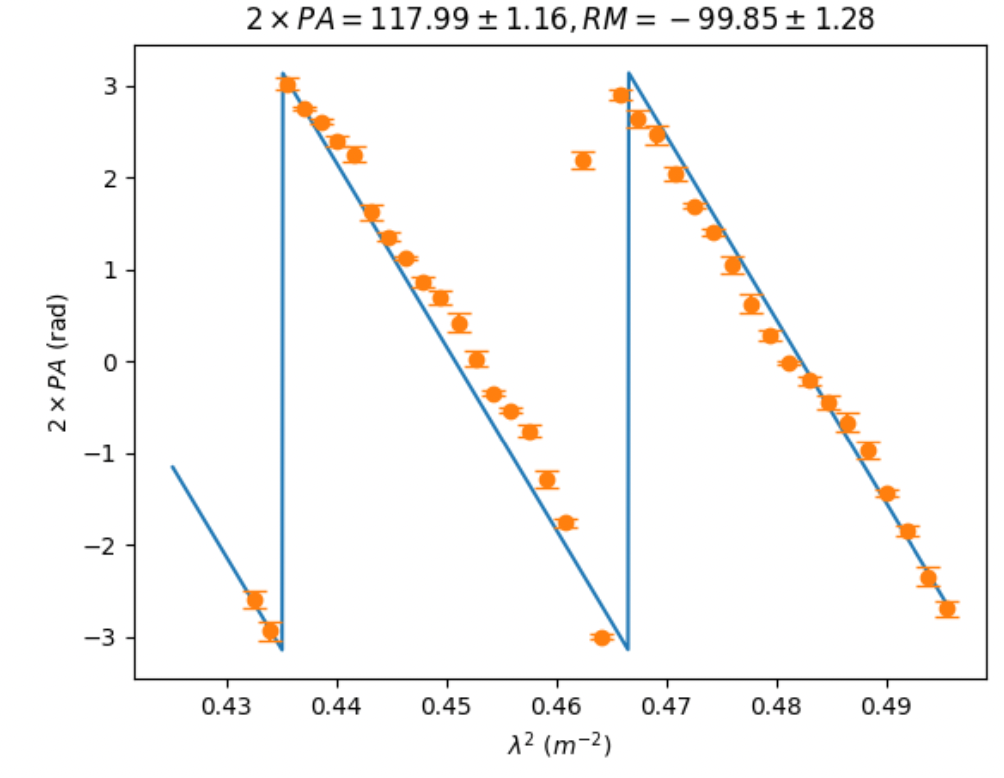}
\caption{RM fit to PSR B0136+57, band 3 LSB data on 31 May 2021. \label{fig:exp4}}
\end{figure}

\begin{itemize}
\item Pulse averaged source polarisation properties (IAU/IEEE):  RM $= -94$ rad/m$^2$, Stokes Q = -62\%, U = -26\%, V = -5\%
\item Observations date and settings: 31 May 2021; Band 3,  LO I : 500 MHz LO, LSB, 200 MHz BW
\item Polarisation leakage and position angle calibration was done using 3C 48 and 3C 138 respectively.
\item Post calibration and post RM fit results:
\begin{enumerate}
\item RMfit = $-99$ rad/m$^2$
\item  $2\chi_0$ ($\chi_0$ = Intrinsic position angle) = 117 degree (Figure \ref{fig:exp4}).
\item Stokes Q $= -$ve, Stokes $U = +$ve, Stokes $V = +$ve.
\item Result: Stokes U and V flipped, and Stokes Q remained the same.
\end{enumerate}
\end{itemize}

%Thus
Hence we performed experiments on two PSRs in four different epochs and using USB and LSB at both band 3 and band 4. In all three experiments, results are similar, after performing polarisation calibration and RM correction. Stokes $Q$ sign remained the same, and Stokes $U$ and $V$ signs flipped. There is no difference between the results from the data taken using LSB and USB in interferometric mode after the proper full polarisation calibration. This result is consistent with a swap of $R$ and $L$, as discussed in Section \ref{subsec:coherent_swap}.
%Thus it suggest the polarisation swap is in R-L.

\begin{figure}
\centering
\includegraphics[width=0.4\textwidth]{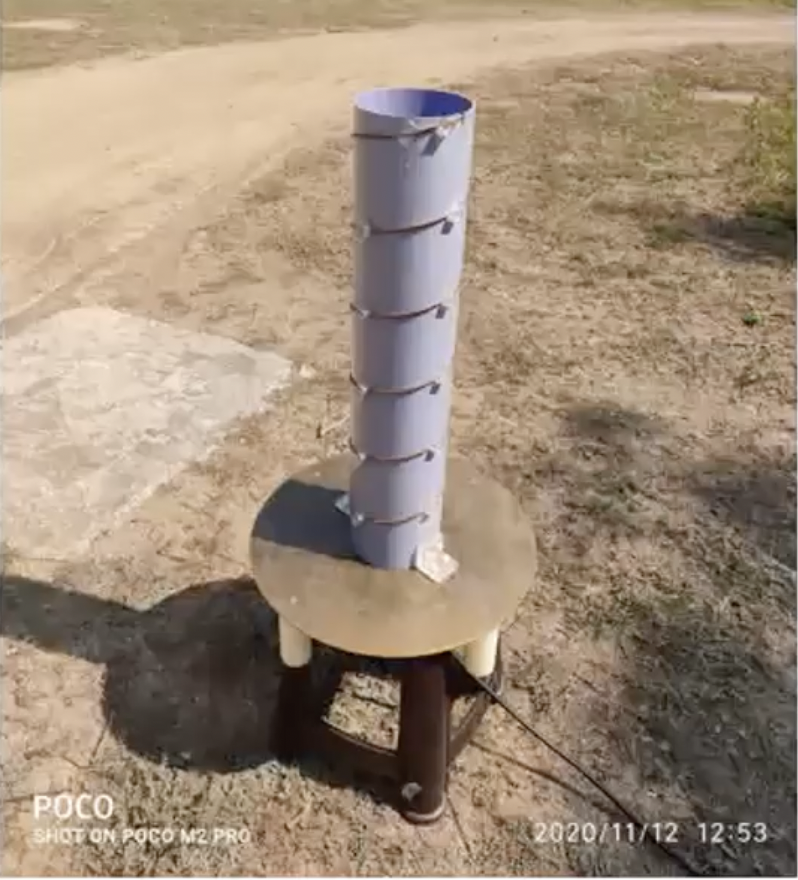}
\includegraphics[width=0.47\textwidth]{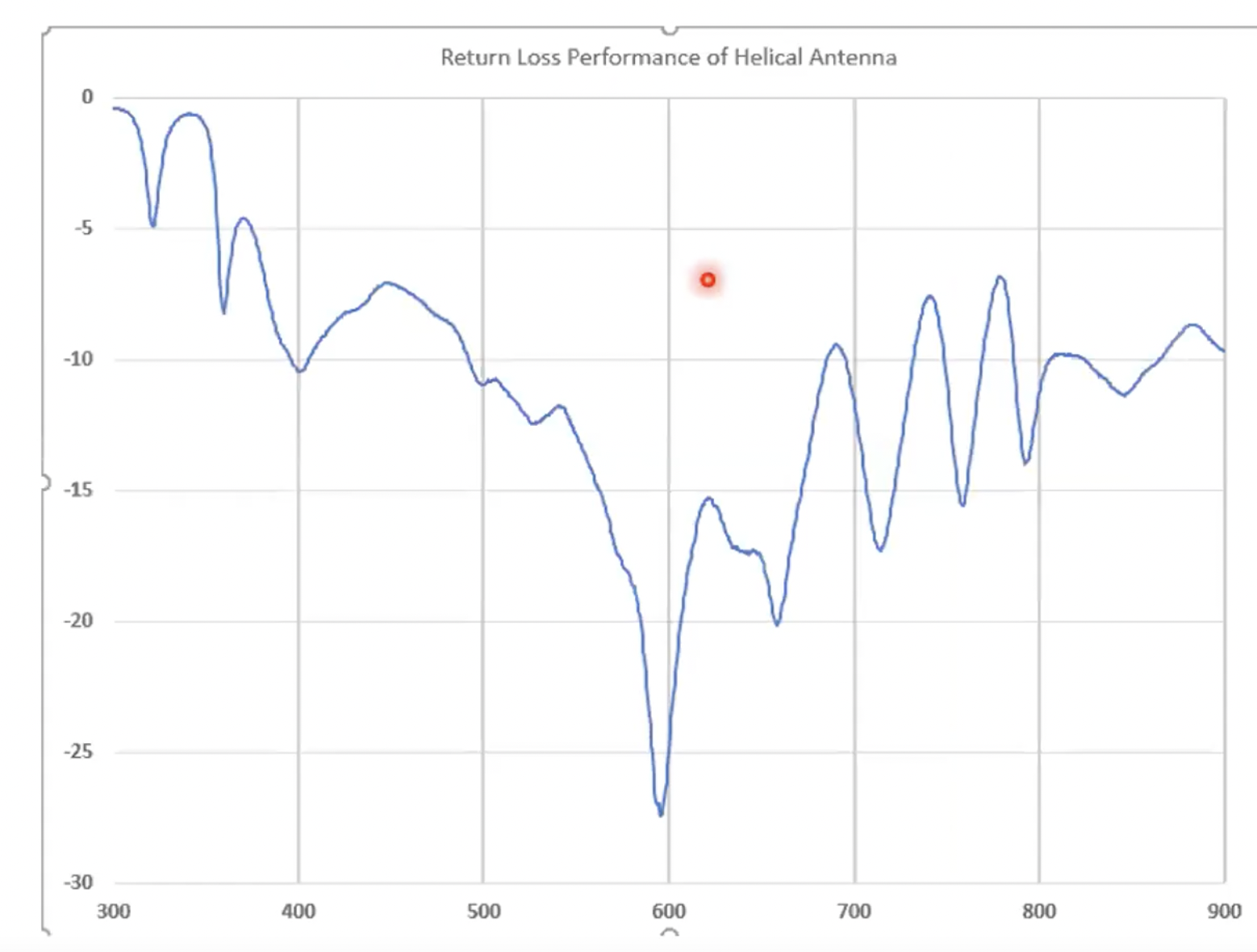}
\caption{A right-handed helical antenna as per IEEE convention.  Return loss performance of the helical antenna. The antenna tunes well within uGMRT band 4. \label{fig:hel}}
\end{figure}

\section{Understanding the effect of the prime focus feed}
So far, all the results have shown that GMRT data can be explained by the swap of $R$ and $L$ except beamformed data in LSB, where polarisation calibration is not done). The swap of $R$ and $L$ is equivalent to the effect of GMRT feeds being the prime focus, i.e. the radiation from the astronomical source falls on the antenna and gets reflected when reaches the feed and in that process, RCP and LCP are interchanged due to reflection. To explicitly check for these effects, we needed to carry out tests with  artificial sources of known polarisation. We built an RCP and an LCP helical antenna in the lab. We also carried out some tests with the satellites of known polarisation.

\subsection{Engineering tests using helical antenna at band 3 and band 4}\label{subsec:helical_ant}
A helical antenna is in the shape of a corkscrew that produces radiation along the axis of the helix antenna. The helical antenna can produce circularly polarised fields. In November 2020 an axial mode helical antenna was built in the lab (Figure \ref{fig:hel}). In a design of a helical antenna, the radiation is fed from below via the wire in the image. The direction of the coil determines the sense of circular polarisation. As per this image, according to IEEE standards, this is a right-hand circularly polarised helical antenna.

\begin{figure}
\centering
\includegraphics[width=0.43\textwidth]{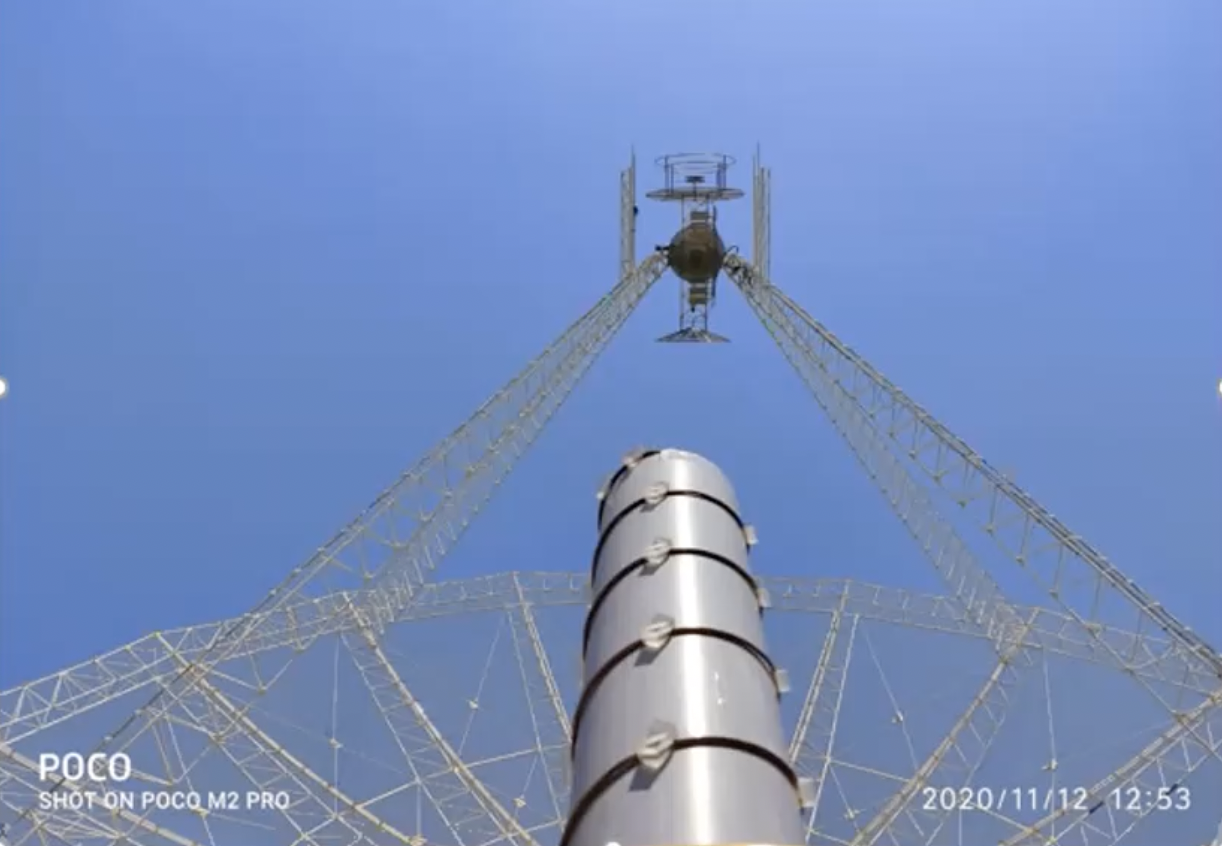}
\includegraphics[width=0.43\textwidth]{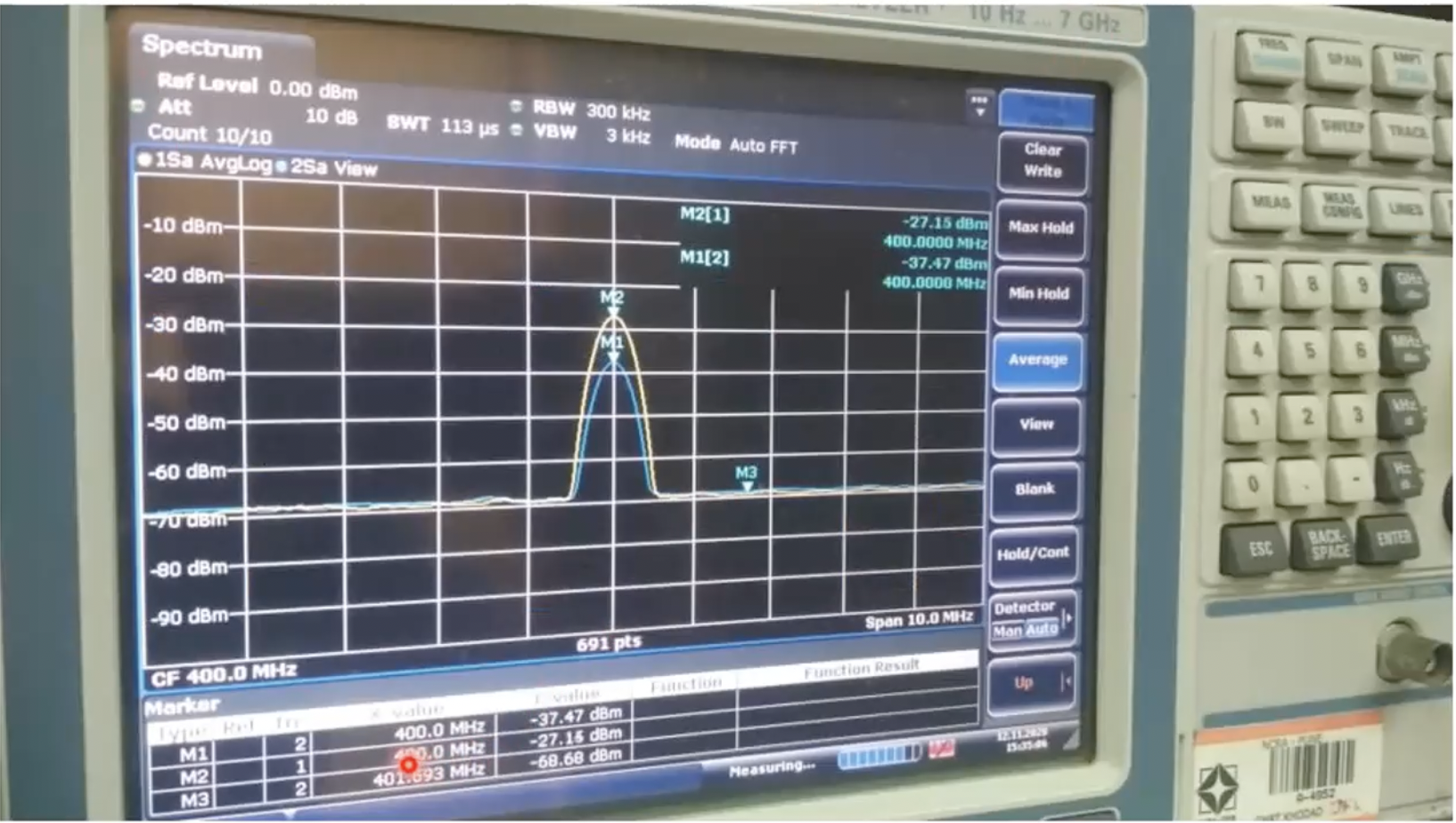}
\includegraphics[width=0.86\textwidth]{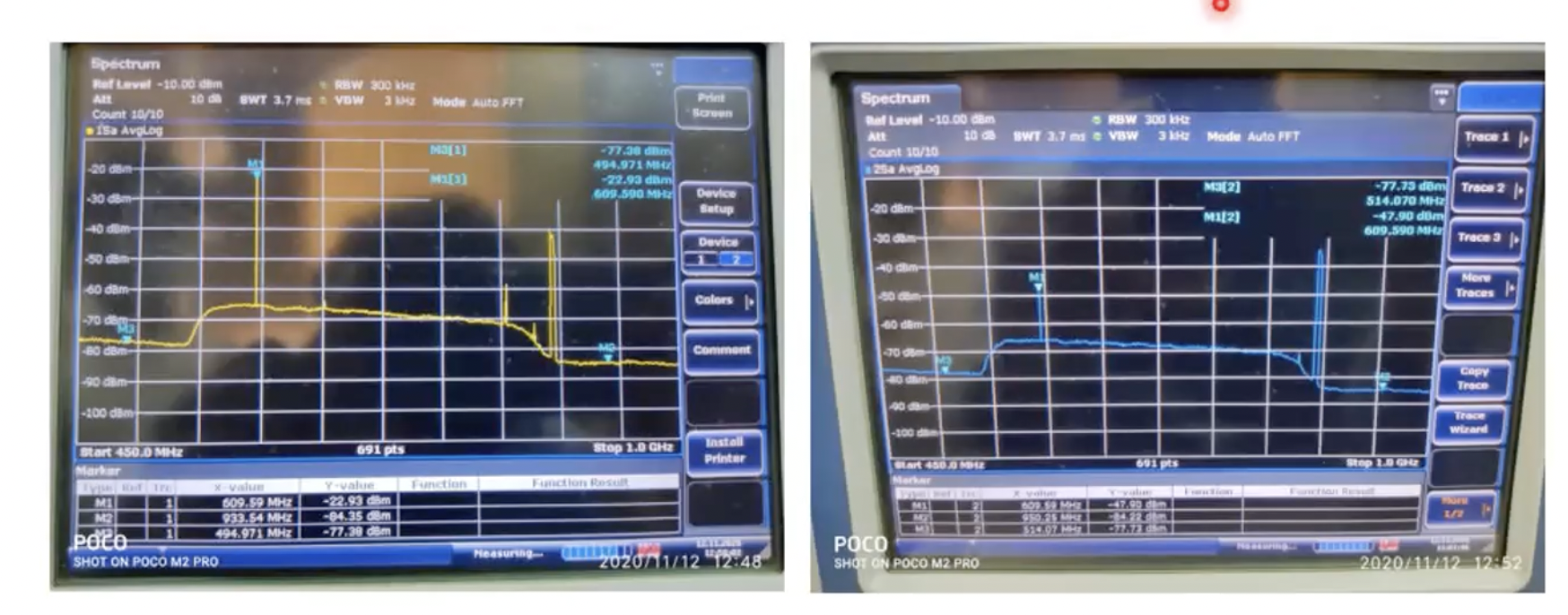}
\caption{{\it Top Left:} The test setup with the RHCP helical antenna as per IEEE convention.  {\it Top right:} Results of band 3 set up. The yellow curve is channel 1 and the blue curve is channel 2. {\it Bottom left and right:} Left one is channel 1 and the right one is channel 2. Higher power seen in channel 1\label{fig:hel2}}
\end{figure}

The antenna was designed for 585 MHz and it was found to be tuning well between $480-700$ MHz, i.e. in the uGMRT band 4  (Figure \ref{fig:hel}). In this test, we brought the GMRT antenna to the lowest possible elevation and focused the band 4 feed to the ground  (Figure \ref{fig:hel2}). The RCP helical antenna was placed on the ground in the line of sight of the feed. Note that the GMRT parabolic dish is not used here, the radiation directly goes to the feed, and there is no reflection of the radiation from the dish. We radiated the EM signal at 610 MHz, and we have seen a strong signal of -22.8 dB in channel 1 and a weaker signature of -47.9 dB in channel 2. The power difference between the two channels is 26 dB, which is quite significant. This ensures that the  channel 1 is RCP when the dish is not involved in the experiment.

Since the helical antenna was also mildly tuning to 400 MHz, we carried the same test in band 3 as well. In Figure \ref{fig:hel2}, channel 1 (yellow curve) receives higher power at $-27.2$ dB, and channel 2 receives a lower power at $-37.5$ dB. This also ensures that for band 3 as well, channel 1 is RCP when the dish is not involved in the experiment.

We also built an LCP helical antenna and carried out the same test. In this case, channel 2 showed higher power in both band 3 and band 4. \textbf{This engineering test ensures} channel 1 is RCP and channel 2 is LCP,  when dish is not involved in the experiment.

\subsection{Engineering test using dipole antenna at band 5}\label{subsec:dipole_ant}
We also attempted to test channels 1 and 2 for band 5 for which the polarisation channels are linear. For this, we used a newly designed cross-dipole feed. The setup is the same as that for the helical antenna. We first brought the antenna dish to the lowest elevation. We then pointed band 5 feed towards the ground and put a linearly polarised horn antenna in the line of sight on the ground.  

\begin{table}
\caption{Test results with the cross dipole feeds at band 5 \label{fig:horn}}
\begin{tabular}{lcc}
\hline
Radiated power& FE output Channel 1 & FE output Channel 2\\
(--40 dBm) polarisation & (dBm) & (dBm)\\
\hline
Vertical & $-18.1$ & $-38.2$\\
Horizontal &  $-38.8$ & $-16.9$\\
\hline
\end{tabular}
\label{table_band5}
\end{table}

When we radiate vertical polarisation via the vertical probe of the horn antenna and terminate the horizontal polarisation by placing 50 $\Omega$ resister, channel 1 gets higher power. Now we rotate the element and radiate through the horizontal probe and terminate the vertical component by putting 50 $\Omega$ resister, the channel 2 gets higher power (Table \ref{table_band5}). This means channel 1 is X and channel 2 is Y.

\vspace{0.5cm}
\noindent In both tests, the effect of the reflection due to the GMRT antenna is avoided since the feed is directly receiving radiation. \textbf{These tests also verified that Channel 1 and Channel 2 are labeled corrected from the feed to the final output.}

\section{Tests using satellites}
A satellite is an artificial spacecraft for communication that relays and amplifies telecommunications signals with a transponder. It can be considered as a middleman between a transmitter and a receiver. The satellites are used for television, telephone, radio, internet, and military applications. In satellite telecommunication, a downlink (the link from a satellite down to one or more ground stations or receivers), and an uplink (the link from a ground station up to a satellite) are used.

Many of these satellites are circularly polarised and they work at the GMRT frequency range.
To understand the effect of the antenna dish, we chose circularly polarised satellites with downlink  ranges between bands 2 and 3 frequencies\footnote{Thanks to Ankur, Sanjeet, Pravin, Imran, Gaurav, Technical support staff, NCRA Workshop, and Control Room.} In addition, we carried out
some tests in band 5 as well.

\begin{figure*}
\centering
\includegraphics[width=0.9\textwidth]{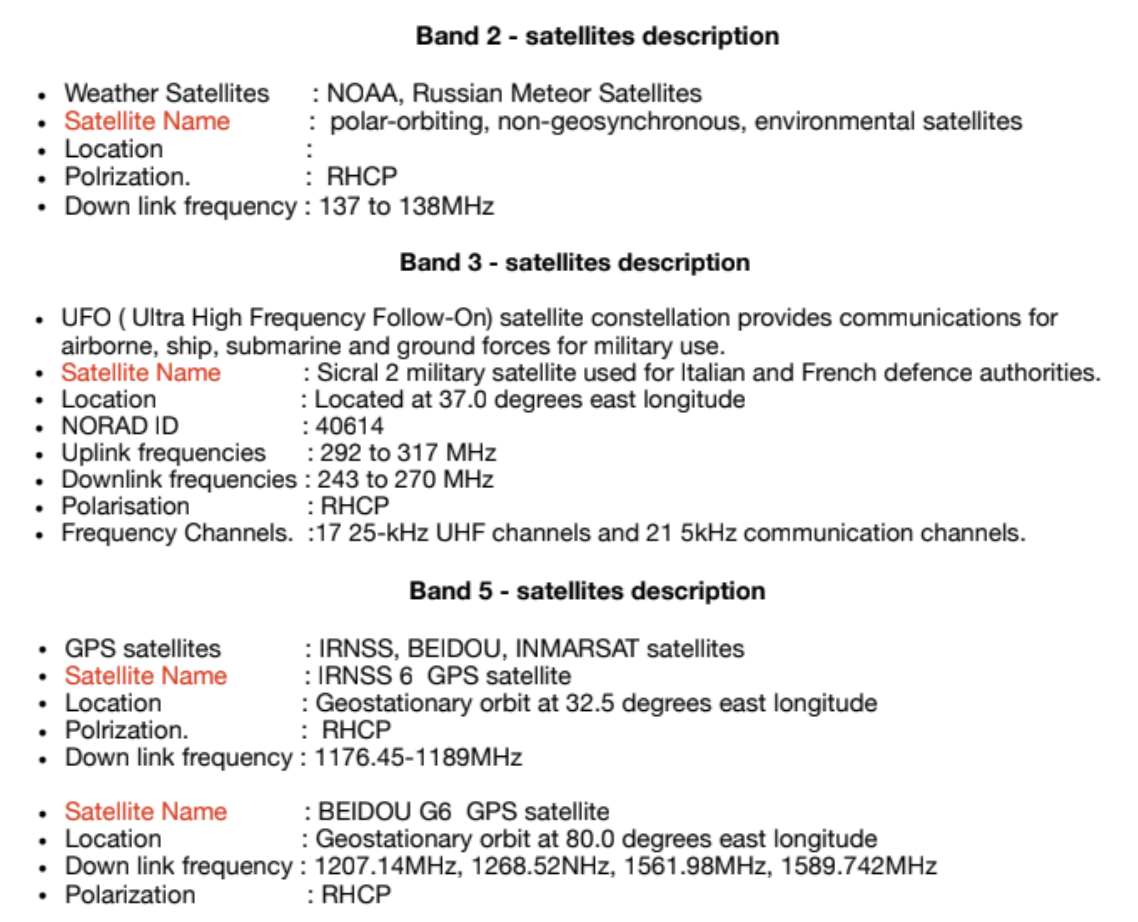}
\caption{Description of satellites used for various tests.}
\end{figure*}

\subsection{Tests at band 2}\label{subsec:sat_band2}
For band 2, we used weather satellites, which are polar-orbiting, non-geosynchronous, environmental satellites such as NOAA, and Russian Meteor satellites. They have a downlink frequency of 137 to 138 MHz and are right-handed circularly polarised (downlink polarisation)\footnote{A downlink is a link from a satellite down to one or more ground stations or receivers, and an uplink is a link from a ground station up to a satellite. }.
Skynet-5B as well as Sicral 1B satellites were used for band 2  tests. 

Skynet -5B satellite uses downlink frequency in the 240-270 MHz band. However, this band is visible in band-2 at the furthest right side of our 120- 250MHz band and hence can be used for band 2 testing. The satellite is RCP (the downlink signal) as per IAU/IEEE convention. 

Sicral 1B is an Italian military communications satellite. It is equipped with a communications package that includes three frequency bands, the extreme-high frequency (EHF), the ultra-high frequency (UHF), and super-high frequency (SHF) transponders. The satellite is located at 37.0 degrees east latitude. The uplink and downlink frequencies are 292 - 317 MHz and 243 - 270 MHz, respectively. This is also RCP as per IAU/IEEE convention. The satellite has 17 25-kHz UHF channels and 21 5-KHz communication channels.

\begin{figure}[h]
\centering
\includegraphics[width=0.9\textwidth]{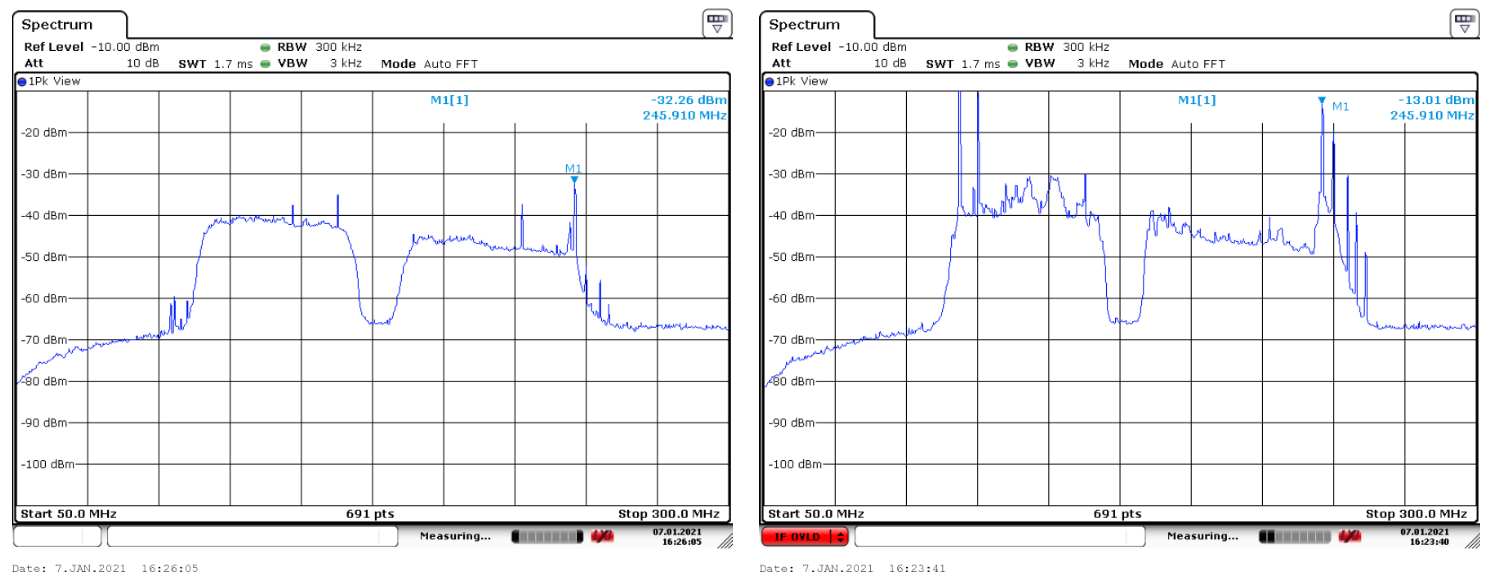}
\includegraphics[width=0.9\textwidth]{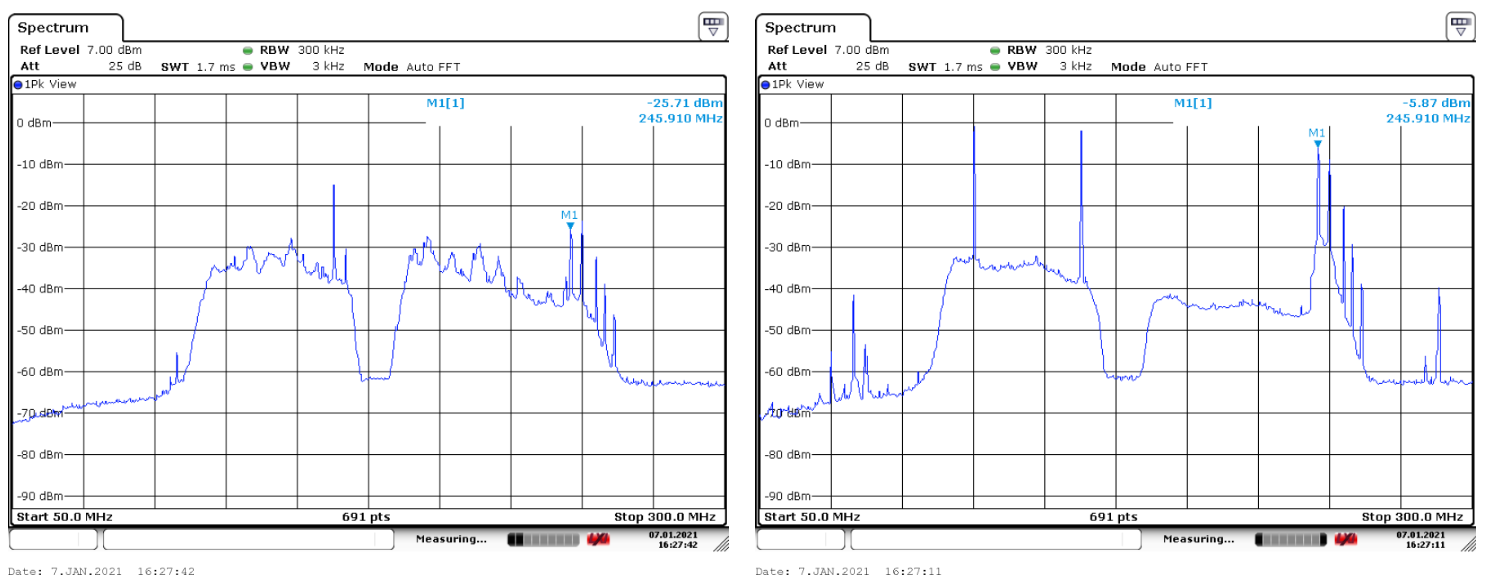}
\includegraphics[width=0.9\textwidth]{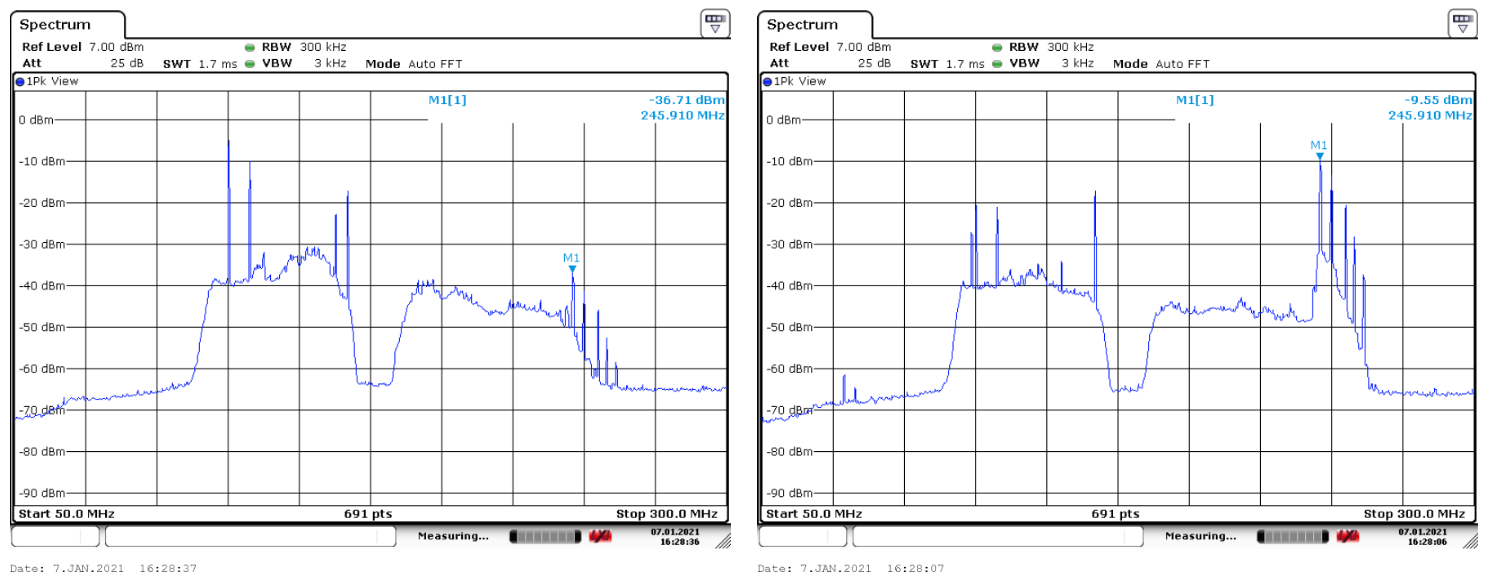}
\caption{Satellite tests at band 2. The top panel is the tests with C11 antenna, the middle panel with E3, and the bottom panel with
S2 \label{fig:b2}}
\end{figure}

These tests were done with the GMRT antennas C11, E03, and S02. The satellites were tracked and the data were recorded at the receiver room and looked at the fibre optics receiver output monitoring port. We note that channel 2 shows high power reception with $\sim -20$ dB power and channel 1 power is down by around 20--27 dB at a level of $-40$ dB. Hence for this RCP satellite, channel 2 receives higher power (Figure \ref{fig:b2}).

\subsection{Test at band 3}
\begin{figure}[h]
\centering
\includegraphics[width=0.45\textwidth]{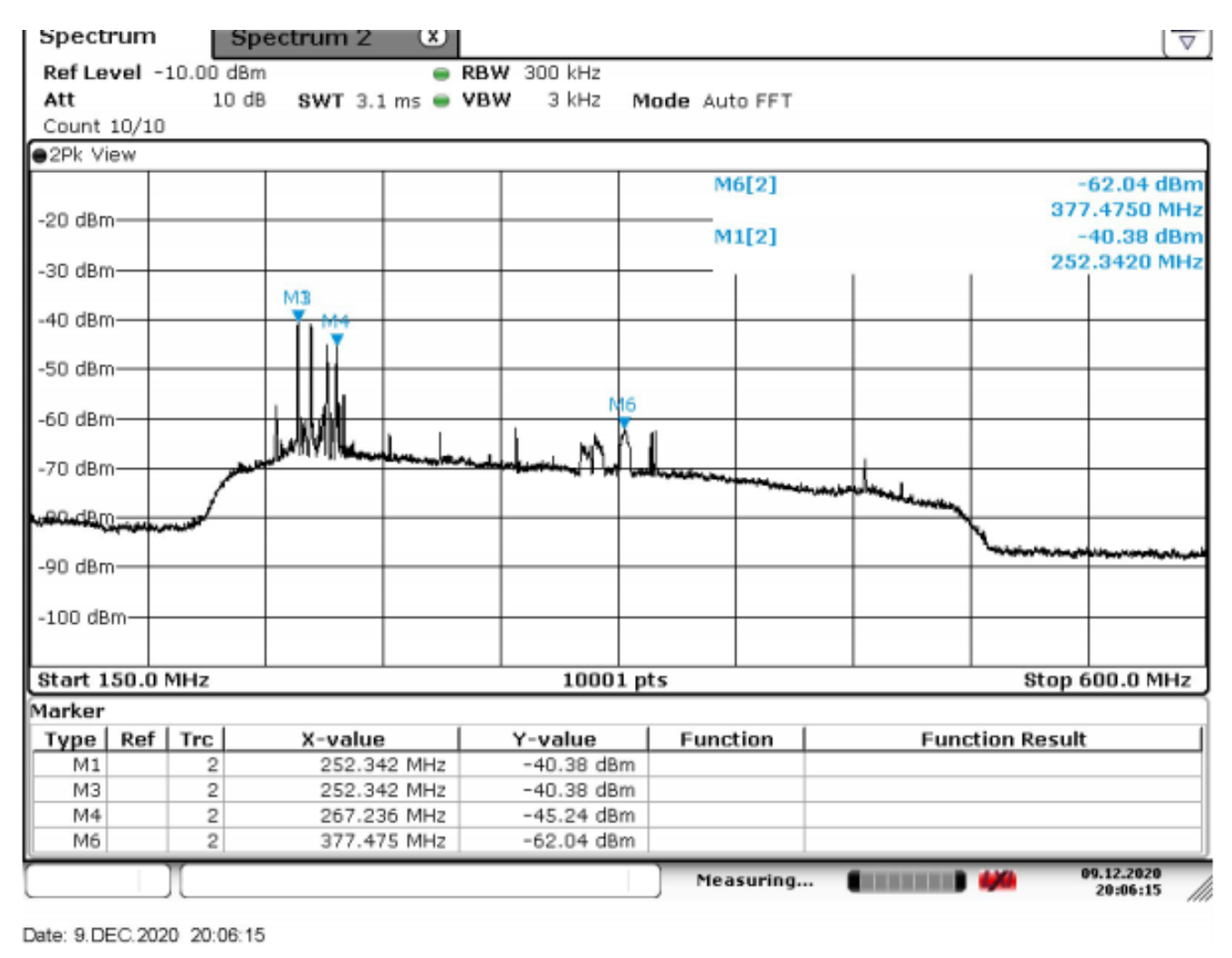}
\includegraphics[width=0.45\textwidth]{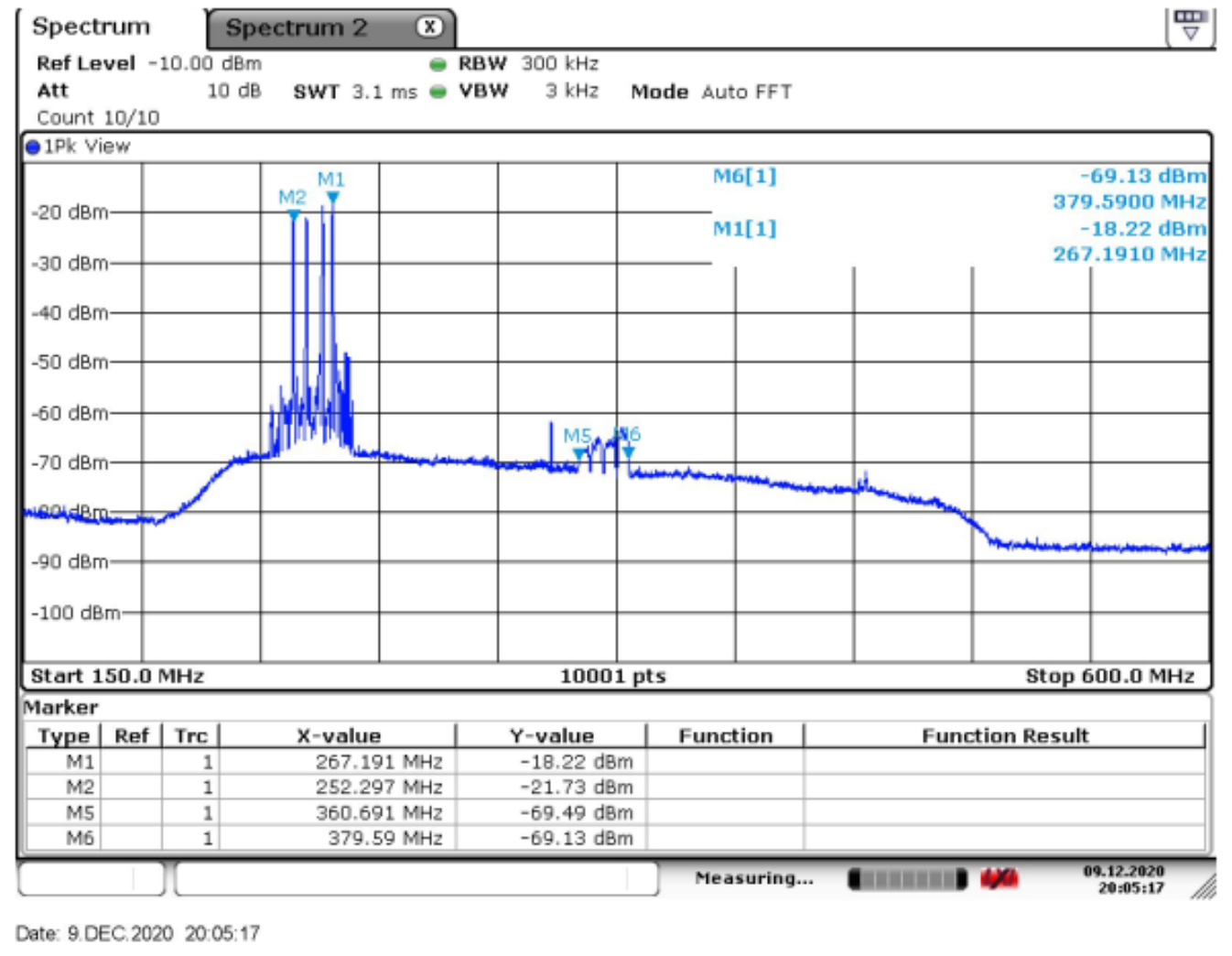}
\includegraphics[width=0.65\textwidth]{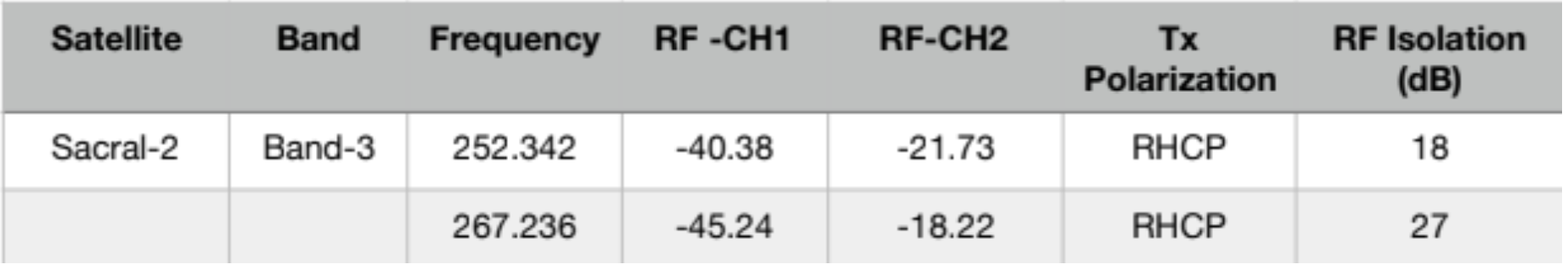}
\caption{Satellite tests at band 3 at frequencies 252.342 and 267.236 MHz. Channel 2 receives higher power in both cases.
 label \label{fig:b3}}
\end{figure}

The test was carried out by Sicral 1B and Sicral 2 satellites. We did the test with C10 antenna and the data were recorded at 252 and 267 MHz frequencies. In both bands, channel 2 received higher power with -21.7 and -18.2 dB power at both frequencies, respectively. Channel 1 received lower power between -40 and -45 dB causing the RF isolation of 18-27 dB, confirming that for this RCP {\color{red}satellite}, channel 2 receives higher power (Figure \ref{fig:b3}). 

\subsection{Tests at band 5}
\begin{figure}
\centering
\includegraphics[width=0.45\textwidth]{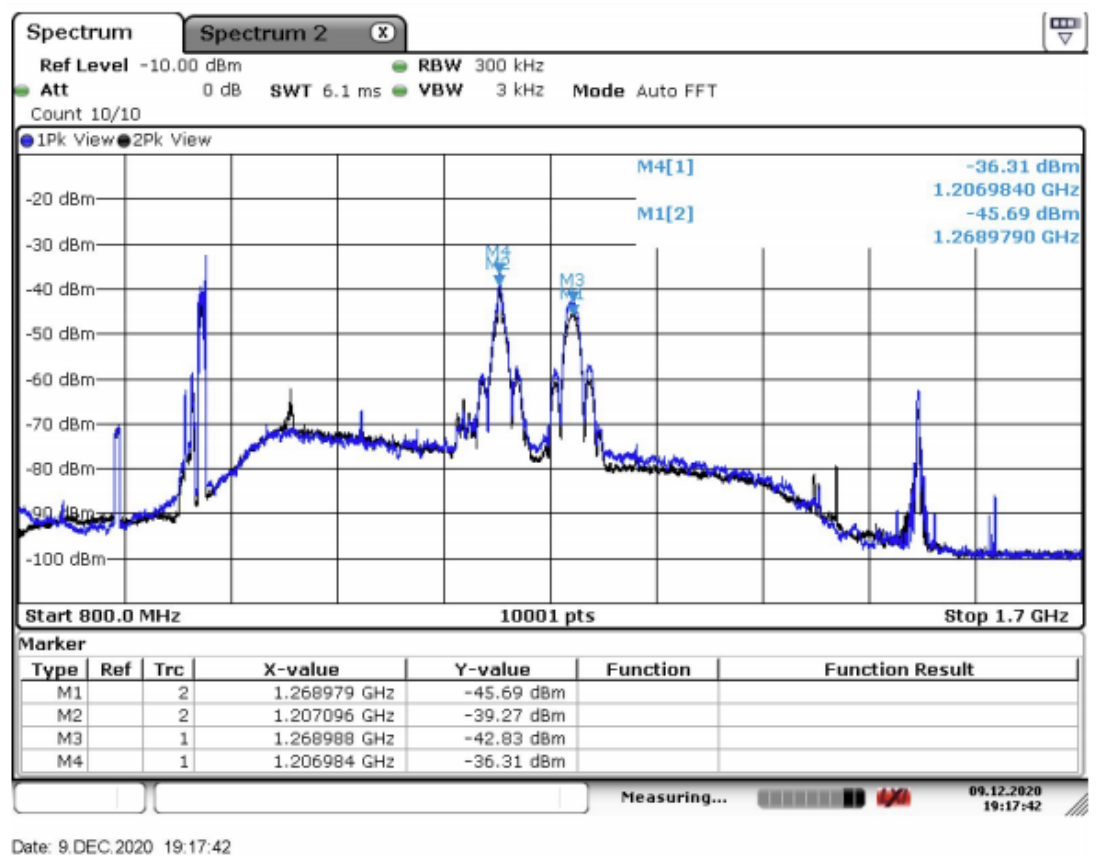}
\includegraphics[width=0.45\textwidth]{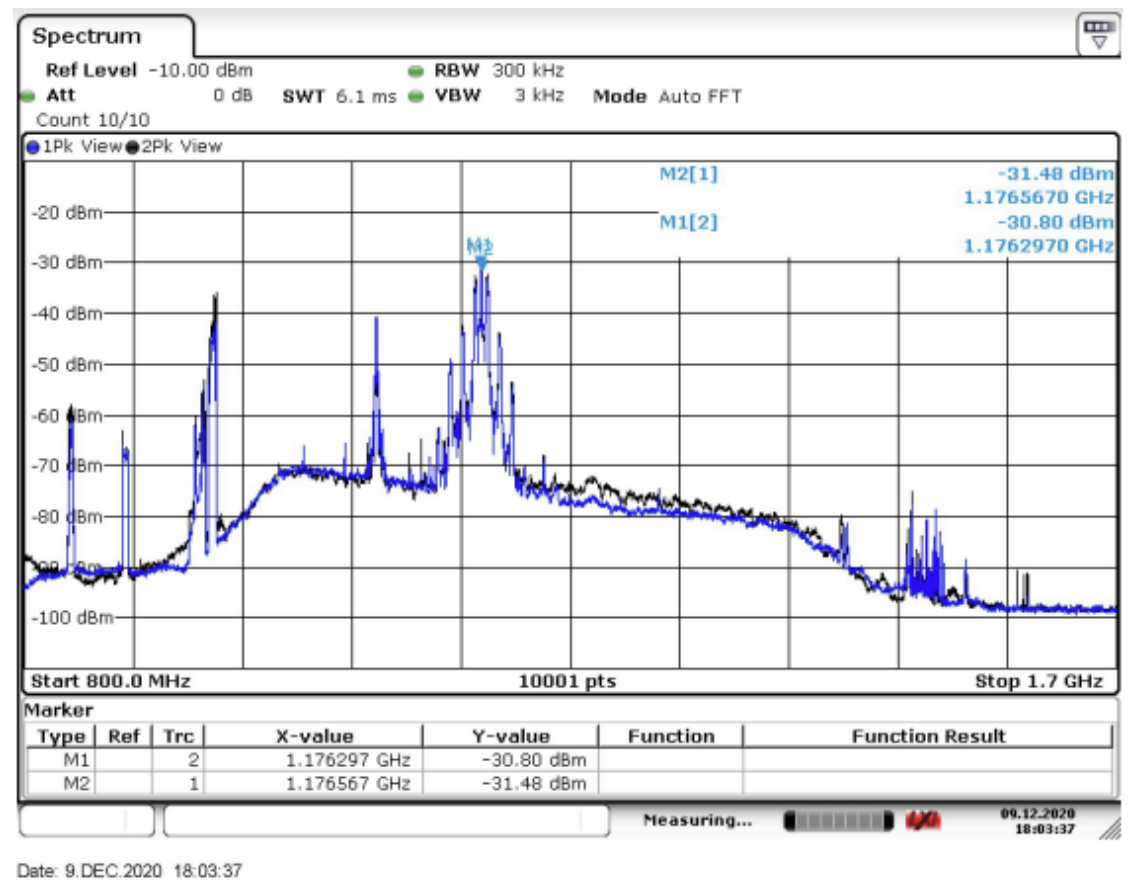}
\includegraphics[width=0.75\textwidth]{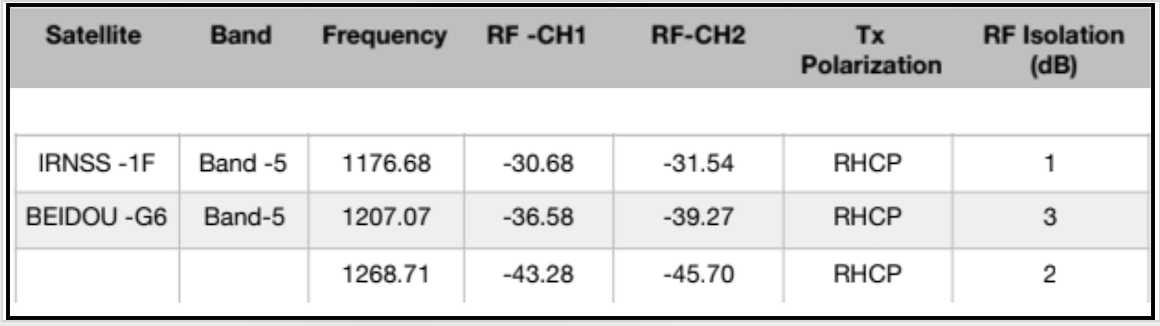}
\caption{Tests with IRNSS-1F and BEIDOU-G6 in band 5. Both channels receive equal power. \label{fig:sat1}}
\end{figure}

For band 5, we used GPS satellites. These were IRNSS and BEIDOU satellites. IRNSS 6 used was at the Geostationary orbit at 32.5 degrees east longitudes. The downlink frequency of the satellite is 
1176.45-1189 MHz. BEIDOU is a G6 satellite, which is 80 degrees east in the geostationary orbit with downlink frequencies of 1207.14 MHz, 1268.52 MHz, 1561.98 MHz, and 1589.742 MHz. Both are RCP satellites as per IAU/IEEE convention. Since GMRT band 5 feeds are dipolar, we should not receive a power difference. This is indeed the case as can be seen in figure and table in \ref{fig:sat1}, both channels receive equal power.

\subsection{Tests with and without the GMRT antenna}
\begin{figure}[h]
\centering
\includegraphics[width=0.9\textwidth]{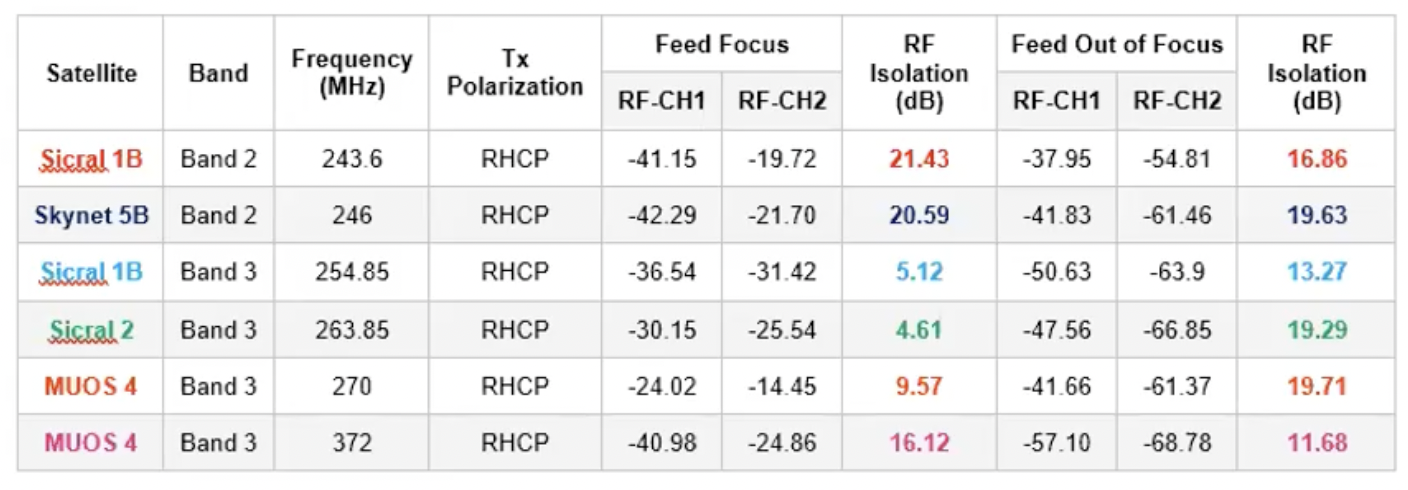}
\caption{Tests with the satellite with the feed in focus (i.e. default setting with radiation first falling onto the antenna and then reflected and received by the feed)   and feed out of focus (i.e. feed directly receiving the radiation from the satellite). \label{fig:sat3}}
\end{figure}

Tests with helical antenna were without the reflection from the antenna dish as the feeds directly received the radiation. The tests showed channel 1 to be RCP and channel 2 to be LCP. To reconfirm the effect of dish, we carried out tests with feed in focus (i.e. default setting with radiation first falling onto the antenna dish and then reflected and received by the feed) and feed out of focus (i.e. feed directly receiving the radiation from the satellite).  We used Sicral 1B and Skynet 5B for band 2, Sicral 1B, Sicral 2, and MUOS 4 for band 3. All of these satellites are RCP in nature. As seen in Table \ref{fig:sat3}, with feed in focus, in all settings, channel 2 receives higher power. However, channel 1 receives higher power when the feed is directly receiving the radiation.

This result suggests that channel 1 receives higher power when the RCP radiation directly falls on the feed. However, when it is reflected from the antenna, it reverses its polarisation and channel 2 receives higher power. This experiment concludes that the reflection at the dish causes a reversal in polarisation between $R$ and $L$, which is consistent with the results from astronomical observation as described in Section \ref{sec:astro_test}.

\section{Conclusions}\label{sec:conclusion}
This brings us to the end of this study. Below are our main takeaway points.
\begin{enumerate}
\item GMRT antenna Azimuth 0 points to South. The H dipole is parallel to the feed-rotating axis which is along the East-West direction, V dipole is perpendicular to the feed-rotating axis, which is along the North-South direction. Hence as per IAU/IEEE convention, H is equivalent to the Y component of the electric field and V is equivalent to the X component of the electric field.

\item Engineering tests have shown that channel 1 receives V polarisation without any phase delay and channel 2 receives V polarisation with 90 degrees phase delay. Similarly, channel 2 receives H polarisation without any phase delay and channel 1 receives H polarisation with 90 degrees phase delay.  This means channel 1 $=V+iH$ and channel 2 $=H+iV$. Translating it into the cartesian coordinate system, Channel 1 is $X+iY$ and channel 2 is $Y+iX$, which according to IAU/IEEE convention are RCP (North component ($X$) of the electric vector leading East component $Y$) and LCP (East component ($Y$) of the electric vector leading North component $X$), respectively.  

\item The above results are also confirmed by the tests performed using the helical antenna.  When this directly radiates towards the feeds, channel 1 receives higher power for the right circularly polarised helical antenna, and channel 2 receives higher power for the left circularly polarised helical antenna. \textbf{This confirms that the Channel 1 is RCP ($R$) and channel 2 is LCP ($L$) from feed to the final data products}. 

\item The observations have revealed that when the data are fully polarised calibrated (flux and phase calibration, bandpass calibration, and polarisation calibration) for pulsars with well-established polarisation properties with known Stokes $I$, $Q$, $U$, $V$, the GMRT Stokes $U$ and $V$ always have a mismatch. This can be explained if $R$ and $L$ are swapped. This solves the incomprehensible result obtained from the initial test (Section \ref{sec:initial}).

\item The helical antenna tests and the lab tests did not involve the reflection of the GMRT antenna. \textbf{The observations involved reflection of the antenna and indicated a swap of $R$ and $L$, which is equivalent to ignoring the effect of reflection, which converts RCP to LCP and vice versa.}

\item To incorporate the effect of reflection by the GMRT antenna, we carried out tests with satellites of known polarisation with feed in focus and feed out of focus. In bands 2 and 3, for RCP satellites, with feed in focus (i.e. radiation first falling on the antenna dish and then reflecting to reach the feed), channel 2 showed higher power and channel 1 showed lower power. With feeds directly receiving the radiation, channel 1 showed higher power for RCP satellites. This is consistent with observations, which were showing channel 2 to receive higher power (which is done with feed in focus). \textbf{This test again establishes the polarisation reversal due to the reflection by the antenna dish.}

\item All of these results (tests with astronomical source in interferometric mode, tests using helical antennas, tests using satellites keeping the feed in focus to the dish and directly to the satellite) are consistent with each other. 

\textbf{The study indeed proves that GMRT channels 1 and 2 are true $R$ and $L$. However, since GMRT is a prime focus instrument, the reflection due to the dish reverses the sense of polarisation and converts RCP into LCP and vice versa. This has not been taken into account and because of that reason, a right circularly polarised source behaves like a left circularly polarised source upon reflection and produces higher power in channel 2.}

\item Since astronomical observations are always done with feed in focus ((i.e. radiation first falling on the antenna dish and then reflected to reach the feed), one needs to reassign channel 1 to $L$ and channel 2 to $R$ for all circular feeds, i.e. bands 2, 3 and 4 (however, see next section for caveats in band 2) to obtain correct polarisation of the source.
\end{enumerate}

\section{Caveats and future studies}
\begin{enumerate}
\item While there is no difference in the interferometric data whether it is taken in the USB mode vs LSB mode,  and consistent with the engineering tests, the beamformed data does not show this result. In beam data, Stokes $U$ (thus position angle) are consistently inverted between USB vs LSB data. This needs to be understood. As of now, parallactic angle corrected beam data in the USB mode with $R$ and $L$ swapped will give the correct results. One thing that needs to be noted here is that like the interferometric analysis, no robust polarisation calibration is performed for beamformed data.

\item The above tests have been done only with bands 3 and 4. The uGMRT band 2 polarisation calibration is still in progress. Once that is done, the observational tests have to be done in bands 2 as well\footnote{Preeti Kharb and the team is working on the band 2 calibration}. However, the engineering and satellite tests have shown that the band 2 behavior is the same as bands 3 and 4. Hence even without the observational confirmation, one can successfully apply our results.

\item We have no conclusive tests for band 5 which have dipolar feeds. These tests have to be carried out separately and determined for band 5. Hence the results in band 5 remain inconclusive. Engineering tests have shown that band 5 channels 1 and 2 are $X$ and $Y$. The calculations show that the reflection effect due to GMRT being a prime focus instrument must result in $X$ and $Y$ becoming $X$ and $-Y$. Its effects on Stokes parameters need to be tested via  astronomical observations and polarisation calibrations.
\end{enumerate}

{\bf Our main conclusion is GMRT Stokes V and U signs need to be reversed by the observer to make them consistent with the IAU/IEEE convention. The cause of this is R \& L swap incurred by the reflection suffered by sky signals at the antenna dish before entering the feed at the prime focus. This objective can be achieved by reassigning channel 1 to $L$ and channel 2 to $R$ for all circular feeds, i.e. bands 2, 3, and 4. The study  remains inconclusive for band 5, which has linear feeds.}

\section*{Acknowledgements}
We thank the staff of the GMRT that made these observations possible. The GMRT is run by the National Centre for Radio Astrophysics of the Tata Institute of Fundamental Research. We thank Yashwant Gupta for his useful insights. B.D. thanks, Minhajur Rahaman for suggesting the pulsar B1702--19 to use as the polarisation standard. B.D. and P.C. thank Jayaram N. Chengalur for useful discussions. S.K. thanks Kadaladi Pavankumar for helping with beam polarisation data. P.C. thanks, Dipanjan Mitra for clarifying some important polarisation concepts. We thank the GMRT engineering team for help on multiple occasions. We especially acknowledge Ankur, Pravin Raybole, Sanjeet Rai, Imran Khan, Gaurav Parikh and Vilas Bhalerao for the engineering tests carried out in this study. This research has used NASA's Astrophysics Data system.

\bibliographystyle{mnras}

\end{document}